\documentclass[]{book}
\newcommand*{\INCLUDEMETRIC}{}
\newcommand*{\INCLUDEAPPS}{}

\usepackage[]{graphicx} 
\usepackage{mathrsfs}
 \expandafter\let\csname equation*\endcsname\relax
 \expandafter\let\csname endequation*\endcsname\relax 
\usepackage{amsmath}
\usepackage{amsthm}
\usepackage{amssymb}
\usepackage{alphalph}
\usepackage{cancel}
\usepackage[svgnames,dvipsnames,hyperref]{xcolor}
\usepackage[hyphens]{url}
\usepackage[hyperfootnotes=true]{hyperref}
\usepackage{breakurl}
\hypersetup{colorlinks=true, linktoc=page}
\hypersetup{citecolor=RoyalBlue, linkcolor=cyan, urlcolor=PineGreen, filecolor=green}
\usepackage[noadjust]{cite}
\usepackage{diagrams}
\usepackage{longtable}
\usepackage{fancyhdr}
\renewcommand{\chaptermark}[1]{\markboth{\thechapter.\ #1}{}}

\newenvironment{abstract}{ \chapter*{\centering \large Abstract} }{}
\renewenvironment{quote}
 {\list{}{\rightmargin\leftmargin}%
 \item\relax\em}
 {\endlist}
\renewcommand{\thechapter}{\Roman{chapter}}
\renewcommand{\thesection}{\arabic{section}}

\newcounter{savesection}
\newcounter{savefigure}
\newcommand{\newchapter}[1]{
\setcounter{savesection}{\value{section}}
\setcounter{savefigure}{\value{figure}}
\chapter{#1}\chaptermark{#1}
\setcounter{section}{\value{savesection}}
\setcounter{figure}{\value{savefigure}}
}
\newtheoremstyle{break}  % follow `plain` defaults but change HEADSPACE.
  {\topsep}   % ABOVESPACE
  {\topsep}   % BELOWSPACE
  {\upshape}  % BODYFONT
  {0pt}       % INDENT (empty value is the same as 0pt)
  {\textit} % HEADFONT
  {.}         % HEADPUNCT
  {5pt plus 1pt minus 1pt}  % HEADSPACE. `plain` default: {5pt plus 1pt minus 1pt}
  {}          % CUSTOM-HEAD-SPEC
\theoremstyle{break}
\newtheorem{para}{}[section]
\newtheorem{parastar}[para]{*\!\!}
\newcommand{\labcount}[1]{\newcounter{#1}\setcounter{#1}{\value{section}}\newcounter{#1_sub}\setcounter{#1_sub}{\value{subsection}}}
\newcommand{\group}{G} %{\mathcal{G}}
\newcommand{\groupderiv}{\partial_\group}
\newcommand{\G}{\group} %!TEX encoding = UTF-8 Unicode{\mathcal{T}}
\newcommand{\g}{g} %{t}
\newcommand{\timederiv}{{\partial}_\G}
\newcommand{\MC}{\df{\theta}_\n{\scriptscriptstyle MC}}
\newcommand{\wMC}{\vec{w}_\n{\scriptscriptstyle MC}}
\newcommand{\genMC}{\Wedge{\df{\theta}}_\n{\scriptscriptstyle MC}}
\newcommand{\genwMC}{\Wedge{\vec{w}}_\n{\scriptscriptstyle MC}}
\newcommand{\lapse}{N}
\newcommand{\shift}{\vec{N}}
\newcommand{\shiftform}{\df{\nu}}
\newcommand{\vdual}{\df{\upsilon}}
\newcommand{\Wedge}{\mathsf{\Lambda}\hspace{0.5pt}}
\newcommand{\cont}[1]{\n{i}_{#1}}
\newcommand{\mult}[1]{\n{e}_{#1}}
\newcommand{\lcont}[1]{\n{i}(#1)}
\newcommand{\lmult}[1]{\n{{e}}(#1)}
\newcommand{\signop}{\n{n}}

\newcommand{\bp}{\begin{para}\rm}
\newcommand{\bpstar}{\begin{parastar}\rm}
\newcommand{\bph}[1]{\bp{\it #1.\hspace{4pt}}\rm }
\newcommand{\bphstar}[1]{\bpstar{\it #1.\hspace{4pt}}\rm }
\newcommand{\ep}{\end{para}}
\newcommand{\epstar}{\end{parastar}}

\renewcommand{\vec}[1]{\mathbf{#1}}
\newcommand{\n}[1]{{\mathrm{#1}}}
\newcommand{\hhat}[1]{{\displaystyle\mathop{\vspace{-24pt}#1}^{\scriptscriptstyle\Delta}}\,\!}
\newcommand{\tw}[1]{#1\textwidth}

\newcommand{\issue}{\color{red}}
\newcommand{\dfs}[1]{{{\cal F}^{#1}}}
\newcommand{\ves}[1]{{{\cal X}^{#1}}}
\newcommand{\df}[1]{{\boldsymbol{#1}}}
\newcommand{\lie}[1]{\mathcal{L}_{#1}}
\newcommand{\pd}{\n{pd}}
\newcommand{\pdsys}[1]{#1}
\newfont{\ar}{cmss10 scaled 1000}%font for \pdim
\newfont{\sar}{cmss10 scaled 800}%font for \oned
\newfont{\ssar}{cmss10 scaled 800}%font for \pdim
\newfont{\sssar}{cmss10 scaled 600}%font for \oned
\newcommand{\pdim}[1]{\mbox{\ar #1}}%physical dimension
\newcommand{\oned}{\pdim{1}_{\mbox{\sar D}}}%neutral element of multiplicative dimension module
\renewcommand{\arraystretch}{1.25}
\usepackage[english]{babel}
\newcommand{\Xspace}{\mathcal{A}} %{\mbox{\textswab{U}}}
\newcommand{\Yspace}{\mathcal{B}} %{\mbox{\textswab{V}}}
\newcommand{\Zspace}{\mathcal{C}} %{\mbox{\textswab{W}}}
\newcommand{\xgen}{\vec{a}} %{\mbox{\textswab{u}}}
\newcommand{\ygen}{\vec{b}} %{\mbox{\textswab{v}}}
\newcommand{\zgen}{\vec{c}} %{\mbox{\textswab{w}}}
\newcommand{\affineform}{\df{\Psi}} %{\mathcal{A}}
\newcommand{\coordtime}{t}
\newcommand{\specialwedge}{\barwedge}
\newcommand{\twist}{\times}
\newcommand{\prototimetwist}{{+}}
\newcommand{\timetwist}{{\scriptscriptstyle\prototimetwist}}
\newcommand{\protodoubletwist}{\protect{\times\!\prototimetwist}}
\newcommand{\doubletwist}{\protect{\times\!\timetwist}}
\newcommand{\curv}{{\mathscr C}}
\newcommand{\surf}{{\mathscr A}}
\newcommand{\vol}{{\mathscr V}}
\newcommand{\interval}{{\mathscr H}}
\newcommand{\atest}{k}
\newcommand{\btest}{\ell}
\newcommand{\lapseprod}{\xi}
\newcommand{\proxy}{\n{P}}
\begin{document}
\frontmatter
\title{Observers and Splitting Structures in Relativistic Electrodynamics}
\author{Bernhard Auchmann\footnote{CERN, TE-MPE, CH-1211 Geneva 23, Switzerland,\newline E-mail:~\textsf{bernhard.auchmann@cern.ch}.}, Stefan Kurz\footnote{Robert Bosch GmbH, Corporate Sector Research and Advance Engineering,
$\;\;\;\;\;\;\;\;\;\;$D-70465 Stuttgart, Germany, \newline E-mail:~\textsf{stefan.kurz2@de.bosch.com}.}}
%\address{$^1$ CERN, TE-MPE, CH-1211 Geneva 23, Switzerland}
%\address{$^2$ Tampere University of Technology, Department of Electrical Engineering -- Electromagnetics, P.O.~Box 692, FI-33101 Tampere, Finland}
%\eads{\mailto{bernhard.auchmann@cern.ch}, \mailto{stefan.kurz@tut.fi}}
%\makeatletter\@openrightfalse
%\hypersetup{linkcolor=Crimson}
%%%%%%%%%%%%%%%%%%%%%%%%%%%%%%%%%%%%%%%%%%%%%%
%%%%%%%%%%%%%%%%%%%%%%%%%%%%%%%%%%%%%%%%%%%%%%
\maketitle 

%\part{A Frame- and Coordinate-Free Relativistic Splitting}
\begin{abstract}
We introduce a relativistic splitting structure as a means to map fields and equations of electromagnetism from curved four-dimensional space-time to three-dimensional obser\-ver's space. We focus on a minimal set of mathematical structures that are directly motivated by the language of the physical theory. Space-time, world-lines, time translation, space platforms, and time synchronization all find their mathematical counterparts. The splitting structure is defined without recourse to coordinates or frames. This is noteworthy since, in much of the prevalent literature, observers are identified with adapted coordinates and frames. Among the benefits of the approach is a concise and insightful classification of splitting structures that is juxtaposed to a classification of observers. The application of the framework to the Ehrenfest paradox and Schiff's ``Question in General Relativity'' further illustrates the advantages of the framework, enabling a compact, yet profound analysis of the problems at hand.
\end{abstract}
%\pacs{53Z05, 78A25, 83C50}
\tableofcontents
\mainmatter
\newchapter{Introduction} 
\section{Goals and context}
We use the term observer in a literal way, a person or entity experiencing electro\-magnetic phenomena, making measurements or doing experiments. A relativistic splitting structure, or short splitting, is a mathematical structure that allows to
\begin{enumerate}
\item relate electromagnetic quantities in space-time to measurable quantities in the three-dimensional observer's space;
\item formulate initial-value problems in order to explain or predict electromagnetic phenomena.
\end{enumerate}
In {\sc R. Geroch}'s 1985 book ``Mathematical Physics'' \cite[p.~1]{geroch} we read:
\begin{quote}\sloppy What one often tries to do in mathematics is to isolate some given structure for concentrated, individual study: what constructions, what results, what definitions, what relationships are available in the presence of a certain mathematical structure - and only that structure? But this is exactly the sort of thing that can be useful in physics, for, in a given physical application, some particular mathematical structure becomes available naturally, namely, that which arises from the physics of the problem. Thus mathematics can serve to provide a framework within which one deals only with quantities of physical significance, ignoring other, irrelevant things. [...] Such a body of knowledge, once established, can then be called upon whenever it makes contact with the physics.\end{quote}
What is the structure that arises naturally from the theoretical understanding of relativistic observers? We propose that \cite{Fecko1997, Giulini2006, Guzman2003, Kocik1998, Minguzzi2003, Nouri-Zonoz2003} 
\begin{itemize}
\item space-time is a four-dimensional differentiable manifold with Lorentzian metric; 
\item a space-time that is densely filled with world-lines is a {\em fiber bundle}; 
\item time translation motivates to use the {\em 1-D Lie group of translations} for the typical fiber in the above fiber bundle;
\item an observer's space is modeled by the bundle's base manifold; 
\item a time synchronization is modeled by a section of the bundle; 
\item and the splitting of space-time into spatial and temporal subspaces is an {\em Ehresmann connection}. 
\end{itemize}
In the spirit of the above quote, we make an abstraction and study what constructions, results, definitions, and relationships are available in the presence of {\em an Ehresmann connection on a fiber bundle with fibers that are diffeomorphic to a Lie group}. The definition of the splitting structure ensues. In this paper we develop the theory for the Abelian Lie group of 1-dimensional translations, that is, time translations. A generalization to arbitrary, non-Abelian Lie groups is available for the entire pre-metric framework of Chapter~\ref{sec:premetric}. The generalized framework is, however, not within the scope of this work. We only mention the fact here to motivate the relativistic splitting structure as an instance of a more general mathematical theory of splitting structures.

The formulation of the splitting uses neither frames nor coordinates, even though in much of the prevalent literature observers and their splittings are identified with adapted coordinates and frames. We provide a translation of the major concepts in Section~\ref{amoreadapt}. In {\sc C.~Misner, K.~Thorne, and J.~Wheeler}'s 1973 book ``Gravitation" \cite[Box 3.2]{misner}, a mathematical language without frames, coordinates, and index calculus is called `geometric'. In this sense, the presented splitting is geometric. Moreover, we privilege exterior calculus over tensor calculus, seek to employ metric only where it is essential, and assign a physical dimension to physical quantities. In contradistinction to the prevalent literature, all observed physical fields are modeled on the observer's three-dimensional space, rather than in four-dimensional space-time. We believe that the presented model subsumes, characterizes, and extends various approaches to space-time splitting.
The rigorous distinction between metric and pre-metric concepts and the focus on coordinate- and frame-free descriptions has practical relevance for the formulation of numerical schemes. Grid-based techniques naturally distinguish between topological information, encoded in the connectivity of the grid, and metric information, that enters the formulation in the constitutive relations \cite{Auchmann2010,Stern:2014fr}.

We do not attempt to give a complete historical overview of splitting techniques and observer models, referring the reader, for example, to \cite{Jantzen2012}. Instead, we give an account of the papers and books that have had a direct influence on our work, in chronological order: 
In {\sc B. Mashhoon}'s 1990 paper \cite{Mashhoon1990} and in the 2003 continuation \cite{Mashhoon2003} we have found an instructive discussion of the hypothesis of locality. %It motivates the statement that only the threading observer model produces measurable physics. 
Some of the terminology in the observer classification is taken from {\sc T.~Matolcsi}'s 1993 book \cite{Matolcsi}, see also \cite{Matolcsi1998}.
Of great consequence was the reading of {\sc M.~Fecko}'s 1997 paper \cite{Fecko1997} where a coordinate-free splitting is introduced based on only the four-velocity and its metric Riesz dual in the spirit of the theory of connections on principal bundles. The formal links to that theory are discussed in the paper's Appendix~G: if the four-velocity is a complete vector field, then it generates an action of a Lie group and turns space-time into a principal bundle. Four-velocities are generally incomplete in the presence of space-time singularities; see \cite[Ch.~8]{hawking}. Moreover, the Ehresmann connection induced by the Riesz dual of four-velocity is not principal, in general.
{\sc J.~Kocik} uses in the 1998 paper \cite{Kocik1998} Ehresmann connections on fiber bundles and the Fr\"olicher-Nijenhuis bracket to formalize the splitting framework of M. Fecko and to discuss its properties. He makes a point of the usefulness of Ehresmann connections in mathematical physics. 
%
%The definition of a physical dimension system is due to {\sc A.~Prechtl}, 2001 \cite{prechtldim}. He is also the author of comprehensive lecture notes on relativistic electrodynamics, in which we found instructive the discussion of different definitions of energy-momentum tensors \cite{Prechtl2007}.
%
The 2002 paper by {\sc G.~Rizzi} and {\sc M.L.~Ruggiero} \cite{Rizzi2002} defines the observer metric on the base manifold of a fiber bundle. From their (editors) 2004 book \cite{Rizzi2004}, in partiular {\sc O. Gron}'s contribution \cite{gron}, we have learned a great deal on the Ehrenfest paradox and the history of debates on the topic of rotating observers that have been ongoing up to our time.
{\sc J.~J.~Cruz~Guzm\'an} and {\sc Z.~Oziewicz} in 2003 \cite{Guzman2003} revisit Kocik's 1998 work. The paper provides a rigorous algebraic perspective and discusses nonorthogonal splittings.
{\sc E.~Minguzzi} in the 2003 paper \cite{Minguzzi2003} employs an Ehresmann connection on a principal bundle in his frame-based approach and points out that the mathematics involved is closely related to that of gauge theories. 
The 2003 book by {\sc F.~Hehl} and {\sc Y.~Obukhov} \cite{hehlbook} has deepened our conviction of the necessity to distinguish between pre-metric and metric concepts in a theory. It is one of the few works that consider physical dimensions of mathematical objects throughout. Our paragraphs on the splitting of the energy-momentum balance are largely based on this book. 
The definition of space-time is taken from {\sc E.~Minguzzi} 2008 \cite{Minguzzi2008}.
The 2012 monograph by {\sc R.T.~Jantzen, P.~Carini,} and {\sc D.~Bini} \cite{Jantzen2012} gives a valuable historical overview of splitting techniques. Our discussion of kinematic parameters is based on this work. 
A comprehensive list of references was compiled by {\sc D.~Bini and R.T.~Jantzen} in \cite{references}.\par
The paper is self-contained. Several standard textbook definitions, as those of fiber bundles, connections, and the Lie algebra, are repeated in a compact form in order to have at our disposal all the tools that come with them, for example, transition functions, Maurer-Cartan form, fundamental field map, etc. Paragraphs that present standard textbook material are indicated by a star. 
As for our mathematical references, the main sources are {\sc Y.~Choquet-Bruhat, C.~DeWitt-Morette,} and {\sc M.~Dillard-Bleick}, ``Analysis, Manifolds, and Physics'' \cite{amp}, and {\sc I. Kol{\'a}{\v r}, P.~W. Michor,} and {\sc J. Slov{\'a}k}, ``Natural Operations in Differential Geometry'' \cite{kolar}. For algebraic concerns we consulted {\sc W.H.~Greub} \cite{greub1,greub2}.
%
%\newpage
%%%%%%%%%%%%%%%%%%%%%%%%%%%%%%%%%%%%%%%%%%%%%%
\section{Notation, conventions, multi-linear algebra}\label{notation}\labcount{notation}
\bphstar{Differential forms, multi-vector fields, and vector-valued fields}\label{fieldsforms}
We denote $\ves{k}(M)$ and $\dfs{k}(M)$ the smooth $k$-vector fields and differential $k$-forms on a smooth differentiable manifold $M$, respectively. $\mathcal{F}^k_\twist(M)$ denotes the twisted differential $k$-forms on $M$; see \ref{twisted}. Let $V$ denote a finite-dimensional vector space. $\ves{k}(M;V)=\ves{k}(M)\otimes V$ and $\dfs{k}(M;V)=\dfs{k}(M)\otimes V$ denote multi-vector fields and forms that take values in $V$. For a more compact notation in the algebraic operations of \ref{algebra} we use the direct-sum spaces $\ves{}(M)=\bigoplus_{k=0}^{n}\ves{k}(M)$ and $\dfs{}(M)=\bigoplus_{k=0}^{n}\dfs{k}(M)$, $n=\dim(M)$.
\epstar
\bphstar{Twisted differential forms}\label{twisted}
Let $M$ be an orientable $n$-dimensional manifold. A twisted differential form $\df{\gamma}_\twist\in\mathcal{F}^k_\twist(M)$ can be represented as a pair $(\df{\gamma},\df{\kappa})\in\dfs{k}(M)\times\dfs{n}(M)$, where $\df{\kappa}$ is an everywhere nonzero {\em volume form} on $M$. There is an equivalence relation
\begin{align*}
(\df{\gamma}^\prime,\df{\kappa}^\prime)\sim(\df{\gamma},\df{\kappa})\;:\;\df{\kappa}^\prime=\lambda\,\df{\kappa},\;\df{\gamma}^\prime=\n{sgn}(\lambda)\,\df{\gamma},\;\quad\lambda\in C^\infty(M);
\end{align*}
see \cite[Ch.~3]{Bossavit1991} and \cite[Ch.~28]{burke}. Twisted forms can be pulled back under diffeomorphisms, $\varphi^*\df{\gamma}_\twist=(\varphi^*\df{\gamma},\varphi^*\df{\kappa})$, compare with \ref{extend}. Similar considerations hold for multi-vector fields. The product of two twisted objects yields an ordinary object, by selecting representatives with the same orientation.
\epstar
\bphstar{Sign operator}\label{signop}
The sign operator 
\begin{align*}\signop:\xgen\mapsto (-1)^{\n{deg}(\xgen)}\xgen\end{align*}
is borrowed from \cite{rham}, with $\n{deg}(\xgen)$ the degree of a multi-vector field or differential form $\xgen$.
\ifdefined\LONGVERSION
Another sign operator is introduced for (co)vector-valued fields $w$ in $\ves{}(\,\cdot\,;\Wedge^\ell\mathfrak{g})$ or $\dfs{}(\,\cdot\,;\Wedge^\ell\mathfrak{g})$
\begin{align*}\n{m}:w\mapsto (-1)^\ell w.\end{align*}
\fi
\epstar
\bphstar{Interior, and exterior products}\label{intext}
The generalized interior product or con\-traction by multi-vector fields follows from the standard contraction by vector fields via
\begin{align*}\cont{\vec{a}\wedge\vec{b}}\df{\gamma}=\cont{\vec{b}}\cont{\vec{a}}\df{\gamma}\end{align*}
with $\vec{a}\in\ves{1}(M)$, $\vec{b}\in\ves{k}(M)$, $\df{\gamma}\in\dfs{\ell}(M)$, $\ell\geq k+1$, \cite[5.14.]{greub2}. For convenience, we also use the notation $\lcont{\vec{a}\wedge\vec{b}}\,\df{\gamma}$. Following \cite{Fecko1997,kolar}, we use an exterior product operator\footnote{In \cite{Kurz2009a} we have used the letter `j' for the exterior product.}
\begin{align*} \mult{\vec{a}}\vec{b}=\vec{a}\wedge\vec{b}.\end{align*}
\epstar
\bphstar{Products of scalar-, vector-, and tensor-valued fields}\label{algebra}
In items 1-6 below, we extend the basic algebraic operations of duality pairing, the interior product, and the exterior product to vector- and tensor-valued fields; compare with \cite[1.5, 1.16, 1.21]{greub2}. We have operations involving Lie-(co)algebra-valued fields in mind, as they will appear starting from Section~\ref{sec:liealgebra}. For the sake of brevity, the duality pairing, the interior product, and the exterior product are represented by the bilinear operation
\[
f:\Xspace\times\Yspace\to\Zspace,\quad(\xgen,\ygen)\mapsto \zgen=f(\xgen,\ygen),
\]
where $\Xspace$, $\Yspace$, $\Zspace\in\{C^\infty(M),\ves{}(M),\dfs{}(M)\}$. Let $V$, $V^\prime$ denote finite-dimensional vector spaces.% in $\ves{}(M;V)$ and $\dfs{}(M;V)$. %The dot in the notation is a placeholder for either $P$, $\group$, or, in the case of parametric fields, $(X,\group)$. 
%For the vector space we have $V\in\{\mathbb{R},\mathfrak{g},\mathfrak{g}^*,\mathfrak{g}^*\otimes\mathfrak{g}\}$ in mind, where $\mathfrak{g}$ is the Lie algebra, see \ref{liealgebra}.
%, or tensor product. 
\begin{enumerate}
\item\label{algebra1} The product of a scalar-valued and a vector-valued field is a vector-valued field (idem for covector-valued fields),
\[
f:\left\{\begin{aligned}
\Xspace\times(\Yspace\otimes V)\to\Zspace\otimes V,\quad(\xgen,\ygen\otimes\vec{v})&\mapsto \zgen\otimes\vec{v},\\
(\Xspace\otimes V)\times\Yspace\to\Zspace\otimes V,\quad(\xgen\otimes\vec{v},\ygen)&\mapsto \zgen\otimes\vec{v}.
\end{aligned}\right.
\]
\item \label{algebra2}
We define the following duality pairing between a vector-valued field and a covector,
\[
(\Xspace\otimes V)\times V^*\to\Xspace,\quad(\xgen\otimes\vec{v})(\df{\gamma})\mapsto \xgen\otimes\df{\gamma}(\vec{v}).
\]
\item \label{algebra3}For the product of a vector- and a covector-valued field there are two natural possibilities \cite[Eqs.~(3.18), (3.19)]{Yavari2008}. Either we obtain a scalar-valued field by duality pairing 
%\footnote{Recall that the duality pairing $\df{\eta}(\vec{v})$, $\df{\eta}\in\Wedge^kV^*$, $\vec{v}\in\Wedge^\ell V$ is zero for $k\ne\ell$, in contrast to contraction.}
of the covector with the vector \cite[p.~lvi]{frankelbook},
\[
f:(\Xspace\otimes V)\times(\Yspace\otimes V^*)\to\Zspace,\quad(\xgen\otimes\vec{v},\ygen\otimes\df{\gamma})\mapsto \zgen\otimes\df{\gamma}(\vec{v}).
\]
Alternatively, the product may result in a tensor-valued field by taking the tensor product,
\[
f^\otimes:(\Xspace\otimes V)\times(\Yspace\otimes V^*)\to\Zspace\otimes (V\otimes V^*),\quad(\xgen\otimes\vec{v},\ygen\otimes\df{\gamma})\mapsto \zgen\otimes(\vec{v}\otimes\df{\gamma}).
\]
We distinguish the different extensions of $f$ by attaching a superscript $^\otimes$ to tensor-valued product operators.
\item \label{algebra3a}The product of two (co-)vector-valued fields yields a tensor-valued field,
\[
f^\otimes:(\Xspace\otimes V)\times(\Yspace\otimes V^\prime)\to\Zspace\otimes (V\otimes V^\prime),\quad(\xgen\otimes\vec{v},\ygen\otimes\vec{v}^\prime)\mapsto \zgen\otimes(\vec{v}\otimes\vec{v}^\prime).
\]%
\item\label{algebra4}
The product of a $\binom{1}{1}$-tensor-valued field and a (co)vector-valued field is a (co)vector-valued field, which is obtained by the action of the tensor $\vec{t}\in T=V^*\otimes V^\prime$ on the (co)vector,
\[
f:\left\{\begin{aligned}
(\Xspace\otimes T)\times(\Yspace\otimes V)\to\Zspace\otimes V^\prime,\quad(\xgen\otimes\vec{t},\ygen\otimes\vec{v})&\mapsto \zgen\otimes\vec{t}(\vec{v}),\\
(\Xspace\otimes V)\times(\Yspace\otimes T)\to\Zspace\otimes V^\prime,\quad(\xgen\otimes\vec{v},\ygen\otimes\vec{t})&\mapsto \zgen\otimes\vec{t}(\vec{v}).
\end{aligned}\right.
\]%
Likewise, the product of a symmetric $\binom{0}{2}$-tensor-valued field and a vec\-tor-valued field is a covector-valued field, and the product of a symmetric $\binom{2}{0}$-tensor-valued field and a covector-valued field is a vector-valued field.
\item \label{algebra5}Finally, the product of two $\binom{1}{1}$-tensor-valued fields $\vec{s},\vec{t}\in T$ is a $\binom{1}{1}$-tensor-valued field $\vec{s}\circ\vec{t}\in T$, which is defined by contraction, $(\vec{s}\circ\vec{t})(\vec{v},\df{\gamma})=\vec{s}(\vec{v})\bigl(\vec{t}(\df{\gamma})\bigr)$, $(\vec{v},\df{\gamma})\in V\times V^*$. Therefore,
\[
f:\begin{aligned}
(\Xspace\otimes T)\times(\Yspace\otimes T)\to\Zspace\otimes T,\quad(\xgen\otimes\vec{s},\ygen\otimes\vec{t})&\mapsto \zgen\otimes\vec{s}\circ\vec{t}.
\end{aligned}
\]
Note that in general $\vec{s}\circ\vec{t}\ne\vec{t}\circ\vec{s}$.
\end{enumerate}
\epstar
\bphstar{Pullback and pushforward}\label{extend}
Much of our notation is borrowed from \cite{amp}, such as $\varphi'$ and $\varphi^*$ for the push and pull maps of a diffeomorphism $\varphi$. The operations are extended to multivectors $\vec{u}=\vec{u}_1\wedge\dots\wedge\vec{u}_k$ and differential forms $\df{\gamma}=\df{\gamma}^1\wedge\dots\wedge\df{\gamma}^k$ via exterior compound, that is,
\begin{align*}
\varphi'\vec{u}&=\varphi'\vec{u}_1\wedge\dots\wedge\varphi'\vec{u}_k,\\
\varphi^*\df{\gamma}&=\varphi^*\df{\gamma}^1\wedge\dots\wedge\varphi^*\df{\gamma}^k.
\end{align*}
Likewise, the operations are extended to vector-valued fields and forms, $\vec{v}\in V$ for some vector space $V$, by
\begin{align*}
\varphi'(\vec{u}\otimes\vec{v})&=\varphi'\vec{u}\otimes\vec{v},\\
\varphi^*(\df{\gamma}\otimes\vec{v})&=\varphi^*\df{\gamma}\otimes\vec{v}.
\end{align*}
\epstar
%
%Note that we attempt to use the term {\em time} only in contexts that are directly linked to the metric, as in {\em time directed} or {\em proper time}. Other common uses as in {\em relative time manifold} (axis), or {\em time coordinate} are avoided, since we view these objects as metric-free.
%
%%%%%%%%%%%%%%%%%%%%%%%%%%%%%%%%%%%%%%%%%%%%%%
\section{Preliminaries}
We give a brief overview of the physical dimension system, as well as our working definitions of space-time and world-lines. The latter topics are grouped in this section as they invoke the metric which will not be mentioned explicitly in the following chapter.
\bph{Physical dimension system}\label{pdsyst}
Let $\pdsys{D}$ denote the system of physical dimensions \cite{fleischmann}, and $\pdsys{S}$ the set of physical quantities. Each element of $\pdsys{D}$ represents a type of physical quantity. The minimum structure of $\pdsys{D}$ is that of a multiplicative Abelian group. The associative and commutative group operation is $\pdsys{D}\times\pdsys{D}\to\pdsys{D}:(\pdim{A},\pdim{B})\mapsto\pdim{AB}$, the dimension product. The neutral element reads $\oned$, and the inverse of an element $\pdim{D}\in\pdsys{D}$ is $\pdim{D}^{-1}$. The multiplicative Abelian group naturally includes exponentiation by integers.\footnote{Since the Abelian group does not comprise all prevalent physical-dimension systems, the structure is generalized in the (unfortunately not publicly available) treatise \cite{prechtldim} to a multiplicative module over a ring $R$, where $R$ may be $\mathbb{Z}$, $\mathbb{Q}$, or $\mathbb{R}$, and $R=\mathbb{Z}$ is the Abelian-group case. The operation $R\times\pdsys{D}\to\pdsys{D}$ is given by exponentiation $(q,\pdim{D})\mapsto \pdim{D}^q$. This proposal unifies the Abelian-group approach of \cite{fleischmann} with the multiplicative vector-space approach found, for example, in \cite{carlson}. For an example where the integer numbers are not sufficient, we mention the cgs system, which requires rational exponents.} $r$ denotes the rank of the group. A basis of $\pdsys{D}$ is a set $\pdsys{B}(\pdsys{D})=\{\pdim{D}_1,\pdim{D}_2,\dots,\pdim{D}_r\}\subset\pdsys{D}$ with the property that
\begin{align*}\pdim{D}_1^{q_1}\pdim{D}_2^{q_2}\dots\pdim{D}_r^{q_r}=\oned\quad\Rightarrow\quad q_1=q_2=\dots =q_r=0,\end{align*}
and that
\begin{align*}\n{span}\bigl({\pdsys{B}}(\pdsys{D})\bigr)=\{\pdim{D}_1^{q_1}\pdim{D}_2^{q_2}\dots\pdim{D}_r^{q_r}\,|\,q_i\in\mathbb{Z}\}=\pdsys{D}.\end{align*}
For the purposes of this paper, we follow the IEC/ISO standard \cite{iso}, and use for a basis
\begin{align*} \pdsys{B}(\pdsys{D})=\{\pdim{L, T, M, I}\},\end{align*}
that is, length, time, mass, and electric current.\footnote{Frequently the physical dimension of charge is used as a basis element in favor of current \cite{sommerfeld, hehlbook}.} We also use the derived physical dimensions of electric voltage $\pdim{U}=\pdim{M}\pdim{L}^2\pdim{T}^{-3}\pdim{I}^{-1}$ and action $\pdim{A}=\pdim{U}\pdim{I}\pdim{T}^2=\pdim{M}\pdim{L}^2\pdim{T}^{-1}$. The map $\pd:\pdsys{S}\to\pdsys{D}$ is called the {\em physical-dimension map}.\footnote{The physical-dimension map is sometimes denoted by brackets, for example, in \cite{hehlbook,post,schouten}. Other texts use curly braces for numerical values and brackets for units \cite{wallot}.} A set with constant dimension map is said to be of homogeneous dimension.
%
%It is generally neither surjective nor injective. The physical dimension system is an integral part of the algebraic structures which are used to model physical quantities. For example, an Abelian additive group or a vector space used to model physical quantities must be of homogeneous physical dimension, whereas a field (in the algebraic sense) is necessarily of neutral dimension. Note that a physical dimension system makes no assumptions on the units that may be used to express numerical data. 
\ep
\bph{Algebraic structures of physical quantities}\label{algstrucphys}
Sets of physical quantities are ususally equipped with algebraic structures. For example, consider a structure $(S,+,D,\pd)$, consisting of a set of quantities $S$, a rule of addition $+:S\times S\to S, (x,y)\mapsto x+y$, a system of physical dimensions $D$, and a dimension map. The structure is called an {\em additive group of quantities} if 
\begin{enumerate}
\item $(S,+)$ is an Abelian group and 
\item the dimension map is constant.
\end{enumerate}
Additive groups of quantities are, by definition, of homogeneous dimension. A {\em vector space of quantities} is an additive group $(S,+,D,$ $\pd)$ of quantities together with a field $\mathbb{F}$ and a multiplication $\mathbb{F}\times S\to S : (\lambda,x)\mapsto \lambda x$ fulfilling the vector-space axioms. As for the additive group, the vector space of quantities is equipped with a constant dimension map. In the same way, an algebraic {\em field of quantities} is of homogeneous physical dimension. That dimension, however, must be the neutral dimension, since the inverse of an element of the field must have the same physical dimension as the element itself. To apply this framework, we equip the spaces in the tangent or the exterior bundle of a manifold with a constant dimension map, turning them into vector spaces of quantities. For example, for the electromagnetic field $F\in\dfs{2}(P)$, $\pd(F)=\pdim{UT}$, the spaces ${\sf{\Lambda}}^2T_p^*P$ are assigned the physical dimension of magnetic flux, thus turning them into vector spaces of fluxes.\footnote{This approach was outlined in \cite{prechtldim}. It differs from prevalent ones \cite{schouten, post} in that it does not involve the transformation behavior of component functions of quantities. For this reason algebraic concepts are more in line with the overall goal of this paper, that is, to focus on mathematical structures rather than their coordinate- and frame-based representations.}
% For more on the comparison of approaches see Appendix~\ref{apost}.}
\ep
%\bph{Extended observer}
%An extended observer, or simply observer, may be thought of as a swarm of test particles, sometimes called reference fluid or, with Einstein, a reference mollusk \cite[Sec.~28]{Einstein1920}. Each particle represents a virtual measurement device. For the purposes of this article, we assume that measurements could be made on an arbitrarily small scale. A single test particle is called {\em local observer}.
%\ep
\bphstar{Space-time}\label{spacetime}
A space-time $(P,\mathbf{g})$ is a time-oriented
%\footnote{A time-orientable Lorentz manifold admits a globally defined time-like vector field \cite[Prop.~2.3]{Minguzzi2008}, i.e., it admits the construction of a principal bundle with time-like fibers.\label{spacetimefootnote}} 
connected Lorentz manifold of signature $(+,-,-,-)$; compare with \cite[Def.~2.5]{Minguzzi2008}.\footnote{Note that \cite{Minguzzi2008} uses the signature $(-,+,+,+)$.} For the sake of our purposes, the manifold shall be orientable \cite[B.1.2]{hehlbook}. The metric tensor field $\mathbf{g}$ is sufficiently smooth and has physical dimension $\pd(\mathbf{g})=\pdim{L}^2$. For a discussion of this choice of physical dimension see Appendix~\ref{mpost}. We generally assume that the space-time be globally hyperbolic \cite[3.11]{Minguzzi2008} even though this constraint might be too restrictive for some applications, in which case it could be somewhat relieved; compare with \cite[Sec.~2]{Minguzzi2007}. No other restrictions are posed, that is, our setting is general relativity. It is only for the applications in Chapter~\ref{sec:applications} that we consider Minkowski space-time as an example.
\epstar
\bphstar{World-line} \label{world-line}
A world-line is a time-like, future-directed, sufficiently smooth curve in space-time. Intuitively, it models the existence of a test particle. A bundle of world-lines has been called a reference fluid, reference dust, or -- by Einstein -- a reference mollusk, \cite[Sec.~28]{Einstein1920}.
\epstar
%%%%%%%%%%%%%%%%%%%%%%%%%%%%%%%%%%%%%%%%%%%%%
%%%%%%%%%%%%%%%%%%%%%%%%%%%%%%%%%%%%%%%%%%%%%
\newchapter{Pre-Metric Setting}\label{sec:premetric}
For the purposes of this chapter, we assume the existence of a set of world-lines that densely fills an interesting region of space-time, and `forget' the space-time metric until it will be re-introduced in the next chapter. In Sections~\ref{sec:fiber}-\ref{sec:ehresmann} we introduce mostly textbook definitions, leading up to the discussion of Ehresmann connections on principal bundles. The definitions are given in their usual, general form, and our intentions with regard to the definition of a relativistic splitting structure are stated in the brief introductions. Starting from Section~\ref{sec:transitions} we abandon the general scope of principal bundles with arbitrary structure groups, and continue to develop our framework for the particular Lie group of 1-dimensional (time-) translations. Generalizations to arbitrary Lie groups are available, but not useful for the purposes of this paper. In Sections~\ref{sec:pmobserver} and~\ref{sec:pmemag} the relativistic splitting structure is defined and discussed. 

%%%%%%%%%%%%%%%%%%%%%%%%%%%%%%%%%%%%%%%%%%%%%%
\section{Fiber bundles: world-lines in space-time}\label{sec:fiber}
We consider general fiber bundles. Fiber charts make the bundle manifold locally diffeomorphic to a product space of base manifold and typical fiber. With regard to our application, the bundle manifold is space-time; the fibers are world-lines; the base manifold models space as seen by the observer; and the typical fiber is there to keep track of the sequence of events along the world-lines. A section of the fiber bundle serves as a time synchronization. This convention allows the time-translation of Section~\ref{sec:principal} to act on all events in the section synchronously.
\begin{figure}
\centering
\setlength{\unitlength}{0.08cm}
\hspace{-2cm}\begin{picture}(100,70)
\includegraphics[width=8cm]{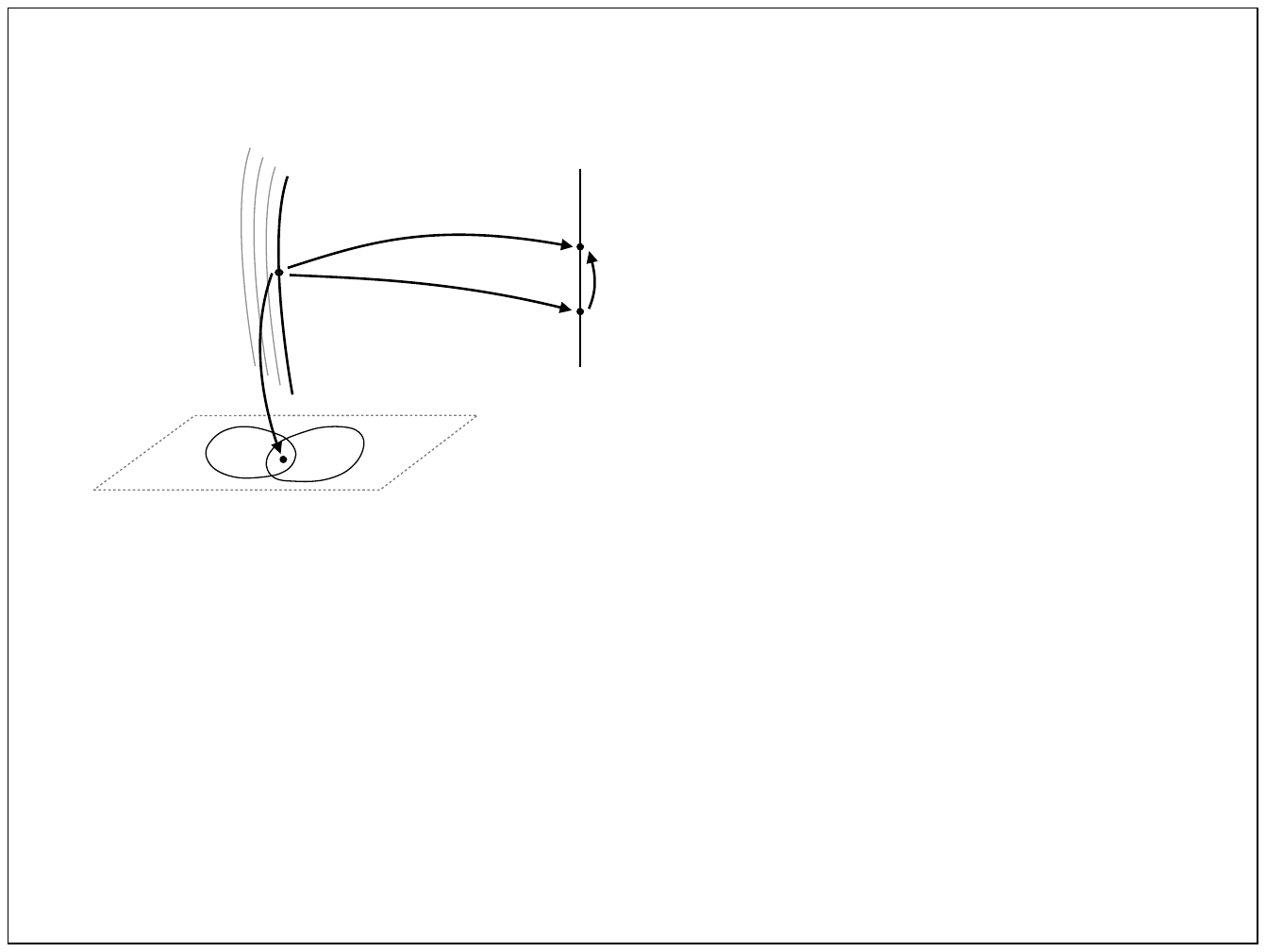}
\put(-7,65){$F$}
\put(-80,60){$P$}
\put(-60,62){$F_x=\pi^{-1}(x)$}
\put(-67,44.5){$p$}
\put(-59.5,4.3){$x$}
\put(-70,21.5){$\pi$}
\put(-35,53.5){$\hhat{\varphi}_{i,x}$}
\put(-30,34){$\hhat{\varphi}_{j,x}$}
\put(-0.5,40){$\varphi_{ij}(x)=\hhat{\varphi}_{i,x}\circ\hhat{\varphi}^{-1}_{j,x}$}
\put(-73.5,8){$U_i$}
\put(-54,8){$U_j$}
\put(-87,4){$X$}
\end{picture}
\caption{Fiber bundle $(P,\pi,X,F)$ with typical fiber $F$, base space $X$, fiber at $x\in X$ $F_x=\pi^{-1}(x)$. The fiber charts $\varphi_i$ and $\varphi_j$ define the transition function at $x$, $\varphi_{ij}(x) = \hhat{\varphi}_{i,x}\circ\hhat{\varphi}_{j,x}^{-1}$.}\label{fig:fiber_bundle}
\end{figure}
\bphstar{Fiber bundle} \label{fiberbundle}
Let $(P,\pi,X,F)$ denote a fiber bundle. It consists of the differentiable manifolds $P$, $X$, $F$, and a smooth mapping $\pi:P\to X$; furthermore it is required that each $x\in X$ has an open neighborhood $U$ such that $\pi^{-1}(U)$ is diffeomorphic to $U\times F$ via a fiber-respecting diffeomorphism $\varphi$:
\begin{diagram}[height=\tw{0.06},width=\tw{0.07}]
\pi^{-1}(U)&&\rTo^{\varphi}&&U\times F\\
&\rdTo_{\pi}&&\ldTo_{\n{pr}_1}&\\
&&U&&
\end{diagram}
Herein, $\pi^{-1}(U)$ is the inverse image of the projection $\pi$ and $\n{pr}_1$ is the canonical projection. $P$ is the total space or {\em bundle manifold} of dimension $n$, $X$ is the {\em base manifold}, $\pi:P\to X$ is a surjective submersion called the {\em projection}, and $F$ is called the standard fiber or {\em typical fiber}. The dimension of the typical fiber is denoted $q$ throughout the text. $\pi^{-1}(x)=F_x$ is called the fiber at $x$. $(U,\varphi)$ as above is called a {\em fiber chart} or a local trivialization of $P$. 
The diffeomorphism $\varphi:\pi^{-1}(U)\to U\times F$ has the form $\varphi=(\pi,\hhat{\varphi})$.
We denote by $\hhat{\varphi}_{x}:F_x\to F$ the restriction of $\hhat{\varphi}$ to a single fiber. A fiber bundle is called {\em trivializable} if there exists a global fiber chart $(X,\varphi)$. compare with \cite[9.1]{kolar} and \cite[p.~125 ff.]{amp}.
\epstar
\bphstar{Transition functions}\label{transfunc}
A collection of fiber charts $\{(U_i,\varphi_i)\}$, such that $\{U_i\}$ is an open cover of $X$, is called a {\em fiber-bundle atlas}. For a given atlas, $\varphi_{ij}:U_i\cap U_j\to \n{Diff}(F):x\mapsto \hhat{\varphi}_{i,x}\circ\hhat{\varphi}_{j,x}^{-1}$ is a diffeomorphism of $F$ for each $x\in U_{ij}=U_i\cap U_j$. The mappings $\varphi_{ij}$ are called the transition functions of the bundle. They satisfy the {\em cocycle condition} $\varphi_{ij}(x)\circ\varphi_{jk}(x)=\varphi_{ik}(x)$ for $x\in U_{ijk}$ and $\varphi_{ii}(x)=\n{Id}_F$; see Fig.~\ref{fig:fiber_bundle}. compare with \cite[9.1]{kolar} and \cite[p.~125 ff.]{amp}. 
\epstar
\bphstar{Vertical fields}\label{verfields}
\label{vertical}The family of vertical vector spaces is defined by $ V_pP=\n{Ker\,}(\pi^\prime)\subset T_pP$ for all $p\in P$. compare with \cite[p.~359]{amp}. Denote 
\begin{align*}\mathcal{X}^1_V( P)=\{\vec{v}\in\mathcal{X}^1(P)\,|\,\vec{v}_p\in V_pP,\,\forall\,p\in P\}\end{align*} 
the space of vertical vector fields. The definition extends to vertical multi-vector fields $\mathcal{X}_V^k(P)=\{\vec{v}\in\ves{k}(P)\,|\,\pi^\prime\,\vec{v}_p=0$, $\forall\,p\in P\}$. 
\epstar
\bphstar{Section}\label{section}
A fiber bundle admits {\em local sections}: For each $p\in P$ there is an open neighborhood $U$ of $\pi(p)$ in $X$ and a smooth mapping $s : U \to P$ with $\pi\circ s = \n{Id}_U$; \cite[2.4]{kolar}, \cite[p.~132]{amp}. A {\em global section} of a fiber bundle is a smooth mapping $s:X\to P$ such that $\pi \circ s=\n{Id}_X$. A fiber bundle is trivializable iff it admits a global section \cite[16.14.5]{Dieudonne1972}. For an example see Fig.~\ref{fig:section}.
\epstar
%
%%%%%%%%%%%%%%%%%%%%%%%%%%%%%%%%%%%%%%%%%%%%%%
\section[Principal bundles: introducing time-translation]{Principal bundles: introducing time-\\translation}
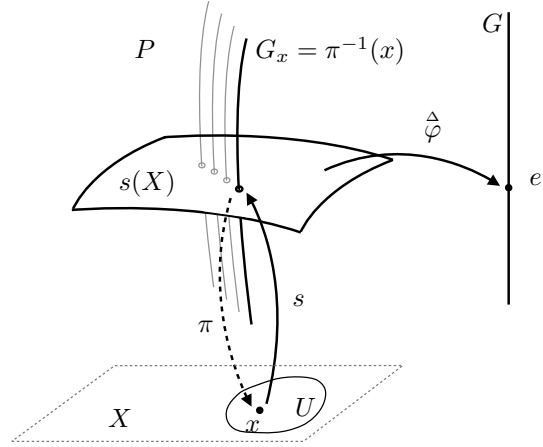
\begin{figure}
\centering
\setlength{\unitlength}{0.07cm}
\begin{picture}(100,87)
\includegraphics[width=7cm]{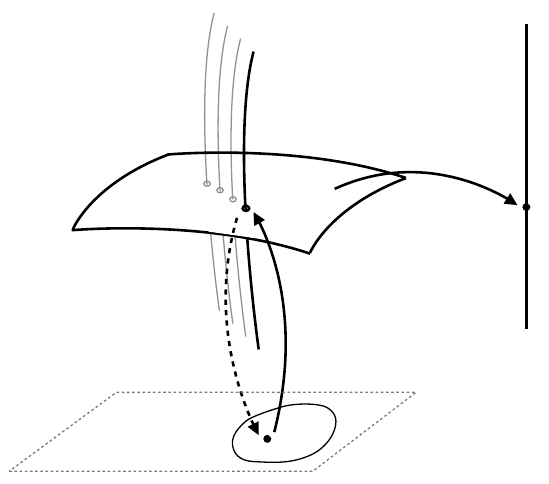}
\put(-78,50){$s(X)$}
\put(-75,75){$P$}
\put(-80,5){$X$}
\put(-9,80){$\group$}
\put(-52,75){$\group_x=\pi^{-1}(x)$}
\put(-45,28){$s$}
\put(-54,4){$x$}
\put(-44.5,7){$U$}
\put(-63,23){$\pi$}
\put(-20,60){$\hhat{\varphi}$}
\put(0,51){$e$}
\end{picture}
\caption{Chart-induced section $s:U\to P$ in the fiber chart $\varphi$ of a principal bundle.}\label{fig:section}
\end{figure}

\begin{figure}
\centering
\setlength{\unitlength}{0.057cm}
\begin{picture}(100,90)
\includegraphics[width=6cm]{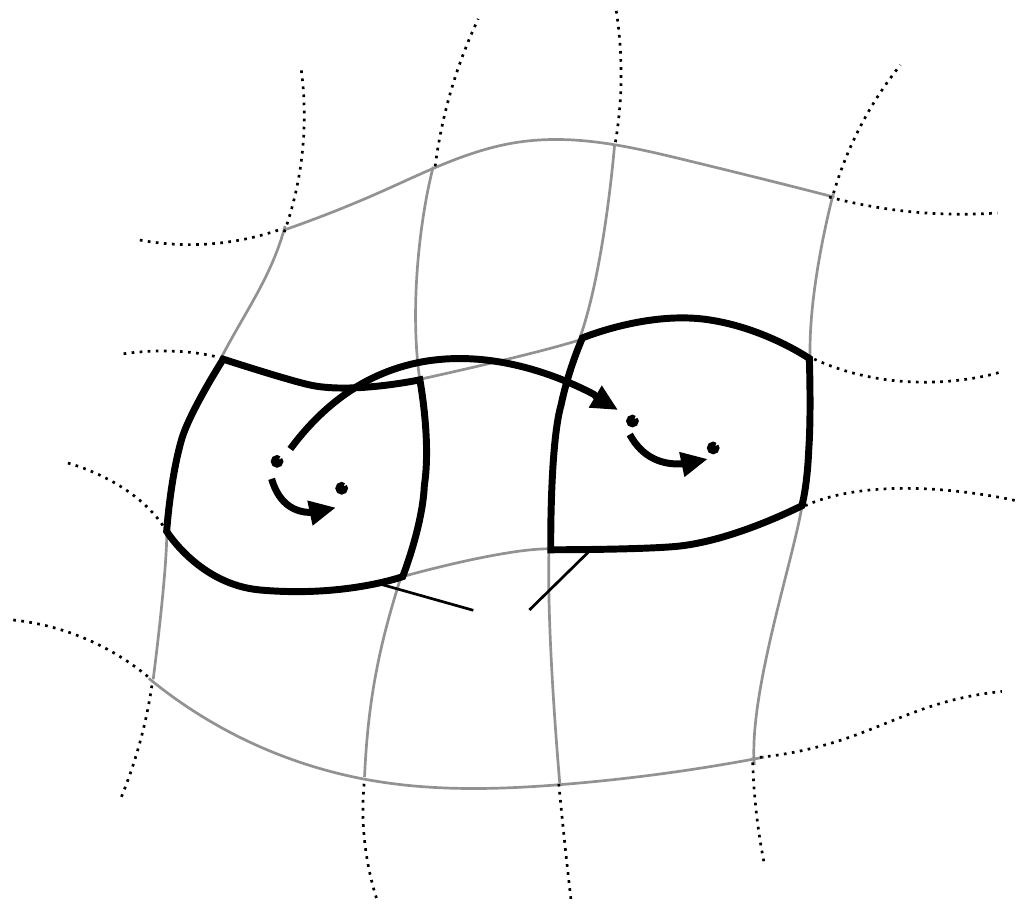}
\put(-84.5,46){$\varphi_j$}
\put(-68.5,43){$\varphi_i$}
\put(-77,37){$g_{ij}$}
\put(-41,53.5){$\varphi_\ell$}
\put(-30.5,48){$\varphi_k$}
\put(-40,42){$g_{k\ell}$}
\put(-58,61){$\varphi_{\ell j}$}
\put(-60,26){Max. $\group$-atlases}
\put(-30,87){Max. $\n{Diff}(\group)$-atlas}
\end{picture}
\caption{Schematic view of two $\group$-atlases in the partition of a $\n{Diff}(\group)$-bundle's maximal fiber-bundle atlas. Transition functions between fiber charts in the same maximal $\group$-atlas are $\group$-actions, whereas transition functions between different maximal $\group$-atlases are general diffeomorphisms of $\group$.}\label{fig:atlases}
\end{figure}
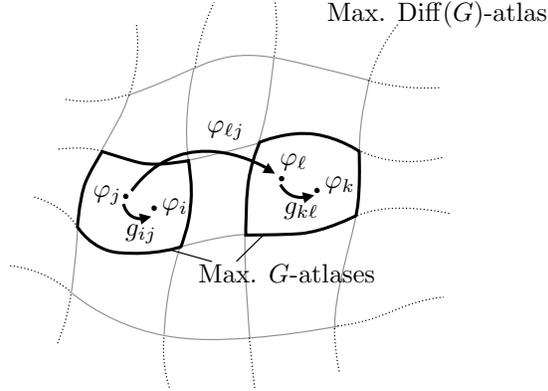

\labcount{principal}\label{sec:principal}
A principal bundle is a fiber bundle i) whose typical fiber is isomorphic to a Lie group and ii) whose fiber charts are compatible with a principal right-action of the group along the fibers. In our application, the fibers are world-lines and the typical fiber is isomorphic to the Lie group $\G$ of 1-dimensional translations. The principal right-action represents time translation along world-lines. We find that a fiber bundle whose typical fiber is isomorphic to a Lie group -- omitting condition ii) above -- partitions naturally into infinitely many principal bundles, which are distinguished by their principal right-action. That implies that a given congruence of world-lines admits infinitely many realizations of time-translation. We have no a priori preference as to which of the time translations is used, although some choices may be more practical than others; see \ref{classobs} and~\ref{premetricEmagDiscussion}.
\begin{figure}
\centering
\setlength{\unitlength}{0.1cm}
\begin{picture}(100,63)
\includegraphics[width=10cm]{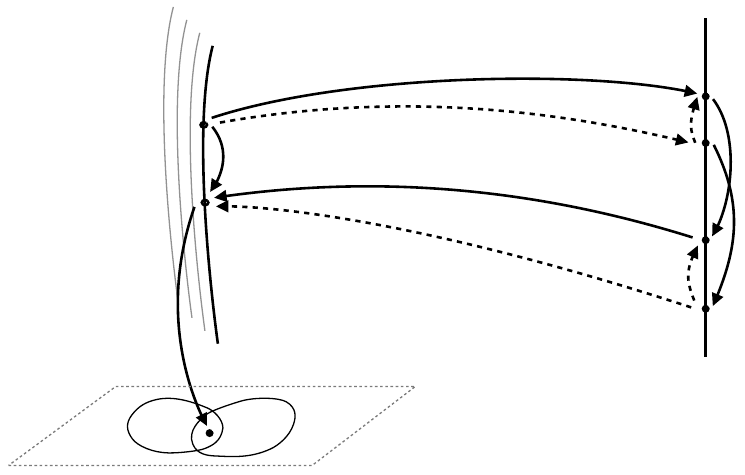}
\put(-69,42){${r}_g$}
\put(-71.5,56){$\group_x$}
\put(-79,16){$\pi$}
\put(-80.5,6){$U_i$}
\put(-66,6){$U_j$}
\put(-90,3){$X$}
\put(-9,60){$\group$}
\put(-9,53){$g_i$}
\put(-9,42){$g_j$}
\put(-48,45){$\hhat{\varphi}_{j,x}$}
\put(-45,54){$\hhat{\varphi}_{i,x}$}
\put(-35,39.){$\hhat{\varphi}^{-1}_{i,x}$}
\put(-37,25){$\hhat{\varphi}^{-1}_{j,x}$}
\put(-17.5,47.3){$g_{ij}(x)$}
\put(-17.5,27){$g_{ij}(x)$}
\put(-1.5,43){$R_g$}
\put(-1,30.5){$R_g$}
\put(-76,47){$p$}
\put(-102,37){${r}_g(p)=({r}_g)_i(p)$}
\put(-93,33){$=({r}_g)_j(p)$}
\put(-89,52){$P$}
\end{picture}
\caption{Definition of the principal right action. For $\varphi_i$ and $\varphi_j$ from the same maximal $\group$-bundle atlas, they both define the same right action along the fibers. Reproduced from \cite[p.~129]{amp} with permission.\label{fig:right_action}}
\end{figure}
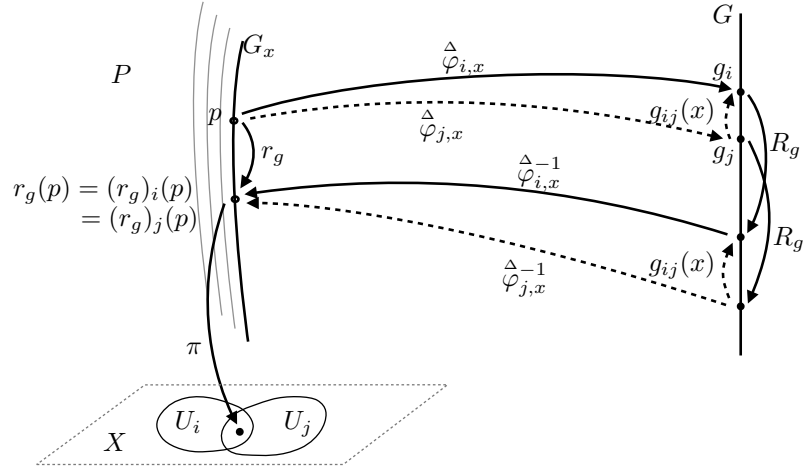
\bphstar{\texorpdfstring{$\group$-bundle}{G-bundle}}\label{Gstructure}
Let $\group$ be a Lie group and $(P,\pi,X,F)$ a fiber bundle. A $\group$-bundle structure on the fiber bundle consists of the following data (verbatim from \cite[10.1]{kolar}): 
\begin{enumerate}
\item\label{gstruc1} A left action $\ell:\group\times F\to F$ of the Lie group on the typical fiber; 
\item\label{gstruc2} A fiber-bundle atlas $\{(U_i,\varphi_i)\}$, the transition functions of which act on $F$ via left action: There is a family of mappings $g_{ij}:U_{ij}\to \group$ satisfying the cocycle condition $g_{ij}(x)g_{jk}(x)=g_{ik}(x)$ for $x\in U_{ijk}$ and $g_{ii}(x)=e$, with $e$ the identity element of $\group$, such that the transition function reads $\varphi_{ij}(x)=\ell\bigl(g_{ij}(x)\bigr)$. 
\end{enumerate}
A fiber-bundle atlas as in~Point~\ref{gstruc2}~is called a {\em $\group$-atlas}. Two $\group$-atlases are said to be equivalent, if their union is also a $\group$-atlas. A {\em maximal $\group$-atlas} contains all equivalent atlases. To be precise, a {\em $\group$-bundle structure} is the pair of a left-action of $\group$ on $F$ (see~Point~\ref{gstruc1}) and a maximal $\group$-atlas. A $\group$-bundle $(P,\pi,X,F,\group)$ is a fiber bundle with a $\group$-bundle structure. $\group$ is called the {\em structure group} of the $\group$-bundle. compare with \cite[p.~126 ff.]{amp}. 
\epstar
\bphstar{\texorpdfstring{Fibre bundles as $\group$-bundles}{Fibre bundles as G-bundles}}\label{fibergbundle}
Every fiber bundle in \ref{fiberbundle} trivially is a $\group$-bundle, setting $\group=\n{Diff}(F)$ and $\ell\bigl(\varphi_{ij}(x)\bigr)=\varphi_{ij}(x)$. 
\epstar
\bphstar{\texorpdfstring{Principal $\group$-bundle}{Principal G-bundle}}\label{princbund}
A principal $\group$-bundle $(P,\pi,X,\group)$ is a $\group$-bundle where the typical fiber is equal to the structure group, and where the left action is just $\group$ acting upon itself from the left; compare with \cite[10.2]{kolar} and \cite[p.~129]{amp}. 
\epstar
\bphstar{Principal right action}\label{principaction}
\label{realization} A principal $\group$-bundle admits a unique right action called the principal right action. Let $\pi(p)\in U_i$. Then
\begin{align*}{r}:\pi^{-1}(U_i)\times \group\to \pi^{-1}(U_i): (p,g)\mapsto {\varphi}_i^{-1} \bigl(\pi(p),R_g\hhat{\varphi}_i(p)\bigr),\end{align*}
 where $R_g:\group\to \group:h\mapsto hg$ is the right group action. As a consequence of the $\group$-bundle structure, $r$ is independent of the fiber chart $(U_i,\varphi_i)$, and is, therefore, defined on all of $P$; see Fig.~\ref{fig:right_action}. The action is transitive and free.
% (from Wikipedia): Every free action on a nonempty set is faithful $\equiv$ effective. The action is regular, if it is both transitive and free.
Its orbits are the fibers of the principal bundle; compare with \cite[10.2]{kolar} and \cite[p.~129 ff.]{amp}.
\epstar
\bphstar{Chart-induced section}\label{canonsec}
A fiber chart $(U_i,\varphi_i)$ canonically induces a local section on a principal bundle via $s_i=\varphi_i^{-1}\circ \overline{\n{Id}}$, where $\overline{\n{Id}}:X\to X\times \group:x\mapsto (x,e)$; see Fig. \ref{fig:section} and compare with \cite[p.~363]{amp}. 
\epstar
\bphstar{\texorpdfstring{$\group$-atlas from local sections and principal right action}{G-atlas from local sections and principal right action}}
A free right action ${r}:\group\times P\to P$ whose orbits are exactly the fibers, and a family of local sections $s_i:U_i\to\pi^{-1}(U_i)$, such that $\{U_i\}$ is an open cover of $X$, together define a unique $\group$-atlas with fiber charts $\varphi_i^{-1}(x,g)={r}\left(s_i(x),g\right)$ for $x\in U_i$, \cite[10.3, 10.4]{kolar}.
\epstar
\bph{Foliation}\label{foliation}
A local section $s:U\subset X\to P$ (for example, a chart-induced section), together with the principal right action, defines a foliation of $\pi^{-1}(U)\subset P$. The leaves are labeled by the group elements and defined as $S_g=r_g\,s(U)$.
\ep
\bph{\texorpdfstring{$\n{Diff}(\group)$-bundle}{Diff(G)-bundle}}
A fiber bundle $\bigl(P,\pi,X,\group,\n{Diff}(\group)\bigr)$ which has the Lie-group $\group$ for the typical fiber, and $\n{Diff}(\group)$ for the structure group, is a $\n{Diff}(\group)$-bundle; compare with \ref{fibergbundle}. 
Note that elements of $\n{Diff}(\group)$ are not generally compatible with the group structure.
\ep
\bph{\texorpdfstring{Partition of the $\n{Diff}(\group)$-bundle}{Partition of the Diff(G)-bundle}}\label{partition}
Let $\sim_\group$ denote an equivalence relation on the maximal fiber-bundle atlas of a $\n{Diff}(\group)$-bundle,\footnote{The notation $\exists!$ reads `there is exactly one ...'.} 
\begin{align*} \varphi_i\sim_\group \varphi_j\Longleftrightarrow \exists!\,\,g_{ij}(x)\in\group:\varphi_{ij}(x)=L\bigl(g_{ij}(x)\bigr).\end{align*}
Two fiber charts are equivalent, if the transition-map can be captured by a group action. The equivalence relation partitions the maximal fiber bundle atlas into maximal $\group$-atlases; see Fig.~\ref{fig:atlases}. It partitions the $\n{Diff}(\group)$-bundle into principal $\group$-bundles. The left-action of the $\group$-bundle structure is $\ell=L$, with $L$ the left-action of $\group$ on itself. We will return to this construct in Section~\ref{sec:transitions}.
\ep
%
%%%%%%%%%%%%%%%%%%%%%%%%%%%%%%%%%%%%%%%%%%%%%%
\section{The fundamental field map: populating the principal bundle}\label{sec:liealgebra}
\labcount{lie}
The Lie algebra of a Lie group is the pair formed by the vector space of left-invariant vector fields on the Lie group, and the Lie-bracket multiplication. In the case of the 1-dimensional translation group $\G$, the Lie-bracket of any two vectors is trivially zero. What is of use for us is the Maurer-Cartan form and its twin, the Maurer-Cartan vector field. In a principal bundle, they can be pushed and pulled to the bundle manifold, yielding objects that are defined independently of a particular choice of a basis on the Lie algebra. They depend only on the principal right-action. The pushed vector field will be used in the definition of the splitting map in
\ifdefined\LONGVERSION
\ref{splitting}.
\else
\ref{psipush}.
\fi
The pulled form represents a chart-associated principal connection that is used in a certain class of splittings; see~\ref{frobenius} and~\ref{classobs}.
\bphstar{Lie algebra}\label{liealgebra}
A vector field $\vec{v}\in\ves{1}(\group)$ is called left invariant if 
\begin{align*}({L_g}^\prime)_h\vec{v}_h=\vec{v}_{g h}\,\forall\,g,h\in \group,\end{align*} 
where $L:\group\times\group\to \group:(g,h)\mapsto g h$ is the left translation; recall that the prime denotes the pushforward. The Lie algebra $\mathfrak{g}$ of a Lie group $\group$ is the space of left invariant vector fields on $\group$, together with Lie bracket multiplication. We have $\n{dim}(\group)=\n{dim}(\mathfrak{g})=q$. Every left invariant $\vec{v}\in\ves{1}(\group)$ is uniquely represented by its vector at identity via $\vec{v}_g=({L_g}^\prime)_e\vec{v}_e$; compare with \cite[4.11]{kolar} and \cite[p.~155]{amp}. We follow the common practice to identify the Lie algebra $\mathfrak{g}$ of the Lie group $\group$ with the tangent space $T_e\group$ of the group at the identity element $e$.
\epstar
\bphstar{Lie co-algebra}
A 1-form $\df{\gamma}\in\dfs{1}(\group)$ is called left invariant if 
\begin{align*}(L_g^*)_h\df{\gamma}_{gh}=\df{\gamma}_h\,\forall\,g,h\in \group.\end{align*} 
The dual of the Lie algebra, that is, the Lie co-algebra $\mathfrak{g}^*$ is the space of left invariant 1-forms on $\group$ \cite[p.~208]{amp}. 
\epstar
\bphstar{Fundamental field map} \label{canonisomorphism}
Let ${r}_p:\group\to P:g\mapsto r(p,g)$. The image of $\hat{\vec{v}}\in\mathfrak{g}$ under the fundamental field map 
\begin{align*}\zeta:\mathfrak{g}\to\mathcal{X}^1_V(P):\hat{\vec{v}}\mapsto\vec{v},\end{align*} 
is defined pointwise by 
\begin{align*}\vec{v}_p&=({{r}_p})^\prime_e\hat{\vec{v}};
\intertext{compare with \cite[5.13]{kolar} and \cite[p.~360]{amp}. For a given fiber chart $\varphi(p)=(x,g)$, the fundamental field map can, equivalently, be written as $\vec{v}_p = (\hhat{\varphi}_{x}^{-1})_g' \,(L_g')_e\hat{\vec{v}}$, and so, fiberwise,}
\left.\vec{v}\right|_{\group_x }&=(\hhat{\varphi}_x^{-1})'\check{\vec{v}},\end{align*} 
where $\check{\vec{v}}$ is the left invariant vector field associated with $\hat{\vec{v}}$.
This definition is independent of the selected fiber chart. The hat in the notation of a Lie-algebra vector is used to indicate that $\hat{\vec{v}}$ is related to a vertical vector field $\vec{v}$ by the fundamental field map.
\ifdefined\LONGVERSION It holds that $r_h'\vec{v}=\zeta\bigl(\n{Ad}(h^{-1})\hat{\vec{v}}\bigr)$; see \ref{afundam}. The adjoint representation $\n{Ad}\,(g):\mathfrak{g}\to\mathfrak{g}:\vec{e}\mapsto L_g' R_{g^{-1}}'\,\vec{e}$ is identity $\n{id}_\mathfrak{g}$ for Abelian Lie groups. compare with \cite[4.24]{kolar} and \cite[p.~166]{amp}.
\fi
\epstar
\bphstar{Maurer-Cartan form}\label{MC}
The Maurer-Cartan form is a Lie-algebra-val\-ued 1-form on $\group$, $\MC\in\dfs{1}(\group;\mathfrak{g})$.
It is defined for any $\vec{v}_g\in T_g\group$ by
\begin{align*}\MC({\vec{v}}_g)=\vec{a},\quad\mbox{where}\quad\vec{v}_g=(L_g)'_e\vec{a},\end{align*}
$\vec{a}\in\mathfrak{g}$; compare with \cite[p.~168 and p.~364]{amp}. $\MC$ is left-invariant by definition. If $\group$ is Abelian all $\group$-invariant forms are closed and, hence, $\n{d}\,\MC=0$; compare with \cite[Eq.~(2.44)]{Morita2001}.
\epstar
\bph{Maurer-Cartan vector field}\label{MC2}
By duality, we introduce the Maurer-Cartan vector field as a Lie-coalgebra-valued vector field on $\group$, $\wMC\in\ves{1}(\group;\mathfrak{g}^*)$. It is defined for any $\df{\gamma}_g\in T^*_gG$ by
\begin{align*}\df{\gamma}_g(\wMC)=\df{\alpha},\quad\mbox{where}\quad\df{\gamma}_g=(L_{g^{-1}})^*_e\df{\alpha},\end{align*}
$\df{\alpha}\in\mathfrak{g}^*$. It holds that
\begin{align*}\MC(\wMC)^\otimes=1\otimes\vec{t}\in C^\infty(\group;\mathfrak{g}\otimes\mathfrak{g}^*),\end{align*}
where $\vec{t}$ is the unit tensor for the pair $(\mathfrak{g},\mathfrak{g}^*)$; compare with \cite[3.14]{greub2}. The tensor-valued duality product is defined in \ref{algebra}, \ref{algebra3}.
\ep
\bph{Fundamental field}\label{MC2b}
%We interpret $\wMC$ as a Lie-coalgebra-valued vector field on $\group$, $\wMC\in\ves{1}(\group;\mathfrak{g}^*)$.
\label{MCpush} Denote 
\begin{align*}\vec{w} = \zeta(\wMC)\end{align*} 
the image of $\wMC$ under the fundamental field map. We call $\vec{w}$ the fundamental field. It follows that $\vec{w}\in\ves{1}(P;\mathfrak{g}^*)$.
The fundamental field represents the fundamental field map 
\begin{align*}\vec{w}(\hat{\vec{v}})=\zeta(\hat{\vec{v}})=\vec{v},\end{align*} 
for all $\hat{\vec{v}}\in\mathfrak{g}$.
\ep
%
%%%%%%%%%%%%%%%%%%%%%%%%%%%%%%%%%%%%%%%%%%%%%%
\section{Ehresmann connection: families of space platforms}\label{sec:ehresmann}
An Ehresmann connection can be defined as a family of horizontal subspaces, decomposing every tangent space of the bundle manifold into a horizontal subspace, and the vertical subspace along the fibers. The definition of horizontal vector fields and vertical forms ensues. In the context of our application, the horizontal subspaces have been called infinitesimal space platforms \cite{Rizzi2002}. The observer's space is represented in space-time by the family of all the space platforms in a given time synchronization.
\bphstar{\texorpdfstring{Ehresmann connection on a principal $\group$-bundle}{Ehresmann connection on a principal G-bundle}}\label{ehresconnonprinc}
An Ehresmann connection on a principal $\group$-bundle $(P,X,\pi,\group)$ is defined in three equivalent ways:
\begin{enumerate}
\item As a {\em horizontal lift} $\sigma_p:T_xX\to T_pP$, $x = \pi(p)$, such that $\sigma_p$ is linear, $\pi^\prime\circ\,\sigma_p=\n{Id}_{T_xX}$, and $\sigma_p$ depends differentiably on $p$.\footnote{Note that $\pi'$ is defined here only as a map between tangent spaces, $T_pP\to T_xX$, since $\pi$ is not a diffeomorphism.} 
\item As a field of {\em horizontal vector spaces} $H_pP\subset T_pP$, where $H_pP$ depends differentiably on $p$. 
\item \label{def3}As a {\em connection 1-form} $\df{\omega}\in\dfs{1}(P;\mathfrak{g})$ such that $\df{\omega}$ depends differentiably on $p$, and
$\df{\omega}(\vec{v})=1\otimes\hat{\vec{v}}$,
$\vec{v}=\zeta(\hat{\vec{v}})\,\forall\hat{\vec{v}}\in\mathfrak{g}$.
\end{enumerate}
The definitions are related by $\n{Im\,}\sigma_p=H_pP=\{\vec{v}\in T_pP\,|\,\df{\omega}_p(\vec{v})=0\}\,\forall\,p\in P$;
\label{principal}
compare with \cite[p.~358 ff.]{amp}. The horizontal lift maps vectors in a tangent-space of the base manifold into horizontal subspaces in the bundle space. Horizontal vectors are in the null-space of the connection 1-form. Definition~\ref{def3}~implies that $\df{\omega}(\vec{w})^\otimes = 1\otimes\mathbf{t}$.
The above connection 1-form $\df{\omega}$ of an Ehresmann connection on a principal $\group$-bundle is related to the {\em connection form $\Upsilon$ of an Ehresmann connection on a general fiber bundle} via $\Upsilon=\vec{w}\otimes\df{\omega}=\zeta\circ\df{\omega},\in\dfs{1}(P;TP)$; compare with \cite[11.4]{kolar}.
\epstar
\bphstar{Principal connection}\label{principalconnect}
An Ehresmann connection on a principal $\group$-bundle is a principal connection if it fulfills
\begin{enumerate}
\item $\sigma_{{{r}_g}(p)}=({{r}_g})^\prime_p\sigma_p$, 
\item $H_{{{r}_g}(p)}(P)=({{r}_g})^\prime_p H_pP$, and
\item ${r}_g^*\,\df{\omega}=\n{Ad\,}(g^{-1})\,\df{\omega}$;
\end{enumerate}
compare with \cite[11.1]{kolar} and \cite[p.~358 ff.]{amp}. The adjoint representation $\n{Ad}(g^{-1})$ is identity for Abelian Lie groups. For a principal connection, the principal right-action defines the evolution of a horizontal subspace in a point $p\in P$ along the fiber through $p$.
\epstar
\bph{Horizontal fields}\label{horfields}
For a given Ehresmann connection on $P$, there is a direct sum decomposition $\ves{k}(P)=\mathcal{X}_V^k(P)\oplus\mathcal{X}_H^k(P)$,
%(see \cite{long}),
where the space of {\em horizontal multi-vector fields} is given by 
\begin{align*}\mathcal{X}_H^k(P)=\{\vec{v}\in\ves{k}(P)\,|\,\vec{v}_p\in\mathsf{\Lambda}^k\n{Im\,}\sigma_p,\,\forall\,p\in P\},
\quad 1\le k\le n-q.
\end{align*}
We define the space of {\em horizontal differential forms} by 
\begin{align*}\mathcal{F}_H^k(P)=\{\df{\gamma}\in\dfs{k}(P)\,|\,
\df{\gamma}_p\in\mathsf{\Lambda}^k\n{Im\,}\pi_p^*,\,\forall\,p\in P\},\quad 1\le k\le n-q.
\end{align*}
For a given Ehresmann connection, the space of {\em vertical differential forms} is defined by
\begin{align*}
\mathcal{F}_V^k(P)=\{\df{\gamma}\in\dfs{k}(P)\,|\,\sigma_p^*\df{\gamma}_p = 0,\,\forall p\in P\},
\quad 1\le k\le q.
\end{align*}
Analogously to vector fields, there is a direct sum decomposition $\dfs{k}(P)=\mathcal{F}_V^k(P)\oplus\mathcal{F}_H^k(P)$. Each of the pairs $\bigl(\mathcal{F}_V^k(P),\mathcal{X}_V^k(P)\bigr)$ and $\bigl(\mathcal{F}_H^k(P),\mathcal{X}_H^k(P)\bigr)$ is dual with respect to the duality pairing inherited from the pair $\bigl(\dfs{k}(P), \ves{k}(P)\bigr)$, that is, from the duality of vectors and covectors in each tangent space. %; see \cite{long}.
\ep
%
%%%%%%%%%%%%%%%%%%%%%%%%%%%%%%%%%%%%%%%%%%%%%%
\section{Parametric fields: mapping fields to the observer's space}\label{sec:parametric}\labcount{parametric}
We introduce parametric fields on the base manifold, as well as maps to transfer horizontal fields on the bundle manifold to parametric fields on the base manifold, and vice versa. The relativistic splitting uses the maps to relate fields in space-time to time-parametric fields in the observer's space. As mentioned earlier, from this section on we restrict our development of the framework to the 1-dimensional Lie group of (time-)translations. A principal bundle with this particular structure group is necessarily trivializable. For convenience and without loss of generality, we, therefore, use global fiber charts unless otherwise stated.
\bph{Parametric fields}\label{paramfields}
Denote $\ves{k}(X,\group)=C^{\infty}\bigl(\group;\ves{k}(X)\bigr)$ and $\dfs{k}(X,\group)=C^{\infty}\bigl(\group;\dfs{k}(X)\bigr)$ the parametric multi-vector fields and parametric differential forms on $X$, that is to say, $C^\infty$-functions on $G$ with values in $\ves{k}(X)$ and $\dfs{k}(X)$, respectively.
\ep
\bph{\texorpdfstring{Twisted parametric forms on $\group$}{Twisted parametric forms on G}}\label{gtwisted}
Since $P$ was assumed orientable (see \ref{spacetime}) and diffeomorphic to $X\times\group$, from the orientability of $\group$ it follows that $X$ is orientable. Therefore, orientation can be discussed on a global level.\footnote{Twisted forms can be defined on nonorientable manifolds as well \cite[Ch.~3]{Bossavit1991}, \cite[Ch.~28]{burke}, but we don't delve into this.} Denote 
\begin{align*}\mathcal{F}^k_\doubletwist(X,\G)=C^\infty_\twist\bigl(\G;\mathcal{F}^k_\twist(X)\bigr)\end{align*} 
the parametric differential forms that are twisted with respect to $X$ and $\G$; compare with \ref{twisted} and \cite[Ch.~3]{Bossavit1991}, \cite[Ch.~28]{burke}. Each $\df{\alpha}_\doubletwist\in\mathcal{F}^k_\doubletwist(X,\G)$ can be represented by $(\df{\alpha},\tilde{\df{\kappa}})\in\dfs{k}(X,\G)\times\dfs{n-1}(X,\G;\mathfrak{g}^*)$, where $n=\dim(P)$ and $\tilde{\df{\kappa}}$ is a Lie-coalgebra-valued volume form on $X$, defining an orientation on $X\times\G$. The equivalence relation reads
\begin{align*}
(\df{\alpha},\tilde{\df{\kappa}})^\prime\sim(\df{\alpha},\tilde{\df{\kappa}})\;:\;\tilde{\df{\kappa}}^\prime=\lambda\,\tilde{\df{\kappa}},\;
\df{\alpha}^\prime=\n{sgn}(\lambda)\,\df{\alpha},\quad\lambda\in C^\infty(X,\G).
\end{align*}
\ep
\bph{\texorpdfstring{Integration on $\group$}{Integration on G}}\label{integralgroup}
We will encounter parametric, Lie-coalgebra valued forms that lie in the image of a splitting map; see \ref{psipush}. Their integrals on a family of compact domains $\interval=\interval(x)\subset\group$ of the Lie group
\begin{align*}
\int_\interval\;:\;\dfs{k}(X,\group;\mathfrak{g}^*)\to\dfs{k}(X)
\end{align*}
are defined with $\df{\gamma}\in\dfs{k}(X,\group;\mathfrak{g}^*)$, $\vec{u}_\alpha\in\ves{1}(X)$, $(\vec{e},\df{\varepsilon})\in\mathfrak{g}\times\mathfrak{g}^*$, $\df{\varepsilon}(\vec{e})=1$ by
\begin{align*}
\big.\Bigl(\int_\interval\df{\gamma}\Bigr)(\vec{u}_1,\dots,\vec{u}_k)\big|_x=\int_{\interval(x)}\!\!\!\bigl(\bigl.\df{\gamma}(\vec{u}_1,\dots,\vec{u}_k;\vec{e})\bigr|_x\bigr)\MC(\df{\varepsilon}).
\end{align*}
The definition is independent of the pair of dual bases $(\vec{e}$, $\df{\varepsilon})$.
\ep
\bph{\texorpdfstring{Integration of twisted parametric forms on $\group$}{Integration of twisted parametric forms on G}}
The integral operator defined in \ref{integralgroup} may be extended by
\begin{align*}
\int_\interval\;:\;\mathcal{F}^k_\doubletwist(X,\G,\mathfrak{g}^*)&\to\mathcal{F}^k_\twist(X,\G)\\
(\df{\alpha},\tilde{\df{\kappa}})&\mapsto\Bigl(\int_\interval\df{\alpha},\int_\interval\tilde{\df{\kappa}}\Bigr)\in
\dfs{k}(X,\G)\times\dfs{n-1}(X,\G).
\end{align*}
The result is independent of the orientation of $\interval$, and the twist with respect to $\G$ has been eliminated by the integration.
\ep
\bph{Parametric maps}\label{paramaps}
\label{bigphi}Consider the maps $\pi^\prime:T_pP\to T_xX$ and $\sigma_p:T_xX\to T_pP$. A fiber chart $(X,\varphi)$, $\varphi:p\mapsto(x,g)$ induces maps
\begin{alignat*}{3}
\Pi:\,&\ves{k}(P)\to\ves{k}(X,\group):\,&&\vec{v}\mapsto\bar{\vec{v}},\quad&&\bar{\vec{v}}(g)_x=
\pi^\prime\,\vec{v}_p,\\
\Sigma:\,&\ves{k}(X,\group)\hookrightarrow\ves{k}(P):\,&&\bar{\vec{v}}\mapsto\vec{v},\quad&&\vec{v}_p=
\sigma_p\,\bar{\vec{v}}(g)_x.
\end{alignat*}
We find {from the properties of $\pi^\prime$ and $\sigma_p$} that:
\begin{center}
\begin{tabular}{rlcrl} 
i)&$\Pi\circ\Sigma=\n{Id}_{\ves{k}(X,\group)}$, &&ii)&$\Sigma$ is injective, \\
iii)&$\Pi$ is surjective, &&iv)& $\n{Ker\,}\Pi=\mathcal{X}_V^k(P)$,\\
v)& $\n{Im\,}\Sigma=\mathcal{X}_H^k(P)$. 
\end{tabular}\\
\end{center}
The dual maps are characterized by
\begin{align*}
\Sigma^*:\,&\dfs{k}(P)\to\dfs{k}(X,\group),\\
\Pi^*:\,&\dfs{k}(X,\group)\hookrightarrow\dfs{k}(P),
\end{align*}
and we find that: 
\begin{center}
\begin{tabular}{rlcrl} 
vi)& $\Sigma^*\circ\Pi^*=\n{Id}_{\dfs{k}(X,\group)}$,&&vii)&$\Pi^*$ is injective,\\
viii)& $\Sigma^*$ is surjective,&&ix)&$\n{Ker\,}\Sigma^*=\mathcal{F}_V^k(P)$,\\
x)& $\n{Im\,}\Pi^*=\mathcal{F}_H^k(P)$.
\end{tabular}\\
\end{center}
The maps naturally extend to (co)vector-valued fields and forms; compare with~\ref{extend}.
\ep
\bph{Chart-associated connection}\label{canonconnec}
Denote $\df{\theta}$ the Lie-algebra valued form that is defined for a fiber chart $(X,\varphi)$ by 
\begin{align*}\df{\theta}=\hhat{\varphi}^*\MC.\end{align*} 
\ifdefined\LONGVERSION
With \cite[p.~169]{amp}, we find that ${r}_g^*\,\df{\theta}=\n{Ad\,}(g^{-1})\,\df{\theta}$.
\fi
It follows from the definitions that $\df{\theta}(\vec{v})=1\otimes\hat{\vec{v}}\in C^\infty(P;\mathfrak{g})$, $\vec{v}=\zeta(\hat{\vec{v}})$, and $\df{\theta}(\vec{w})^\otimes = \MC(\wMC)^\otimes=1\otimes\vec{t}$. For Abelian Lie groups, like the group of 1-dimensional translations, $\n{d}\,\MC=0$ implies $\n{d}\,\df{\theta}=0$. $\df{\theta}$ represents a connection 1-form of a principal connection that we call the chart-associated connection. The horizontal lift for the chart-associated Ehresmann connection is denoted $\phi_p$. It is given by 
\begin{align*}\phi_{p}:T_xX\to T_pP:\vec{v}_x\mapsto\phi_p\vec{v}_x = ({r}_g')_{s(x)} (s')_x\vec{v}_x,\end{align*} 
where $s$ is the chart-induced section in the fiber chart $(X,\varphi)$; see \ref{canonsec}. Here $(s')_x$ maps into a tangent space of the section, and $({r}_g')_{s(x)}$ transports the horizontal space into $p=r_g(s(x))$. The horizontal spaces are given by the tangent spaces to the leafs $S_g$ of the foliation; see \ref{foliation}.
In analogy to the maps $\sigma_p$ and $\Sigma$ we define
\begin{alignat*}{3}
\Phi:\,&\ves{k}(X,\group)\hookrightarrow\ves{k}(P):\,&&\bar{\vec{v}}\mapsto\vec{v},\quad&&\vec{v}_p=\phi_p\bar{\vec{v}}(g)_x.\end{alignat*}
The same properties hold for $\Phi$ and $\Phi^*$ as for $\Sigma$ and $\Sigma^*$. 
\ep
\labcount{verhor}
\bph{Horizontal and vertical maps}\label{par:verhor}
The vertical and horizontal maps which are related to the decomposition $\ves{k}(P)=\mathcal{X}_V^k(P)\oplus\mathcal{X}_H^k(P)$ can be represented as
\begin{alignat*}{5}
\n{hor}:\ves{k}(P)&\to\mathcal{X}_H^k(P)&\;:\;&\vec{v}\mapsto \Sigma\,\Pi\, \vec{v},\\
\n{ver}:\ves{k}(P)&\to\mathcal{X}_V^k(P)&\;:\;&\vec{v}\mapsto \Upsilon(\vec{v}),
\end{alignat*}
where $\n{hor}\oplus\n{ver}=\n{Id}_{\ves{k}(P)}$. The horizontal and vertical maps are independent of the fiber chart $(X,\varphi)$%
%, and are, therefore, defined on all of $P$
; see also \ref{asplit}. The maps naturally extend to (co)vector-valued fields; compare with \ref{extend}.
\ep
\bph{\texorpdfstring{Algebraic expressions for the $\n{hor}$ and $\n{ver}$ maps}{Algebraic expressions for the hor and ver maps}}\label{asplit}
Useful algebraic expressions for the hor and ver maps acting on vector fields in $\ves{}(P)$ follow from~\ref{par:verhor}
\begin{align*}
\n{hor} = \cont{\df{\omega}}\circ\mult{\vec{w}},\quad\quad\n{ver} = \mult{\vec{w}}\circ\cont{\df{\omega}}.
\end{align*}
They are used extensively in the splittings defined in \cite{Fecko1997,Kocik1998}, albeit with the four-velocity vector field and its metric Riesz dual in place of $\vec{w}$ and $\df{\omega}$. As compared to \cite{Fecko1997,Kocik1998}, the above expressions are not limited to orthogonal decompositions of the tangent space; see also the pre-metric, holonomic approach in \cite{hehlbook}.
\ep
\ifdefined\LONGVERSION
\bph{Parametric fields and Double fields on $X$ and $G$}\label{prodmani}
The space of parametric forms $\dfs{}(X,\group;\Wedge\mathfrak{g}^*)$ may be identified with the space of double forms $\dfs{0}(\group)\otimes\dfs{}(X;\Wedge\mathfrak{g}^*)$. Moreover the Maurer-Cartan form and vector fields can be canonically embedded in $\dfs{}(\group)\otimes\dfs{0}(X;\Wedge\mathfrak{g})$ and $\ves{}(\group)\otimes\dfs{0}(X;\Wedge\mathfrak{g}^*)$, respectively. We define the isomorphism
\begin{align*}
\xi:\ves{}(X,\group;\Wedge\mathfrak{g})\to\ves{}(\group)\otimes\ves{}(X)\;:\; \vec{v}\mapsto\lmult{\genwMC}\vec{v}
\end{align*}
with its inverse $\xi^{-1}=\lcont{\genMC}$, and the dual maps
\begin{align*}
\xi^{*}:\dfs{}(\group)\otimes\dfs{}(X)\to\dfs{}(X,\group;\Wedge\mathfrak{g}^*)\;:\; \df{\eta}\mapsto\lcont{\genwMC}\df{\eta}
\end{align*}
and its inverse $\xi^{-*}=\lmult{\genMC}$. The exterior and interior products in the maps are defined on the spaces of double multivector fields and double forms \cite{Labbi2005}, extended by the rules for (co)vector-valued fields of \ref{alg1} and \ref{alg2}.
{\issue exterior algebra needs to be treated before this paragraph, possibly already in introduction.}
\ep
\fi
%%%%%%%%%%%%%%%%%%%%%%%%%%%%%%%%%%%%%%%%%%%%%%
\section{Derivatives in the base manifold: Christoffel form, curvature, and variance}
This section describes derivative operators that arise naturally from a principal bundle equipped with an Ehresmann connection. Later on we will identify them as spatial and temporal derivatives of fields in the image of the splitting map. The pullback of the connection form into the base manifold for a given fiber chart is called the Christoffel form. The study of the derivatives of the Christoffel form allows insights into the geometry of the Ehresmann connection and its compatibility with the principal action. 
%The Christoffel form captures the incompatibility of the Ehresmann connection with a given fiber chart. Only in the absence of curvature, and only for principal connections, are there fiber charts that make the Christoffel form vanish.
% See AMP p. 236 for the definition of integral manifold of a differential system. On p. 243 it is pointed out that (at least locally) the integral manifold is a submanifold.
%
%For the scope of this and the following section, we omit the index of the fiber chart in the notation.
%In this section we extend the theory of principal connections on principal $\group$-bundles to general Ehresmann connections, restricting ourselves, however, to principal $\G$-bundles. 
%In a pre-metric setting, the notion of curvature takes a more abstract meaning. It is nonetheless not wrong to be guided by the metric intuition. As we shall see, the notions will naturally take their familiar form in the next section.
%
%Recall that an Ehresmann connection is a differentiable family of three-dimensional subspaces of tangent spaces in space-time. Flatness is the property that subspaces "patch together", forming three-dimensional submanifolds. A lack of flatness will become apparent in the (3+1)-dimensional equations that are derived form four-dimensional ones for a given observer. The Frobenius integrability conditions characterizes the Ehresmann connection.
%
\bphstar{Christoffel form}\label{gaugepot}
The Christoffel form (see \cite[9.7~and 11.4]{kolar}) or connection form in the base manifold (see \cite[p.~362]{amp}) is given by 
\begin{align*}\df{\Gamma}=\Phi^*\df{\omega}\in\dfs{1}(X,\group;\mathfrak{g}),\end{align*}
where we used $\Phi^*$ defined in \ref{canonconnec}.\footnote{Equivalently, we may define $\df{\Gamma}=-\Sigma^*\df{\theta}$, with $\df{\theta}$ defined in \ref{canonconnec}.}
The Christoffel form expresses the compatibility of the Ehresmann connection with the foliation encoded in the fiber chart; compare with \ref{foliation}. The Christoffel form of the Ehresmann connection is not to be confused with the Christoffel form of the Levi-Civita connection. The latter connection will be used only in Section~\ref{sec:kinec} for the discussion of an observer's kinematic parameters.
\epstar
\bph{Group derivative}\label{groupderiv}
We call group derivative the derivation
\[\groupderiv\;:\;\left\{
\begin{aligned}
\dfs{k}(X,\group)&\to\dfs{k}(X,\group;\mathfrak{g}^*)\\
\dfs{k}(X,\group;\mathfrak{g})&\to\dfs{k}(X,\group;\mathfrak{g}^*\otimes\mathfrak{g}),
\end{aligned}\right.
\]
that is defined in terms of the Lie derivative,
\begin{align*}
\groupderiv = \Sigma^*\circ\lie{\vec{w}}\circ\Pi^*.
\end{align*}
The Lie derivative by the Lie-coalgebra-valued vector field $\vec{w}$
\begin{align*}\lie{\vec{w}}\;:\;\dfs{k}(P)\to\dfs{k}(P;\mathfrak{g}^*)\end{align*}
is given with $\df{\gamma}\in\dfs{k}(P)$, $\vec{u}_\alpha\in\ves{1}(P)$, $\vec{v}\in\mathfrak{g}$ by
\begin{align*}
(\lie{\vec{w}}\df{\gamma})(\vec{u}_1,\dots,\vec{u}_k;\vec{v})=(\lie{\zeta(\vec{v})}\df{\gamma})(\vec{u}_1,\dots,\vec{u}_k).
\end{align*}
It extends to Lie-algebra-valued forms by $\lie{\vec{w}}(\df{\gamma}\otimes\vec{v})=(\lie{\vec{w}}\df{\gamma})\otimes\vec{v}$. 
%The group derivative $\groupderiv$ is equivalent to an exterior derivative if and only if the Lie group is Abelian, i.e., $\groupderiv=\xi^*\circ\n{d}_\group\circ\xi^{-*}$. An additional term 
%\begin{align*}
%\sum_{\substack{\beta,\gamma=0 \\ \beta<\gamma}}^\ell(-1)^{\beta+\gamma}\df{\eta}(\vec{u}_1,\dots,\vec{u}_k;[\vec{v}_\beta,\vec{v}_\gamma]_g,\vec{v}_0,\dots,\cancel{\vec{v}_{\beta}},\dots,\cancel{\vec{v}_{\gamma}},\dots,\vec{v}_\ell)
%\end{align*}
%would turn the group derivative into an exterior derivative on any Lie group.
%The group derivative therefore extends to Lie-algebra-valued forms, by
%\begin{align*}\groupderiv\;:\;\dfs{k}(X,\group;\mathfrak{g})\to\dfs{k}(X,\group;\mathfrak{g}^*\otimes\mathfrak{g}):\df{\gamma}\otimes\vec{v}\mapsto(\groupderiv\df{\gamma})\otimes\vec{v}.
%\end{align*}
The group derivative is closely related to the exterior derivative on $\group$.%
\footnote{The relationship holds only for Abelian Lie groups. An equivalent definition of the group derivative in the base manifold reads
$\big.(\groupderiv\df{\gamma})(\vec{u}_1,\dots,\vec{u}_k)\big|_x=\delta\mbox{\textasciicircum}\!\!\left[\bigl.\df{\gamma}(\vec{u}_1,\dots,\vec{u}_k)\bigr|_x\right].$
The operator $\delta\mbox{\textasciicircum}\!\!:C^\infty(\group)\to C^\infty(\group,\mathfrak{g}^*)$, defined in \cite[3.4]{Michor1987}, can be related, for Abelian Lie groups, to the exterior derivative on $\group$ by linear isomorphism.}
Accordingly, it trivially extends to Lie-coalgebra-valued forms by
\begin{align*}
\groupderiv:\dfs{k}(X,\group;\mathfrak{g}^*)\to 0.
\end{align*}
For a coordinate representation of the group derivative see \ref{coordtime}.
\ep
\bph{Exterior covariant derivative}\label{covderiv}
The exterior covariant derivative on the base manifold is given by 
\begin{align*}\n{D}=\Sigma^*\circ\n{d}\circ\Pi^*\;:\;\dfs{k}(X,\group;V)\to\dfs{k+1}(X,\group;V);\end{align*}
compare with \cite[11.5]{kolar} and \cite[p.~372]{amp}. For the finite dimensional vector space we have $V\in\{\mathbb{R},\mathfrak{g},\mathfrak{g}^*\}$ in mind.
The exterior covariant derivative is an antiderivation of degree one.
%It is shown in \cite{long} that
It can be shown that
\begin{align*}\n{D} &= \n{d}-(\df{\Gamma}\specialwedge\groupderiv).\end{align*}
The operator
$(\df{\Gamma}\specialwedge\groupderiv)$
is defined by 
\begin{gather*}(\df{\Gamma}\specialwedge\groupderiv)\;:\;\dfs{k}(X,\group;V)\to\dfs{k+1}(X,\group;V)\\
\df{\gamma}\otimes\vec{v}\mapsto (\df{\Gamma}\wedge\groupderiv\df{\gamma})\otimes\vec{v}.
\end{gather*}
For $V\ne\mathbb{R}$ the operator $(\df{\Gamma}\specialwedge\groupderiv)$ differs from $\mult{\df{\Gamma}}\circ\groupderiv$.\footnote{For example, for $\df{\gamma}\in\dfs{k}(X,\group;\mathfrak{g}^*)$ we find $\mult{\df{\Gamma}}\groupderiv\df{\gamma}=0$ because $\groupderiv\df{\gamma}=0$, while there holds $(\df{\Gamma}\specialwedge\groupderiv)\df{\gamma}\ne0$ in general.}
A coordinate representation of the exterior covariant derivative can be found in \ref{coordtime}.
\ep
\bph{Differentiation of twisted parametric forms}
The differential operators $\n{d}$, $\n{D}$, and $\groupderiv$ naturally extend to twisted forms, for example
\begin{align*}
\n{d}\;:\;\mathcal{F}^k_\doubletwist(X,\G)\to\mathcal{F}^{k+1}_\doubletwist(X,\G)\;:\;(\df{\alpha},\tilde{\df{\kappa}})\mapsto(\n{d}\df{\alpha},\tilde{\df{\kappa}}).
\end{align*}
\ep
\bph{Variance of the Ehresmann connection}\label{variance}\labcount{variance}
We compute the group derivative of the Christoffel form and obtain the {variance of the Ehresmann connection},\footnote{$\df{\chi}$ is called {\em torque of the connection} in \cite[Eq.~(2.8)]{Kocik1998}.}
\begin{align*}\df{\chi}=\groupderiv\df{\Gamma}\in\dfs{1}(X,\group;\mathfrak{g}^*\otimes\mathfrak{g}).\end{align*}
For $\df{\chi}=0$ the Christoffel form is {\em invariant} under the group action. %; see \cite{long}.
Principal connections are $\group$-{\em equivariant}; compare with \cite[11.1(3)]{kolar}.
\ifdefined\LONGVERSION
We find that $\lie{{\vec{v}}}\df{\omega}=-\n{ad}(\hat{\vec{v}})\,\df{\omega}$, where $\n{ad}=\n{Ad}'$, $\n{ad}(\vec{a})\vec{b}=[\vec{a},\vec{b}]$ for $\vec{a},\vec{b}\in\mathfrak{g}$; see \cite[p.~168]{amp}. It follows with $\vec{v}=\vec{w}(\hat{\vec{v}})=\zeta(\hat{\vec{v}})$ and $\df{\omega}=\df{\theta}+\Pi^*\df{\Gamma}$ that for principal connections
\begin{align*}\df{\chi}=-\n{ad}\,\df{\Gamma}.\end{align*}
\fi
For Abelian Lie groups, invariance equals equivariance, so $\df{\chi}=0$ implies that the connection is principal; see \ref{avariance}.
\ep
\bph{Curvature of the Ehresmann connection}\label{curvature}
We generalize the definition of curvature for principal connections in \cite[11.2]{kolar} and \cite[p.~372]{amp} to nonprincipal connections by
\begin{align*} \df{\Omega} = \n{D}\,\df{\Gamma}\in\dfs{2}(X,\group;\mathfrak{g}).\end{align*}
%It is shown in \cite{long} that this 
This definition of curvature is equivalent to the definition for Ehresmann connections on general fiber bundles in \cite[9.4]{kolar}. An Ehresmann connection is said to be {\em flat} iff $\df{\Omega}= 0$.
\ep
\bph{\texorpdfstring{Relationship between $\n{D}$ and $\groupderiv$}{Relationship between D and dG}}\label{relationship_DG}
From the complex property $\n{d}\circ\n{d}=0$ of the exterior derivative on the bundle manifold it can be inferred with \ref{ddecomp} that
\ifdefined\LONGVERSION
\begin{align*}
\groupderiv\circ\groupderiv &= 0,\\
(\n{D}-\mult{\df{\chi}})\circ\groupderiv &= \groupderiv\circ(\n{D}-\mult{\df{\chi}}),\\
(\n{D}-\mult{\df{\chi}})^2&=(\groupderiv\circ\mult{\df{\Omega}}-\mult{\df{\Omega}}\circ\groupderiv)\circ\n{m},\\
\mult{\df{\Omega}}\circ(\n{D}-\mult{\df{\chi}})&=(\n{D}-\mult{\df{\chi}})\circ\mult{\df{\Omega}},\\
\mult{\df{\Omega}}\circ\mult{\df{\Omega}} &= 0.
\end{align*}
\else
\begin{align*}
\left.
\begin{aligned}
\protect[\n{D},\groupderiv]&=\mult{\df{\chi}}\circ\groupderiv\\
\n{D}^2&=-\mult{\df{\Omega}}\circ\groupderiv
\end{aligned}
\quad\right\}&\text{ on }\dfs{k}(X,\group)\text{ and }\dfs{k}(X,\group;\mathfrak{g}),\\
% Proof for the vector-valued forms by writing the relation for scalar-valued forms in
% a basis representation and considering the tensor product with a Lie algebra vector
% from the left.
\left.
\begin{aligned}
(\n{D}-\mult{\df{\chi}})^2&=-\groupderiv\circ\mult{\df{\Omega}}\\
[\n{D},\mult{\df{\Omega}}]&=-\mult{\df{\Omega}}\circ\mult{\df{\chi}}
\end{aligned}
\quad\right\}&\text{ on }\dfs{k}(X,\group;\mathfrak{g}^*).
\end{align*}
\fi
Note that in general neither do $\n{D}$ and $\groupderiv$ commute, nor does $\n{D}$ enjoy the complex property.
\ep
\bph{Relationship between curvature and variance}
It follows from the definition of the curvature form with \ref{relationship_DG} that % Proof can be found 01.11.12, p.2
\begin{align*} \df{\Omega} = \n{d}\,\df{\Gamma} -\df{\Gamma}\wedge\df{\chi}.\end{align*}
Moreover, by applying the relations for $[\n{D},\groupderiv]$ and $\n{D}^2$ from \ref{relationship_DG} to the Christoffel form, the following differential equations are obtained for the curvature and variance forms: % Proof can be found 20.02.13, p.2
\begin{align*}
\groupderiv\,\df{\Omega}-\n{D}\,\df{\chi}&=0,\\ %-\df{\chi}\wedge\df{\chi},\\
\n{D}\,\df{\Omega}+\df{\Omega}\wedge\df{\chi}&=0.
\end{align*}
For principal connections, we obtain from $\df{\chi}=0$ in \ref{variance} the invariance of the curvature form $\groupderiv\,\df{\Omega}=0$ and the {\em Bianchi identity} $\n{D}\,\df{\Omega}=0$; compare with \cite[p.~375]{amp}.
\ep
\bph{Adapted coordinate chart}\label{adaptchart}
A chart on $\G$ is called adapted if it is a Lie-group homo\-morphism $\phi:\G\to(\mathbb{R},+)$, that is, 
\begin{align*}\phi(gh)&=\phi(g)+\phi(h), 
\intertext{fulfilling} 
\phi(e)&=0.\end{align*}
\ep
\bph{Coordinate time and coordinate-expressions for derivatives}\label{coordtime} 
Coordinate time is the codomain of an adapted chart on $\G$. 
Selecting a basis $\vec{e}\in\mathfrak{\g}$ induces an adapted chart via the exponential map $\exp:\mathfrak{\g}\to \G$; see \cite[p.~160]{amp}, \cite[4.18]{kolar}. The exponential map of the translation group is a global diffeomorphism; compare with \cite[p.~131]{kolar}. Hence we define the adapted chart 
\begin{align*}\phi: \G\to(\mathbb{R},+):\g\mapsto\coordtime,\quad \g=\exp(\coordtime{\mathbf{e}}).
\end{align*}
% Compare Kolar, p. 131, from which it follows that the exponential map is a global diffeomorphism in our case. Note that every Abelian Lie algebra is solvable, see http://en.wikipedia.org/wiki/Solvable_Lie_algebra. A Lie group is solvable if its Lie algebra is solvable.
By introducing also coordinates $(x^i)$ in the base manifold we obtain the following coordinate representations of differential operators acting on $\df{\alpha}\in\dfs{k}(X,G)$:
\begin{align*}
\timederiv\,\df{\alpha}&=\partial_\coordtime\df{\alpha}(x^i,\coordtime)\otimes\n{d}\coordtime,\\
\n{d}\,\df{\alpha}&=\n{d}x^i\wedge\partial_{x^i}\df{\alpha}(x^i,\coordtime),\\
\n{D}\,\df{\alpha}&=\n{d}x^i\wedge(\partial_{x^i}-\Gamma_i\partial_\coordtime)\df{\alpha}(x^i,\coordtime),
\end{align*}
where $\df{\Gamma}=\Gamma_i\n{d}x^i\otimes\partial_\coordtime$ and the summation convention has been used; compare with \ref{relationship_DG}. The covariant exterior derivative is in general a combination of spatial and temporal derivatives. More on adapted coordinates and frames can be found in Appendix~\ref{amoreadapt}.
\ep
\bph{Frobenius integrability condition}\label{frobenius}\labcount{frobenius}
We call an Ehresmann connection inte\-grable if the exterior differential equation $\df{\omega}=0$ is completely integrable; see \cite[pp.~236ff]{amp}. The Frobenius integrability condition \[(\n{d}\df{\omega}\wedge\df\omega)^\otimes=0\] is necessary and sufficient for complete integrability; compare with \cite[p.~245]{amp}.
\ifdefined\LONGVERSION
For $l>1$ the condition $\n{d}\df{\omega}\wedge\df{\omega}^l(\df{\varepsilon}_l)\in\dfs{l+2}(P;\mathfrak{g})=0$ has to be considered. These are $l$ conditions in terms of $(l+2)$-forms, $\df{\varepsilon}_l\in\Lambda^l\mathfrak{g}^*$. This gives exactly the conditions from \cite[p.~245]{amp}.
\fi
The following conditions are equivalent:
\begin{enumerate}
\item The Ehresmann connection is integrable.
\item The Ehresmann connection is flat, $\df{\Omega}=0$; see \ref{afrobenius}.
\item The connection 1-form can be written ${\df{\omega}} = f\n{d}{\nu}$, where $f\in C^\infty(P,\mathfrak{g}^*\otimes\mathfrak{g})$, $f$ nonzero, $\nu\in C^\infty(P,\mathfrak{g})$.\footnote{i) The exterior differential equation $\df{\omega}=0$ is said to be algebraically equivalent to $\n{d}{\nu}=0$. The function $\nu$ is called first integral of the equation $\df{\omega}=0$ \cite[p.~243]{amp}. ii) This construction works globally if $X$ is connected, and the Ehresmann connection is complete, with trivial holonomy group \cite[9.9, 9.10]{kolar}.}
\item \label{case4}Recall the partition of a $\n{Diff}(\group)$-bundle into principal $\group$-bundles of \ref{partition}; there exist fiber charts in the maximal atlas of the $\n{Diff}(\group)$-bundle such that the Christoffel form vanishes. An instance can be obtained by $\hhat{\varphi}:P\to\G:p\mapsto\exp\bigl(\nu(p)\bigr)$.
% Compare also 15.03.13, pp. 1-2, where this situation has been regarded as a change of fiber chart.
% To show that $\varphi_j$ is a fiber chart it must be a diffeomorphism. This can (at least locally) be shown by the inverse function theorem, which says that the tangent map must be invertible in each point, i.e., it has a trivial kernel. In our case it means that $\n{Ker\,}\pi^\prime\cap\n{Ker\,}\hhat{\varphi}_j^\prime=\{0\}$. In other words, $\vec{w}_i$, which spans $\n{Ker\,}\pi^\prime$ must not lie in the kernel of $\hhat{\varphi}_j^\prime$. From 15.03.13, p.~1 (which gives an explicit expression for $\hhat{\varphi}_j^\prime$) it follows that $\n{d}\nu(\vec{w}_i)\ne 0$ is required. This is guaranteed everywhere, since $f$ is nonzero.
% In this chart, $\df{\theta}=\n{d}\nu$.
\end{enumerate}
In case \ref{case4}, the foliation induced by the chart (see \ref{foliation}) is called {\em horizontal foliation} \cite[9.6]{kolar}. Each point $p\in P$ lies on a unique leaf $S_g$, $g=\hhat{\varphi}(p)$, such that $T_pS_g=H_pP$ for each $p\in S_g$. Moreover, the chart-associated connection (see \ref{canonconnec}) agrees with the given Ehresmann connection, and for the connection 1-forms it holds that $\df{\omega}=f\df{\theta}$.
\ep
%%%%%%%%%%%%%%%%%%%%%%%%%%%%%%%%%%%%%%%%%%%%%%
\ifdefined\LONGVERSION
\section{The splitting map}\label{sec:splitting}\labcount{splitting}
In this section we generalize the Maurer-Cartan form and vector field, the related fields $\vec{w}$ and $\df{\theta}$, as well as the connection 1-form $\df{\omega}$, so that they take values in the exterior algebras $\Wedge\mathfrak{g}$ and $\Wedge\mathfrak{g}^*$, rather than in $\mathfrak{g}$ and $\mathfrak{g}^*$. %For example, let $(\vec{e}_1,\vec{e}_2)$ be a basis of a two-dimensional Lie algebra $\mathfrak{g}$, then a basis of $\Wedge\mathfrak{g}$ is given by $(1,\vec{e}_1,\vec{e}_2,\vec{e}_1\wedge\vec{e}_2)$. For $q$ being the dimension of $\mathfrak{g}$, the dimension of $\Wedge\mathfrak{g}$ is $2^q$. 
The generalizations are used in the definition of the splitting map and its inverse, relating multivector fields and forms on the bundle manifold to parametric multivector fields and forms on $X$ that take values in $\Wedge\mathfrak{g}$ or $\Wedge\mathfrak{g}^*$, respectively. We start by laying out a few conventions for algebraic operations involving (co)vector-valued fields.
\bph{Algebraic operations for (co)vector-valued fields}\label{alg1}
Let $V$ denote a vector space, typically $\mathfrak{g}$ or $\mathfrak{g}^*$. The rules for interior and exterior products involving differential forms $\dfs{}(\,\cdot\,)$ and vector-valued fields are given by 
\begin{alignat*}{5}\cont{}&\,:\,\ves{}(\,\cdot\,; \Wedge V)\times\dfs{}(\,\cdot\,)&\to\dfs{}(\,\cdot\,; \Wedge V)&\,:\,&(\vec{m}\otimes v,\df{\eta})&\mapsto\bigoplus_{k=0}^q\cont{\vec{m}^k}\df{\eta}\otimes v^k,\\
\mult{}&\,:\,\dfs{}(\,\cdot\,; \Wedge V)\times\dfs{}(\,\cdot\,)&\to\dfs{}(\,\cdot\,; \Wedge V)&\,:\,&(\df{\mu}\otimes v,\df{\eta})&\mapsto\bigoplus_{k=0}^q\mult{\df{\mu}^k}\df{\eta}\otimes v^k,\end{alignat*}
where $\vec{m}^k\otimes v^k\in\ves{}(\,\cdot\,;\Wedge^kV)$, $\df{\mu}^k\otimes v^k\in\dfs{}(\,\cdot\,;\Wedge^kV)$. %Moreover,
%\begin{alignat*}{5}\cont{}&\,:\,\ves{}(\,\cdot\,; \Wedge V)\times\dfs{}(\,\cdot\,; \Wedge V)&\to\dfs{}(\,\cdot\,; \Wedge V)&\,:\,&(\vec{m}\otimes v,\df{\eta}\otimes w)&\mapsto\bigoplus_{k=0}^q\cont{\vec{m}^k}\df{\eta}\otimes v^k,\\
%\mult{}&\,:\,\dfs{}(\,\cdot\,; \Wedge V)\times\dfs{}(\,\cdot\,; \Wedge V)&\to\dfs{}(\,\cdot\,; \Wedge V)&\,:\,&(\df{\mu}\otimes v,\df{\eta})&\mapsto\bigoplus_{k=0}^q\mult{\df{\mu}^k}\df{\eta}\otimes v^k.\end{alignat*}
%Note that we distinguish between $\dfs{}(\,\cdot\,)$ and $\dfs{}(\,\cdot\,;\Wedge V)$, even though $\dfs{}(\,\cdot\,;\Wedge V^0)\subset \dfs{}(\,\cdot\,;\Wedge V)$ and $\dfs{}(\,\cdot\,)$ may be identified.
The dot in the notation is a placeholder for either $P$, $\group$, or, in the case of parametric fields, $(X,\group)$. We define products between covector-valued forms and vector-valued multivector fields by
\begin{alignat*}{5}
\cont{}&\,:\,\ves{}(\,\cdot\,;\Wedge \mathfrak{g})\times\dfs{}(\,\cdot\,;\Wedge \mathfrak{g}^*)&\to\dfs{}(\,\cdot\,)&\,:\,&(\vec{m}\otimes \vec{v},\df{\mu}\otimes\df{\gamma})&\mapsto\bigoplus_{k=0}^q\cont{\vec{m}^k}\df{\mu}^k\otimes \df{\gamma}^k(\vec{v}^k),\\
\mult{}&\,:\,\dfs{}(\,\cdot\,;\Wedge \mathfrak{g})\times\dfs{}(\,\cdot\,;\Wedge \mathfrak{g}^*)&\to\dfs{}(\,\cdot\,)&\,:\,&(\df{\gamma}\otimes \vec{v},\df{\mu}\otimes\df{\gamma})&\mapsto\bigoplus_{k=0}^q\mult{\df{\gamma}^k}\df{\mu}^k\otimes \df{\gamma}^k(\vec{v}^k);\end{alignat*}
see \cite[1.5, 1.16, 1.21]{greub2}. Dual rules are obtained by interchanging the roles of $\dfs{}$ and $\ves{}$ as well as $\Wedge\mathfrak{g}$ and $\Wedge\mathfrak{g}^*$ in the above definitions. {\issue Problem: The splitting map requires the definition of the exterior product here above, whereas in the splitting of the exterior derivative, for example, for $\mult{\df{\Omega}}$ (see \ref{decompd}), we need
\begin{alignat*}{5}\mult{}&\,:\,\dfs{}(\,\cdot\,;\Wedge \mathfrak{g})\times\dfs{}(\,\cdot\,;\Wedge \mathfrak{g}^*)&\to\dfs{}(\,\cdot\,)&\,:\,&(\df{\gamma}\otimes \vec{v},\df{\mu}\otimes\df{\nu})&\mapsto\bigoplus_{k=0}^q\bigoplus_{\ell=0}^q\mult{\df{\gamma}^\ell}\df{\mu}^k\otimes \cont{\vec{v}^\ell}\df{\nu}^k.\end{alignat*}
Analogue definitions for operations with $\df{\chi}\in\dfs{1}(X,\group;\mathfrak{g}^*\otimes\mathfrak{g})$ should follow.
}
\ep
\bph{Algebraic operations for fields related to unit tensors}\label{alg2}
The exterior powers of vector-valued forms $\MC$, $\df{\theta}$ and $\df{\omega}$ follow from
\begin{alignat*}{6}\mult{}&\,:\,\dfs{}(\,\cdot\,;\Wedge \mathfrak{g})\times\dfs{}(\,\cdot\,;\Wedge \mathfrak{g})&\;\to\;&\dfs{}(\,\cdot\,;\Wedge \mathfrak{g})&\;:\;&(\df{\mu}\otimes\vec{u},\df{\eta}\otimes\vec{v})&\;\mapsto\;& \df{\mu}\wedge\df{\eta}\otimes\vec{u}\wedge\vec{v},
\intertext{and for covector-valued multivector fields $\wMC$ and $\vec{w}$}
\mult{}&\,:\,\ves{}(\,\cdot\,;\Wedge \mathfrak{g}^*)\times\ves{}(\,\cdot\,;\Wedge \mathfrak{g}^*)&\;\to\;&\ves{}(\,\cdot\,;\Wedge \mathfrak{g}^*)&\;:\;&(\vec{u}\otimes\df{\mu},\vec{v}\otimes\df{\eta})&\;\mapsto\;& \vec{u}\wedge\vec{v}\otimes\df{\mu}\wedge\df{\eta};\end{alignat*}
compare with \ref{unittens}. We define the interior product between the above covector-valued multivector fields and vector-valued forms by
\begin{align*}\cont{}\,:\,\ves{}(\,\cdot\,;\Wedge \mathfrak{g}^*)\times\dfs{}(\,\cdot\,;\Wedge \mathfrak{g})&\to\dfs{}(\,\cdot\,;\Wedge\mathfrak{g}^*\otimes\Wedge\mathfrak{g})\\(\vec{m}\otimes \df{\nu},\df{\mu}\otimes\vec{v})&\mapsto\bigoplus_{k=0}^q\cont{\vec{m}^k}\df{\mu}^k\; \df{\nu}^k\otimes\vec{v}^k;
\end{align*}
compare with \ref{Qop}. 
\ep
\bph{Generalized Maurer-Cartan form}\label{MC3}
Recall from \ref{MC}-\ref{MC2} the characterization of the Maurer-Cartan form and vector field via the unit tensors $\MC\in(\mathfrak{g}^*,\mathfrak{g})$ and $\wMC\in(\mathfrak{g},\mathfrak{g}^*)$, respectively. The generalization to unit tensors $\genwMC$ and $\genMC$ of $\Wedge(\mathfrak{g},\mathfrak{g}^*)$ and $\Wedge(\mathfrak{g}^*,\mathfrak{g})$, respectively, is defined in \ref{unittens} and \ref{Qop}. For example, let $\group$ be a two-dimensional Lie group, and $(\vec{e}_1,\vec{e}_2)$ and $(\df{\varepsilon}^1,\df{\varepsilon}^2)$ dual bases of its Lie algebra $\mathfrak{g}$ and Lie coalgebra $\mathfrak{g}^*$, respectively. Then 
\begin{align*}
\wMC&=\vec{e}_1\otimes\df{\varepsilon}^1+\vec{e}_2\otimes\df{\varepsilon}^2
\intertext{and}
\genwMC &= (1\otimes 1,\vec{e}_1\otimes\df{\varepsilon}^1+\vec{e}_2\otimes\df{\varepsilon}^2,\vec{e}_1\wedge\vec{e}_2\otimes\df{\varepsilon}^1\wedge\df{\varepsilon}^2).\end{align*} 
\ep
\bph{Generalized chart-associated form and fundamental field}
Along the same lines as in \ref{MC2a} and \ref{MC2b}, we interpret $\Wedge\wMC$ as a multivector field, $\Wedge\wMC\in\ves{}(\group;\Wedge\mathfrak{g}^*)$, and $\Wedge\MC$ as a form, $\Wedge\MC\in\dfs{}(\group;\Wedge\mathfrak{g})$, and proceed to generalize: Denote $\Wedge{\df{\theta}}_i\in\dfs{}(P;\Wedge\mathfrak{g})$ the form that is defined for a fiber chart $(U_i,\varphi_i)$ by 
\begin{align*}\Wedge{\df{\theta}}_i=\hhat{\varphi}_i^{*}\genMC.\end{align*} 
$\Wedge\df{\theta}$ denotes the generalized chart-associated connection that is obtained from the individual $\Wedge\df{\theta}_i$ with the partition-of-unity argument of \ref{canonconnec}.
Moreover, \label{MCpush1}denote $\Wedge{\vec{w}}\in\ves{}(P;\Wedge\mathfrak{g}^*)$ the image under the fundamental field map,
\begin{align*}\Wedge{\vec{w}} = \zeta(\genwMC).\end{align*} 
It follows that $\Wedge{\df{\theta}}(\Wedge{\vec{w}}_p) =\genMC(\genwMC)=\genMC$ for all $p\in P$.%, where $q=\n{dim}(\mathfrak{g})$ and $2^q=\n{dim}(\Wedge\mathfrak{g})$.
%We note that $\vec{w}(\hat{\vec{v}})=\vec{v}=\zeta(\hat{\vec{v}})$, for all $\hat{\vec{v}}\in\mathfrak{g}$, i.e., $\vec{w}$ maps Lie-algebra vectors into flow fields of the principal right action on the bundle manifold.
\ep
\bph{Generalized connection form}\label{gencon}
In analogy to \ref{unittens}, we also generalize the connection 1-form $\df{\omega}\in\dfs{1}(P;\mathfrak{g})$, setting
\begin{alignat*}{2}
\Wedge^k\df{\omega}&=\frac{1}{k!}\underbrace{\df{\omega}\wedge\dots\wedge\df{\omega}}_{\times k},&\quad\quad\Wedge^k\df{\omega}&\in\mathcal{F}^{k}_V(P;\Wedge^k\mathfrak{g}),\\
\Wedge\df{\omega}&=\bigoplus_{k=0}^q\Wedge^k\df{\omega},&\quad\quad\Wedge\df{\omega}&\in\mathcal{F}_V(P;\Wedge\mathfrak{g}),
\end{alignat*}
where $\df{\omega}^0=1\otimes 1$.
\ep
\bph{Splitting map}\label{gensplitting}
For a given Ehresmann connection, the splitting map of vector fields $\n{S}:\ves{}(P)\to\ves{}(X,\group;\Wedge\mathfrak{g})$
is defined by
\begin{align*} \n{S}&={\Pi}\circ\lcont{\Wedge{\df{\omega}}}.
\intertext{It respects the grading in that it maps }
\n{S}\,:\,\ves{k}(P)&\to \bigoplus_{i+j=k}\ves{i}(X,\group;\Wedge^j\mathfrak{g}).
\intertext{The inverse is given by (for a proof see \cite{long})}
\n{S}^{-1}&=\lmult{\Wedge{\vec{w}}}\circ{\Sigma}\;,
\intertext{where $\lmult{\Wedge{\vec{w}}}:\ves{}(P;\Wedge\mathfrak{g})\to\ves{}(P)$. For differential forms, the dual map reads $\n{S}^{*}:\dfs{}(X,\group;\Wedge\mathfrak{g}^*)\to\dfs{}(P)$,
}
\n{S}^{*}&=\lmult{\Wedge{\df{\omega}}}\circ\Pi^*,
\intertext{and its inverse is given by}
 \n{S}^{-*}&=\Sigma^*\circ\lcont{\Wedge{\vec{w}}}.
 \end{align*}
\ep
\bph{Algebraic expressions for the $\n{hor}$ and $\n{ver}$ maps}\label{genasplit}
Algebraic expressions for the $\n{hor}$ and $\n{ver}$ maps acting on vector fields in $\ves{}(P)$ are given by \cite{long}
\begin{align*}
\n{hor}=\lcont{\Wedge^q\vec{w}}\circ\lmult{\Wedge^q\df{\omega}},\quad\quad
\n{ver}=\sum_{j=1}^q(-1)^{j+1}\lmult{\Wedge^j\vec{w}}\circ\lcont{\Wedge^j\df{\omega}}.
\end{align*}
\ep
\bph{Integrals on $\group$ of split forms}
Integrals of parametric, Lie-coalgebra valued forms $\df{\eta}\in\dfs{}(X,\group;\Wedge\mathfrak{g}^*)$ on the Lie group are defined as
\[
\int_{\interval}\df{\eta}\stackrel{\n{def}}{=}\int_{\interval}\xi^{-*}\,\df{\eta},\quad H\subset\group.
\]
\ep
\bph{Splitting of the exterior derivative}\label{decompd}
The operation $\n{S}^{-*}\circ\n{d}\circ\n{S}^{*}$ yields a splitting of the exterior derivative in terms of operators acting on parametric fields:
\begin{align*}\n{S}^{-*}\circ\n{d}\circ\n{S}^{*}=(\groupderiv+\n{D}+\mult{\df{\chi}})\circ\n{m}+\mult{\df{\Omega}};\end{align*}
compare with \ref{signop} for the definition of the sign operator $\n{m}$. {\issue The algebraic definition of $\mult{\df{\chi}}$ and $\mult{\df{\Omega}}$ is discussed in \ref{alg1}.}
\ep
\fi
%
%%%%%%%%%%%%%%%%%%%%%%%%%%%%%%%%%%%%%%%%%%%%%%
\section{The relativistic splitting structure}\label{sec:pmobserver}
We define a relativistic splitting structure with its splitting map. The structure is an instance of the above theory, where the bundle manifold is a four-dimensional space-time, and the fibers are world-lines. Note that splitting structures are useful tools for dimensional reduction in general. In \ref{axialsplit} we use an axial splitting structure to obtain a two-dimensional field problem from a three-dimensional one with axial symmetry.
\bph{Relativistic splitting structure}\label{pmobserver}
A relativistic splitting structure -- or short {\em splitting} -- on a space-time $P$ consists of the following data:
\begin{enumerate}
\item\label{pmobserver1}\sloppy A principal $\group$-bundle in the partition of the fiber bundle $\big(P,\pi,X,\G,$ $\n{Diff}(\G)\big)$; see \ref{partition}. The fibers of the bundle are world-lines; see \ref{world-line}.
\item\label{pmobserver2}An Ehresmann connection.
\item\label{pmobserver3}A global section.
\end{enumerate}
Elements \ref{pmobserver1} and \ref{pmobserver3} select a fiber chart from the principal $\group$-bundle's maximal atlas. All principal $\group$-bundles from the partition of the $\n{Diff}(\G)$-bundle, and all global sections are equivalent; see Section~\ref{sec:transitions}. Some splittings may, however, be more practical than others; see \ref{classobs}. 
\ep
\bph{Time synchronization, time translation, and simultaneity structure}\label{timesynch}
The global section of the splitting structure
%chart-induced section induced by a fiber chart 
models a time synchronization; compare with \ref{canonsec}. 
The principal action of the principal $\G$-bundle, translating events along world-lines, is called time translation.
A simultaneity structure is given by the foliation of~\ref{foliation}.
\ep
\bph{Splitting of vector fields and forms}\label{psipush}
\ifdefined\LONGVERSION
The exterior algebra $\Wedge\mathfrak{\g}$ is two-dimensional, and so the splitting map of \ref{gensplitting} applied to a $k$-vector field produces a pair of a horizontal parametric $k$-vector field, and a vertical parametric Lie-algebra-valued $(k-1)$-vector field,
\else
The splitting map and its dual relate multivector fields and forms on the bundle manifold to parametric multivector fields and forms on $X$, respectively, that take values in $\mathbb{R}$, $\mathfrak{g}$ or $\mathfrak{g}^*$. For a given splitting structure, the splitting map of vector fields is defined by
\fi
\begin{align*}
\n{S}:\ves{k}(P)\xrightarrow{\,\,\sim\,\,}\ves{k}(X,\G)\times\ves{k-1}(X,\G;\mathfrak{\g}):
\vec{v}\mapsto(\vec{\atest},\tilde{\df{\btest}})=(\Pi\,\vec{v},\Pi\,\cont{\df{\omega}}\vec{v});
\end{align*}
see \ref{ehresconnonprinc} and \ref{bigphi}. We denote Lie-algebra-valued vector fields in the image of $\n{S}$ with a tilde (idem for forms). The inverse is given by
\begin{align*} \n{S}^{-1}:(\vec{\atest},\tilde{\df{\btest}})\mapsto\vec{v}=\Sigma\,\vec{\atest}+\vec{w}\wedge\Sigma\,\tilde{\df{\btest}}.\end{align*}
\label{psipull}On the dual side, the splitting map of differential $k$-forms reads
\begin{align*}\n{S}^{-*}:\dfs{k}(P)\xrightarrow{\,\,\sim\,\,}\dfs{k}(X,\G)\times\dfs{k-1}(X,\G;\mathfrak{\g}^*):
\df{\gamma}\mapsto(\df{\alpha},\tilde{\df{\beta}})=(\Sigma^*\df{\gamma},\Sigma^*\cont{\vec{w}}\df{\gamma}).\end{align*}
The inverse is given by
\begin{align*} \n{S}^{*}:(\df{\alpha},\tilde{\df{\beta}})\mapsto\df{\gamma}=\Pi^*\df{\alpha}+\df{\omega} \wedge\Pi^*\tilde{\df{\beta}};\end{align*}
see \ref{MCpush} and \ref{bigphi}. 
\ep
\bph{Splitting of Lie-algebra-valued forms}
The splitting map extends to Lie-algebra-valued forms,
\begin{align*}
\n{S}^{-*}:\dfs{k}(P;\mathfrak{\g})\xrightarrow{\,\,\sim\,\,}\dfs{k}(X,\G;\mathfrak{\g})\times\dfs{k-1}(X,\G;\mathfrak{\g}^*\otimes\mathfrak{\g}):
\df{\gamma}\mapsto(\Sigma^*\df{\gamma},\Sigma^*\cont{\vec{w}}^\otimes\df{\gamma}),
\end{align*}
and it can be shown that \[(\df{\Omega},\df{\chi}) = \n{S}^{-*}(\n{d}\,\df{\omega}).\] 
\ep
\bph{Splitting of twisted forms}\label{twistsplit}
% In more detail: A principal bundle with the translation group as structure group is trivializable. $P$ being the bundle space, it follows that $P$ is diffeomorphic to $X\times\group$ by a global chart.
The splitting map extends to twisted differential forms by
\begin{gather*}
\n{S}^{-*}\;:\;\mathcal{F}^k_\twist(P)\to\mathcal{F}^k_\doubletwist(X,\G)\times\mathcal{F}^{k-1}_\doubletwist(X,\G;\mathfrak{g}^*)\\
(\df{\gamma},\df{\kappa})\mapsto\bigl((\df{\alpha},\tilde{\df{\kappa}}),(\tilde{\df{\beta}},\tilde{\df{\kappa}})\bigr),
\end{gather*}
where $(\df{\alpha},\tilde{\df{\beta}})=\n{S}^{-*}\df{\gamma}$, $(0,\tilde{\df{\kappa}})=\n{S}^{-*}\df{\kappa}$, and $\df{\kappa}$ is a volume form on $P$.
\ep
\bph{Change of Ehresmann connection}\label{changeconnec}
A change of the Ehresmann connection from the connection 1-form $\df{\omega}_\alpha$ to $\df{\omega}_\beta$, while retaining the fiber chart, yields a new splitting; the effect on the split forms is given by
\begin{align*}
\n{S}^{-*}_\beta\circ\n{S}^*_\alpha=
\begin{pmatrix}\n{Id}&\lmult{\df{\Gamma}_\alpha-\df{\Gamma}_\beta}\\0&\n{Id}\end{pmatrix},
\end{align*}
where $\df{\Gamma}_\alpha$, $\df{\Gamma}_\beta$ are the Christoffel forms; compare with \ref{gaugepot}. For convenience, we write the domain of $\mathrm{S}^*$ as a column vector. % Proof can be found 24.03.12, p.1
\ep
\ifdefined\LONGVERSION
\bph{Group derivative and time-variance form}\label{propcurvature}
Recall from \ref{groupderiv} the definition of the group derivative. In the {\issu splitting structure}, it can be written 
\begin{align*}\timederiv=\Sigma^*\circ\lie{\vec{w}}\circ\Pi^*.\end{align*}
In this context, a vanishing group derivative of the Christoffel form
\begin{align*}\df{\chi}=\timederiv\,\df{\Gamma}\in\dfs{1}(X,\G;\mathfrak{t}^*\otimes\mathfrak{t}),\end{align*}
indicates a principal connection.\footnote{$\df{\chi}$ is called {\em torque} in \cite[Eq.~(2.8)]{Kocik1998}.} The Ehresmann connection is {\em principal} iff $\df{\chi} = 0$; see \ref{avariance}. 
\ep
\fi
\bph{Splitting of the exterior derivative}\labcount{splitop}\label{splitd}
The operation $\n{S}^{-*}\circ\n{d}\circ\n{S}^{*}$ yields a splitting of the exterior derivative in terms of operators acting on parametric differential forms in the base manifold. The splitting is given by\footnote{A similar result for scalar-valued forms in the bundle manifold can be found in \cite[Eq.~(5.2)]{Fecko1997}. Note that rows and columns of the matrices are arranged differently.}
\begin{align*}\n{S}^{-*}\circ\n{d}\circ\n{S}^{*}
=\begin{pmatrix}{{\n{D}}}&\mult{{\df{\Omega}}}\\\timederiv&\;\mult{\df{\chi}}- {\n{D}}\end{pmatrix}.\end{align*}
We notice that for holonomic splittings $\df{\Omega}=\df{\chi}=0$. Moreover, in the natural splitting we have $\n{D} = \n{d}$; the splitting of the exterior derivative takes the simple form\footnote{After introduction of a basis of the Lie algebra this coincides with \cite[Eq.~(B.1.28)]{hehlbook}.}
\begin{align*}\n{S}^{-*}\circ\n{d}\circ\n{S}^{*}
=\begin{pmatrix}\n{d}&0\\\timederiv&\;-\n{d}\end{pmatrix}.\end{align*}
More on the splitting of operators can be found in Appendix~\ref{splitop} and \ref{ddecomp}. 
\ep
\bph{Classification of splitting structures}\label{classobs}
\begin{itemize}
\item A splitting is {\em flat} or {\em integrable} if $\df{\Omega} = 0$ and, equivalently, the Frobenius integrability condition is fulfilled.
\item A splitting is {\em principal} if the Ehresmann connection is a principal connection, $\df{\chi} = 0$.
\item A {\em holonomic} splitting\footnote{In frame-based approaches, {\em holonomity} is a property of a frame field. In one version, holonomity is defined via a coframe basis $(\df{\varepsilon}^\mu)$ that is closed, that is, $\n{d}\,\df{\varepsilon}^\mu=0$. In our setting we are concerned only with the temporal direction, see Appendix \ref{amoreadapt}. Note that in \cite{Kocik1998} holonomity is used as a synonym for integrability.} is both, flat and principal. Its connection 1-form is closed, $\n{d}\,\df{\omega}=0$, hence, $\df{\chi}=\df{\Omega}=0$.
\item A {\em natural} splitting is characterized by $\df{\Gamma}=0$, and, equivalently, by $\df{\omega}=\df{\theta}$. This implies the holonomic property.
\end{itemize}
Given an integrable splitting, there exists always an equivalent (see \ref{pmobserver}) natural splitting; compare with \ref{frobenius}, condition \ref{case4}.%
\footnote{Using a fiber chart that yields $\df{\Gamma}=0$ for a flat Ehresmann connection is in analogy to using Cartesian coordinates for flat Euclidean space, so that the Christoffel symbols vanish.}
For a nonintegrable splitting there does not always exist an equivalent principal splitting, though; compare with \cite[Eq.~(39)]{Minguzzi2003}.
\ep%
%
%%%%%%%%%%%%%%%%%%%%%%%%%%%%%%%%%%%%%%%%%%%%%%
\section{Equivalent splitting structures: transitions between fiber charts}\label{sec:transitions}
We have stated in \ref{pmobserver} that, with regard to relativistic splittings, any principal $\group$-bundle in the partition of the fiber bundle $\bigl(P,\pi,X,G,\n{Diff}(G)\bigr)$ is equally valid. Moreover, all global sections and, hence, all fiber charts in the atlas of a principal $\group$-bundle are equivalent. We, therefore, have to study the transformation of parametric fields and operators on the base manifold under changes between any two charts in the maximal $\n{Diff}(\group)$-atlas; compare Fig.~\ref{fig:atlases}. In the context of relativistic splittings, this operation has sometimes been called a {\em gauge transformation} \cite{Jantzen1992}. 
We use indices $i$ and $j$ for all operators and fields that are defined with respect to the fiber charts $(X,\varphi_i)$ and $(X,\varphi_j)$, respectively. 
\labcount{ttranslation}
\bph{Transition-function and the Lie-algebra}\label{translie}
Consider the transition function between fiber charts $\varphi_i$ and $\varphi_j$ in the form
$\varphi_{ij}:X\times\group\to\group$;
compare with \ref{transfunc}. We have $\varphi_{ij}(x,\,\cdot\,):\group\to\group$ and $\varphi_{ij}(\,\cdot\,,g):X\to\group$.
A transition function induces in each point $(x,g_j)$ an invertible linear map $\check{\varphi}_{ij}$ from $\mathfrak{g}$ onto itself,
\begin{align*}
\check{\varphi}_{ij}:X\times\group\to {\n{Lin}}(\mathfrak{g},\mathfrak{g}):(x,g_j)\mapsto(L^\prime_{g_i^{-1}})_{g_i}\circ\varphi_{ij}(x,\,\cdot\,)^\prime_{g_j}\circ(L^\prime_{g_j})_e,
\end{align*}
where $g_i=\varphi_{ij}(x,g_j)$, such that 
\begin{align*}
\check{\varphi}_{ij}(x,g_j):\mathfrak{g}\xrightarrow{\,\,\sim\,\,}\mathfrak{g}:\vec{v}_j\mapsto\vec{v}_i.
\end{align*}
If $\varphi_{ij}$ describes a change of fiber charts within a principal $\group$-bundle we find $\check{\varphi}_{ij}=\n{Id}$. We extend $\check{\varphi}_{ij}(x,g_j)$ to covectors by 
\begin{gather*}
\check{\varphi}_{ij}(x,g_j):\mathfrak{g}^*\xrightarrow{\,\,\sim\,\,}\mathfrak{g}^*:\df{\gamma}_j\mapsto\df{\gamma}_i,\quad\df{\gamma}_i(\vec{v}_i)=\df{\gamma}_j(\vec{v}_j)\;\forall\vec{v}_j\in\mathfrak{g},\\
\intertext{and finally to $\binom{1}{1}$-tensors,}
\check{\varphi}_{ij}(x,g_j):\mathfrak{g^*}\otimes\mathfrak{g}\xrightarrow{\,\,\sim\,\,}\mathfrak{g}^*\otimes\mathfrak{g}:\df{\gamma}_j\otimes\vec{v}_j\mapsto\df{\gamma}_i\otimes\vec{v}_i;
\end{gather*}
compare with \cite[3.1.6]{Fecko}.
The purpose of this map is to encode the effect of the push- and pull-operations of the transition function $\varphi_{ij}(x,\cdot)$ in an operation on the Lie-(co)algebra. The map is used in the following paragraphs.
\ep
\bph{Connection form and fundamental field}\label{pullwomega}% Proof can be found 21.12.13, p.1
Under chan\-ges of fiber charts between different principal $\group$-bundles both the fundamental field $\vec{w}$ and the connection 1-form $\df{\omega}$ undergo a transformation, while the connection form $\Upsilon$ (see \ref{ehresconnonprinc}) remains invariant. Under changes of fiber charts within the same principal bundle, all three are invariant. Consider the map
\[
\check{\varphi}_{ij}\circ\varphi_j:P\to\n{Lin}(V,V),\quad V\in\{\mathfrak{g},\mathfrak{g}^*\},
\]
and let 
\begin{alignat*}{2}\vec{v}_i&=\check{\varphi}_{ij}\bigl(\varphi_j(p)\bigr)\,\vec{v}_j,&\qquad\df{\gamma}_i&=\check{\varphi}_{ij}\bigl(\varphi_j(p)\bigr)\,\df{\gamma}_j, \end{alignat*}
for $p\in P$, $\vec{v}_i,\vec{v}_j\in\mathfrak{g}$, and $\df{\gamma}_i,\df{\gamma}_j\in\mathfrak{g}^*$. Then the transformation of the fundamental field and the connection 1-form is given by
\begin{alignat*}{2}\vec{w}_j\big|_p(\vec{v}_j)&=\vec{w}_i\big|_p(\vec{v}_i),&\qquad
\df{\omega}_j\big|_p(\df{\gamma}_j)&=\df{\omega}_i\big|_p(\df{\gamma}_i).
\end{alignat*}
\ep
\bph{Transition functions and parametric fields}\label{pulliecoalg}
The maps $\varphi_{ij}$ and $\check{\varphi}_{ij}$ induce transition maps for parametric fields in the base manifold. Consider $\Xspace\in\{C^\infty(X,\group),$ $\ves{}(X,\group),\dfs{}(X,\group)\}$ and define
\begin{align*}
\Phi_{ij}^*:\Xspace\xrightarrow{\,\,\sim\,\,}\Xspace:\xgen_i\mapsto \xgen_j,\quad \bigl.\xgen_j(g_j)\bigr|_x=\bigl.\xgen_i\bigl(g_i\bigr)\bigr|_x.
\end{align*}
An object anchored in point $(x,g_i)$ is pulled along the fiber into point $(x,g_j)$. The definition of $\Phi_{ij}^*$ is extended to vector-valued fields with $V\in\{\mathfrak{g},\mathfrak{g}^*,\mathfrak{g}^*\otimes\mathfrak{g}\}$ 
\begin{align*}
\Phi_{ij}^*:\Xspace\otimes V\xrightarrow{\,\,\sim\,\,}\Xspace\otimes V:\xgen_i\otimes\vec{v}_i\mapsto \xgen_j\otimes\vec{v}_j,\end{align*}
by
\begin{align*}
\bigl.(\xgen_j\otimes\vec{v}_j)(g_j)\bigr|_x=\bigl.\xgen_i\bigl(g_i\bigr)\bigr|_x\otimes\check{\varphi}_{ji}(x,g_i)(\vec{v}_i).
\end{align*}
\ifdefined\LONGVERSION
A transition function $\varphi_{ij}(X)$ between fiber charts $\varphi_i$ and $\varphi_j$ induces a change of splitting.
\begin{align*} \n{S}_{j}^{-*} = \Phi_{ij}^*\circ\n{S}_{i}^{-*},\end{align*}
where $\Phi_{ij}^*=\xi^*\circ\varphi_{ij,X}^*\circ\xi^{-*}$ and the following diagram commutes
%\begin{diagram}[height=\tw{0.08},width=\tw{0.1}]
%&&\dfs{}(P)&&\\
%&\ldTo^{\n{S}_i^{-*}}&&\rdTo^{\n{S}_j^{-*}}&\\
%\dfs{}(X,\group;\Wedge\mathfrak{g}^*)&&\rTo^{\Phi^*_{ij}}&&\dfs{}(X,\group;\Wedge\mathfrak{g}^*)\\
%\dTo<{\xi^{-*}}&&&&\dTo>{\xi^{-*}}\\
%\dfs{}(\group)\otimes\dfs{}(X)&&\rTo^{\varphi^*_{ij,X}}&&\dfs{}(\group)\otimes\dfs{}(X)
%\end{diagram}
The operator $\varphi_{ij,X}^*$ is a different pullback $\varphi_{ij,x}^*:\dfs{}(\group)\to\dfs{}(\group)$ for every $x\in X$. It does not involve derivatives of $\varphi_{ij}$ in $X$. Analogously, we obtain
\begin{align*}\n{S}_j=\Phi'_{ji}\circ\n{S}_i,\end{align*}
with $\Phi'_{ji}=\xi^{-1}\circ\varphi'_{ij,X}\circ\xi$.
\fi
The object's value $\vec{v}_i$ is transformed with the linear map $\check{\varphi}_{ji}(x,g_i)$, to yield $\vec{v}_j$. The formulas encompass transitions between different principal $\group$-bun\-dles, as well as changes of charts within a principal $\group$-bundle. 
\ep
\bph{Christoffel-, curvature-, and variance forms}\label{canonshift}
We introduce the Lie-algebra valued form
\begin{align*}\affineform_i(x,g)=\varphi_{ij}\bigl(\,\cdot\,,\varphi_{ji}(x,g)\bigr)^*\MC\in\dfs{1}(X,\group;\mathfrak{g}),\end{align*}
and notice that
\begin{align*}\Phi_{ij}^*\affineform_i=-\affineform_j\end{align*}
holds. The form $\affineform_i$ describes the affine part in the transformation of the Christoffel form,
\begin{align*}
\Phi_{ij}^*\df{\Gamma}_i=\df{\Gamma}_j+\affineform_j;
\end{align*}
%see \cite{long},
see \cite[9.7]{kolar}, and for principal connections \cite[p.~364-365]{amp}. It follows that
\begin{align*}
\Phi_{ij}^*\df{\chi}_i=\df{\chi}_j+\groupderiv\affineform_j.
\end{align*}
For a change of fiber charts within a principal $\group$-bundle $\groupderiv\affineform_j=0$, hence $\Phi_{ij}^*\df{\chi}_i=\df{\chi}_j$. The transformation rule of the curvature form reads for all changes of fiber charts %\cite{long},
\begin{align*}
\Phi_{ij}^*\df{\Omega}_i=\df{\Omega}_j.
\end{align*}
\ep
\bph{The splitting map}
Under a change of fiber charts, splitting and transition maps commute:
\begin{diagram}[height=\tw{0.06},width=\tw{0.07}]
&&\df{\gamma}&&\\
&\ldTo^{\n{S}^{-*}_i}&&\rdTo^{\n{S}^{-*}_j}&\\
(\df{\alpha},\tilde{\df{\beta}})_i&&\rTo^{\Phi^*_{ij}}&&(\df{\alpha},\tilde{\df{\beta}})_j
\end{diagram}
Note, however, that, despite $(\df{\Omega},\df{\chi}) = \n{S}^{-*}(\n{d}\,\df{\omega})$, the commutativity $\Phi^*_{ij}(\df{\Omega},\df{\chi})_i$ $=(\df{\Omega},\df{\chi})_j$ holds only under changes of fiber charts within a principal $\group$-bundle. Transitions between different principal $\group$-bundles alter $\df{\omega}$; see \ref{pullwomega}. This gives rise to an extra term in the transformation rule for $\df{\chi}$; see \ref{canonshift}.

\begin{figure}
\centering
\setlength{\unitlength}{0.08cm}
\begin{picture}(100,73)
\includegraphics[width=8cm]{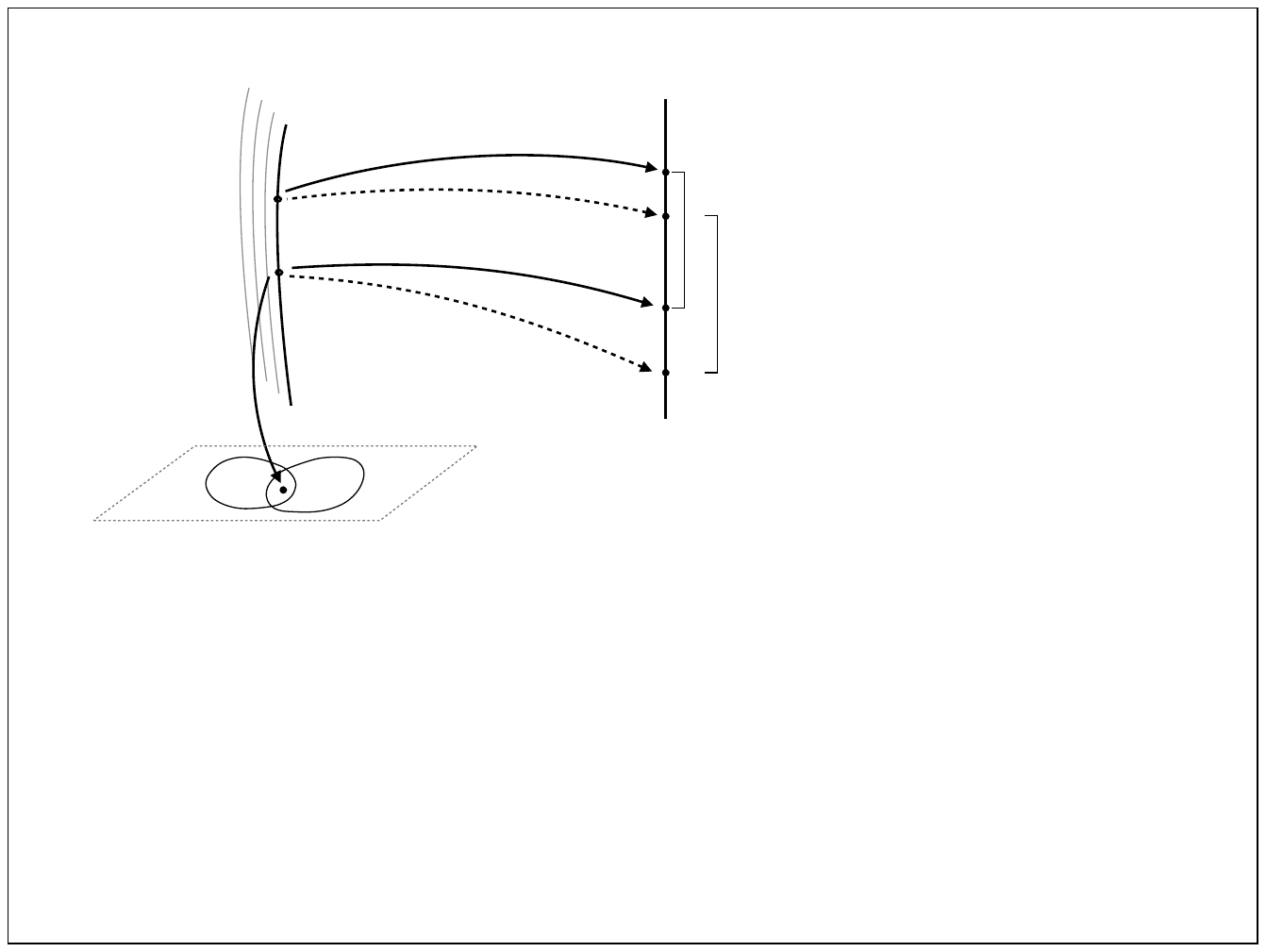}
\put(-68,60){$\group_x$}
\put(-78,20){$\pi$}
\put(-79,6){$U_i$}
\put(-63.4,6.5){$U_j$}
\put(-90,3){$X$}
\put(-7,66){$\group$}
\put(-48,48){$\hhat{\varphi}_{j,x}$}
\put(-45,61){$\hhat{\varphi}_{i,x}$}
\put(-35,42.5){$\hhat{\varphi}_{i,x}$}
\put(-37,27.){$\hhat{\varphi}_{j,x}$}
\put(-6,43){$\interval_i$}
\put(1,35){$\interval_j=\varphi_{ij}(x,\interval_i)$}
\put(-67.3,4){$x$}
\put(-89,52){$P$}
\end{picture}
\caption{Notation and definitions for a change of fiber charts and its impact on the intervals on the fibers of a $\n{Diff}(\group)$-bundle, where $\group$ is the 1-dimensional group of translations.}\label{fig:integral}
\end{figure}

\ep
\bph{Derivatives}\label{derivtrafo}
The group derivative commutes with the transition map,
\begin{align*}\groupderiv\circ \Phi_{ij}^* = \Phi_{ij}^*\circ\groupderiv.\end{align*}
For the exterior derivative and covariant exterior derivative, respectively, the following relations hold,
\begin{alignat*}{2}
&\left.
\begin{aligned}
\n{d}\circ\Phi_{ij}^*&=\makebox[0mm][l]{$\Phi_{ij}^*\circ\bigl(\n{d}+(\affineform_i\specialwedge\groupderiv)\bigr)$}
\phantom{\Phi_{ij}^*\circ\bigl(\n{d}+\lmult{\groupderiv\affineform_i}+(\affineform_i\specialwedge\groupderiv)\bigr)}\\
\n{D}_j\circ\Phi_{ij}^*&=\Phi_{ij}^*\circ\n{D}_i
\end{aligned}
\quad\right\}&&\text{ on }\dfs{k}(X,\group),\\[0.25\baselineskip]
&\left.
\begin{aligned}
\n{d}\circ\Phi_{ij}^*&=\Phi_{ij}^*\circ\bigl(\n{d}+\lmult{\groupderiv\affineform_i}+(\affineform_i\specialwedge\groupderiv)\bigr)\\
\n{D}_j\circ\Phi_{ij}^*&=\Phi_{ij}^*\circ\bigl(\n{D}_i+\lmult{\groupderiv\affineform_i}\bigr)
\end{aligned}
\quad\right\}&&\text{ on }\dfs{k}(X,\group;\mathfrak{g}^*),\\[0.25\baselineskip]
&\left.
\begin{aligned}
\n{d}\circ\Phi_{ij}^*&=\Phi_{ij}^*\circ\bigl(\n{d}-\lmult{\groupderiv\affineform_i}+(\affineform_i\specialwedge\groupderiv)\bigr)\\
\n{D}_j\circ\Phi_{ij}^*&=\Phi_{ij}^*\circ\bigl(\n{D}_i-\lmult{\groupderiv\affineform_i}\bigr)
\end{aligned}
\quad\right\}&&\text{ on }\dfs{k}(X,\group;\mathfrak{g}).
\end{alignat*}
% See 10.12.12, p. 4. I ommitted the transpose of the (1,1)-tensor, since it has no impact on the action of the tensor as defined in
% Appendix A. Moreover, according to Greub 3.17, commutativity between certain tensors can be established.
The operator $(\affineform_i\specialwedge\groupderiv)$ is similar to the operator $(\df{\Gamma}\specialwedge\groupderiv)$ which has been defined in \ref{covderiv}. 
In general, the exterior derivatives do not commute with the transition map. The covariant exterior derivative commutes with the transition map if either the argument is scalar-valued or the transition map is within a principal $\group$-bundle. Note that we have to distinguish between $\n{D}_i$ and $\n{D}_j$. The operators $\n{d}$ and $\groupderiv$ are intrinsically defined on $\dfs{k}(X,\group;\,\cdot\,)$, while $\n{D}$ depends on the Christoffel form and therefore on the fiber chart; compare \ref{coordtime}. The above rules with \ref{canonshift} ensure that
\begin{align*} \begin{pmatrix}{{\n{D}_j}}&\mult{{\df{\Omega}_j}}\\\timederiv&\;\mult{\df{\chi}_j}- {\n{D}_j}\end{pmatrix}\circ\Phi_{ij}^*
=\Phi_{ij}^*\circ\begin{pmatrix}{{\n{D}_i}}&\mult{{\df{\Omega}_i}}\\\timederiv&\;\mult{\df{\chi}_i}- {\n{D}_i}\end{pmatrix}.\end{align*}
% Demonstration can be found 07.03.13
\ep
\bph{Integrals}\label{transint}
Integrals on the Lie group are independent of the respective fiber chart,
\begin{align*}\int_{\interval_j}\circ\;\Phi_{ij}^*=\Phi_{ij}^*\circ\,\int_{\interval_i}\quad\text{ on }\dfs{k}(X,\group;\mathfrak{g}^*),\end{align*}
where $\interval_i=\varphi_{ij}(x,\interval_j)$ and $\interval_i,\interval_j\subset\group$; see Fig.~\ref{fig:integral}.
\ep
%
%%%%%%%%%%%%%%%%%%%%%%%%%%%%%%%%%%%%%%%%%%%%%%
\section{Electromagnetism in the pre-metric setting}\label{sec:pmemag}
In this section we establish the equations of pre-metric electrodynamics that are obtained by splitting Maxwell's equations in four dimensions.
\bph{Maxwell's equations}\label{max}
The splitting of the space-time fields $A$, $F$, $H$, and $J$ yields the parametric fields in three dimensions,
\begin{alignat*}{4}
\n{S}^{-*}A&=(a,-\tilde{\varphi}),&\quad\quad\n{S}^{-*}F&=(b,-\tilde{e}),\\
\n{S}^{-*}H&=(d,\tilde{h}),&\quad\quad\n{S}^{-*}J&=(\rho,-\tilde{\jmath}).
\end{alignat*}
See \cite{Itin2004} for a comprehensive discussion of the signs. Under a change of fiber charts, all parametric fields exhibit a covariant tensorial transformation behavior, which is described by the transition map $\Phi_{ij}^*$; see \ref{pulliecoalg}. Their properties are summarized in the table below. Recall the physical dimension system of \ref{pdsyst} and note that the splitting map is of neutral dimension. Therefore both fields in its image carry the same physical dimension. This result differs from splittings in other texts, for example, \cite{hehlbook}. Another way to look at it is that to eliminate the Lie-coalgebra-value from the forms marked with tilde, a measure of time would be required -- a step that we describe at a later point in the metric setting; see Section~\ref{sec:proxy}.
\begin{center}\small
\renewcommand{\tabcolsep}{1mm}\renewcommand{\arraystretch}{1}
\begin{tabular}[t]{r@{\,}c@{\,}llr@{\,}c@{\,}lp{3.5cm}c}\hline\noalign{\smallskip}
\multicolumn{3}{c}{Space-time\protect\footnotemark} & Name & \multicolumn{3}{c}{Parametric} & Name\protect\footnotemark & Physical\protect\footnotemark\\
\multicolumn{3}{c}{fields} & & \multicolumn{3}{c}{fields} & & dimension\\
\noalign{\smallskip}\hline\noalign{\smallskip}
$A$ &$\in$& $\dfs{1}(P)$ & \raisebox{0pt}[0pt][0pt]{\parbox[t]{1.4cm}{Electro-\\magnetic\\potential}} &
$a$ &$\in$& $\dfs{1}(X,\G)$ & Magnetic vector\newline potential & $\pdim{UT}$\\
&&&& $\tilde{\varphi}$ &$\in$& $\dfs{0}(X,\G;\mathfrak{\g}^*)$ & Electric potential\\
\noalign{\smallskip}\hline\noalign{\smallskip}
$F$ &$\in$& $\dfs{2}(P)$ & \raisebox{0pt}[0pt][0pt]{\parbox[t]{1.4cm}{Electro-\\magnetic\\field}} &
$b$ &$\in$& $\dfs{2}(X,\G)$ & Magnetic flux density & $\pdim{UT}$\\ \\
&&&& $\tilde{e}$ &$\in$& $\dfs{1}(X,\G;\mathfrak{\g}^*)$ & Electric field strength\\
\noalign{\smallskip}\hline\noalign{\smallskip}
$H$ &$\in$& $\mathcal{F}^2_\twist(P)$ & \raisebox{0pt}[0pt][0pt]{\parbox[t]{1.4cm}{Electro-\\magnetic\\excitation}} &
$d$ &$\in$& $\mathcal{F}^2_\doubletwist(X,\G)$ & Electric flux density & $\pdim{IT}$\\ \\
&&&& $\tilde{h}$ &$\in$& $\mathcal{F}^1_\doubletwist(X,\G;\mathfrak{\g}^*)$ & Magnetic field strength\\
\noalign{\smallskip}\hline\noalign{\smallskip}
${J}$ &$\in$& $\mathcal{F}^3_\twist(P)$ & \raisebox{0pt}[0pt][0pt]{\parbox[t]{1.4cm}{Electric\\charge\\current}} &
$\rho$&$\in$& $\mathcal{F}^3_\doubletwist(X,\G)$ & Electric charge density & $\pdim{IT}$\\ \\
&&&& $\tilde{\jmath}$ &$\in$& $\mathcal{F}^2_\doubletwist(X,\G;\mathfrak{\g}^*)$ & Electric current density\\
\noalign{\smallskip}\hline
\end{tabular}
\end{center}
\addtocounter{footnote}{-2}
\footnotetext{For completeness, we mention that $L=-(F\wedge H)/2\in\mathcal{F}^4_\twist(P)$ qualifies as Lagrangian of the electromagnetic field \cite[Eq.~(E.1.11)]{hehlbook}. It splits according to $\n{S}^{-*}L=(0,\tilde{l})$, where $\tilde{l}\in\mathcal{F}^3_\doubletwist(X,\G;\mathfrak{\g}^*)$ is the related parametric Lagrangian in three dimensions, $\pd(L)=\pd(\tilde{l})=\mbox{\ssar A}$.}
\addtocounter{footnote}{1}
\footnotetext{Names of fields follow the IEC/ISO standard \cite{iso}.}
\addtocounter{footnote}{1}
\footnotetext{\label{aqfoot}Following~\cite[Ch.~II, \S 3]{post}, the Faraday fields $A$ and $F$ and their images in the splitting map carry the physical dimension of action per charge $\mbox{\ssar A}\mbox{\ssar Q}^{-1}$ (see~\ref{pdsyst}), and the Amp\`ere-Maxwell fields $H$ and $J$ and their images carry the physical dimension of charge $\mbox{\ssar Q}$. Any product of Faraday and Amp\`ere-Maxwell fields, therefore, carries the physical dimension of action.}
The flux conservation law $\n{d}\, F = 0$ yields the magnetic Gauss law and the Faraday law,
\begin{alignat*}{2}
\n{D}\,{b} &= \df{\Omega}\wedge\tilde{e},\\
\n{D}\,\tilde{e} &=-\timederiv\,{b}+\df{\chi}\wedge\tilde{e}.
\end{alignat*}
From the law $\n{d}\,H=J$ we obtain the electric Gauss law and the Amp\`ere-Maxwell law,
\begin{alignat*}{2}
\n{D}\,d&=\rho-\df{\Omega}\wedge \tilde{h},\\
\n{D}\,\tilde{h}&= \tilde{\jmath}+\timederiv d+\df{\chi}\wedge \tilde{h}.
\end{alignat*}
Finally, with $F=\n{d}\,A$ and $\n{d}\,J=0$, the definitions of the potentials and the law of charge conservation read
\begin{alignat*}{2}
\n{D}\,a&=b+\df{\Omega}\wedge\tilde{\varphi},\\
-\n{D}\,\tilde{\varphi}&=\tilde{e}+\timederiv a-\df{\chi}\wedge\tilde{\varphi},\\
\n{D}\,\tilde{\jmath}&=-\timederiv\rho+\df{\chi}\wedge \tilde{\jmath}.
\end{alignat*}
We wish to emphasize that these equations are defined without any recourse to coordinates, frames, metric tensor or basis of the Lie algebra. A change of fiber charts preserves their form, as can be readily seen by applying the transformations from \ref{canonshift} and \ref{derivtrafo}.
\ep
\bph{Discussion}\label{premetricEmagDiscussion}
\begin{itemize}
\item Maxwell's equations take their usual form for holonomic splittings, except for the covariant exterior derivative. It becomes a standard exterior derivative $\n{D}=\n{d}$ for natural splittings.
\item Parametric electromagnetic fields are called {\em static} with respect to a principal splitting if they are independent of $\g$. With $\timederiv=0$ and $\df{\chi}=\timederiv\df{\Gamma}=0$ the equations simplify accordingly. Still, the curvature 2-form $\df{\Omega}$ is present, and retains a coupling between electrostatic and magnetostatic fields. Only in the case of vanishing curvature there is a complete decoupling into electrostatic and magnetostatic subsystems. This might easily be overlooked, giving rise to paradoxical results, as discussed in \cite{Schiff1939} and Section~\ref{sec:schiff}.
\end{itemize}
\ep
\bph{Alternative formulation}\label{maxalt}
An equivalent set of equations in terms of the field quantities, the Christoffel form, the exterior derivative, and the group derivative is obtained by the factorization in \ref{ddecomp};
\begin{align*}
\n{d}\,(b-\df{\Gamma}\wedge\tilde{e}) &=0,\\
\n{d}\,\tilde{e} &=-\timederiv (b-\df{\Gamma}\wedge\tilde{e}),\\
\n{d}\,(d+\df{\Gamma}\wedge\tilde{h}) &=(\rho-\df{\Gamma}\wedge\tilde{\jmath}),\\
\n{d}\,\tilde{h}&= \tilde{\jmath}+\timederiv(d+\df{\Gamma}\wedge\tilde{h}).
\end{align*}
The equations lend themselves to an integral formulation via the Stokes' theorem in the base manifold. The integral equations thus obtained constitute a metric-free version of what was reported in \cite{Bini2001} as an integral formulation of Maxwell's equations.%
\footnote{For example, compare their equations (31) and (33), with $\nu(\hat{U},u)=0$ and $\gamma(\hat{U},u)=1$, which means that the domain of integration is not moving. The factor $N\gamma(\hat{U},n)^{-1}$ is a lapse function.} In \cite[Sec.~5]{Bini2001} the authors point out that the usual integral formulation of Maxwell's equations is only valid for a vanishing Christoffel form. It should be noted, however, that the above equations already have the status of Maxwell's equations in a natural splitting; the transformation is completed by substituting the expressions in parentheses by the respective fields $b$, $d$, and $\rho$ of the natural splitting. This can be seen by applying the change-of-connection map to Maxwell's equations reported in \ref{max}; see \ref{changeconnec} with $\df{\omega}_\alpha=\df{\omega}$, $\df{\omega}_\beta=\df{\theta}$.\footnote{The same map was used to derive the factorization in \ref{ddecomp}.} 
\ep
\bph{Integral quantities}\label{intquant}
Denote by $\curv$, $\surf$, $\vol$ one-, two-, and three-di\-mensional compact inner-oriented subdomains of $X$, respectively. Outer-orien\-ted subdomains are denoted by $\curv_{{\!\times}}$, $\surf_{{\!\times}}$, while $\vol_{{\!\times}}$ does not bear orientation. Likewise, $g\in\G$ is an instant in time, and $\interval$ is a compact inner-oriented domain in $\G$. Moreover $g_\timetwist=(g,\df{\varepsilon})$, $\mathfrak{g}^*\ni\df{\varepsilon}\ne 0$, is an outer-oriented instant in time, and $\interval_\timetwist$ does not bear orientation. Integrating the electromagnetic fields over suitable subdomains in space and time gives rise to various integral quantities that are sumarized in the table below. By construction, all spatial integrals of parametric forms depend on the selected time synchronization.
\begin{table}[ht!]
\centering\small
\renewcommand{\tabcolsep}{1mm}\renewcommand{\arraystretch}{1}
\begin{tabular}{r@{\,}c@{\,}l@{\,}llcp{2.4cm}@{}c}\hline\noalign{\smallskip}
\multicolumn{4}{c}{Integration} & Name & Integration & Name\protect\footnotemark & Physical\\
\multicolumn{4}{c}{in space} & & in time & & dimension\\
\noalign{\smallskip}\hline\noalign{\smallskip\smallskip}
$\Phi$ & $=$ & $\int_\surf b$ & $\in C^\infty(\G)$ & Magnetic flux & $\Phi(g)$ & Magnetic flux\newline at instant $g$ & $\pdim{UT}$\\\noalign{\smallskip}
$\tilde{U}$ & $=$ & $\int_\curv\tilde{e}$ & $\in C^\infty(\G;\mathfrak{g}^*)$ & Electric voltage & $\int_\interval\tilde{U}$ & Electric voltage\newline impulse\footnotemark\\
\noalign{\smallskip}\hline\noalign{\smallskip\smallskip}
$\Psi$ & $=$ & $\int_{\surf_{{\!\times}}}\!d$ & $\in C^\infty_\twist(\G)$ & Electric flux & $\Psi(g_\timetwist)$ & Electric flux\newline across instant $g_\timetwist$ & $\pdim{IT}$\\\noalign{\smallskip}
$\tilde{V}$ & $=$ & $\int_{\curv_{{\!\times}}}\!\tilde{h}$ & $\in C^\infty_\twist(\G;\mathfrak{g}^*)$ & Magnetic voltage & $\int_{\interval_\timetwist}\!\!\tilde{V}$ & Magnetic voltage\newline impulse\addtocounter{footnote}{-1}\footnotemark & \\
\noalign{\smallskip}\hline\noalign{\smallskip\smallskip}
$Q$ & $=$ & $\int_{\vol_{{\!\times}}}\!\rho$ & $\in C^\infty_\twist(\G)$ & Electric charge & $Q(g_\timetwist)$ & Electric charge\newline across instant $g_\timetwist$ & $\pdim{IT}$\\\noalign{\smallskip}
$\tilde{I}$ & $=$ & $\int_{\surf_{\!\times}}\!\tilde{\jmath}$ & $\in C^\infty_\twist(\G;\mathfrak{g}^*)$ & Electric current & $\int_{\interval_\timetwist}\!\!\tilde{I}$ & Electric charge\newline flow\addtocounter{footnote}{-1}\footnotemark & \\
\noalign{\smallskip}\hline
\end{tabular}
\end{table}
\addtocounter{footnote}{-1}\footnotetext{Concerning the ``Name'' column, note that instants in time correspond with hypersurfaces in space-time by the fiber chart. Therefore, we use the terms ``at instant $g$" and ``across instant $g_\timetwist$".}
\addtocounter{footnote}{1}\footnotetext{The terms ``voltage impulse'' and ``charge flow'' are taken from \cite[Tab.~5]{Tonti2001}. A similar compilation of integral quantities (called ``global variables") can be found in \cite[Tab.~10.3]{tontibook}. Note that in \cite{Tonti2001} outer-oriented time elements are assigned to the Amp\`ere-Maxwell fields, while in \cite{tontibook} they are assigned to the Faraday fields. Our model suggests the first approach.}
\ep
%%%%%%%%%%%%%%%%%%%%%%%%%%%%%%%%%%%%%%%%%%%%%%%%%%%%%%%%%
\section{Main concepts in their component representation}
\label{amoreadapt}
This section continues the discussion about adapted coordinates of \ref{adaptchart} and \ref{coordtime}. Its purpose is mainly to establish a comparison with the coordinate- and frame-based formalisms that can, for example, be found in \cite{gourgbook,bladelbook} and \cite{hehlbook}, where the latter serves as our main reference. We establish coordinates on space-time that are adapted to a given fiber chart. Next we construct a frame field that is adapted to the Ehresmann connection. Except for the case of a flat principal connection this frame field will be anholonomic. We write Maxwell's equations in components with respect to this frame field and show that the results agree with literature. The component representation of vacuum constitutive relations is given in \ref{componenthodge} and \ref{componentconst}, respectively.
\bph{Adapted coordinate charts}
Consider adapted coordinates $(t,x^i)$, where ${\phi}:\G\to(\mathbb{R},+):g\mapsto t$ is a Lie-group homomorphism (compare with \ref{adaptchart}), while $\psi:X\to\mathbb{R}^3:x\mapsto(x^i)$ is a coordinate chart in the base manifold; see \ref{coordtime}. The dual bases of the Lie algebra are given in the domain of chart $\phi$ by the canonical dual bases $(\partial_t,\n{d}t)$. By combining charts ${\phi}$ and $\psi$ with the fiber chart $\varphi:P\to X\times\G$, a coordinate chart of space-time is obtained, $\tilde{\phi}:P\to\mathbb{R}^4:p\mapsto(x^\mu)$, where $\tilde{\phi}=({\phi}\circ\hhat{\varphi},\psi\circ\pi)$. Herein, we have identified $\mathbb{R}^4$ with $\mathbb{R}\times\mathbb{R}^3$, $(x^\mu)=(t,x^i)$. All objects that are defined in the manifolds $P$, $X$, $G$, respectively, can now be represented in the domains of these charts. We do not make a distinction in their notation, though.
\ep
\bph{Fundamental field, connection, and anholonomic frames}
The fundamental field, the connection 1-form, and the horizontal lift can be written with respect to the coordinate frame $(\partial_{x_\mu})$ and a co-frame $(\n{d}x^\nu)$,
\begin{gather*}
\vec{w}=\partial_{x^0}\otimes\n{d}t,\qquad
\df{\omega}=(\n{d}x^0+\Gamma_i\n{d}x^i)\otimes\partial_t,\\[0.5\baselineskip]
\Sigma:\,\partial_{x^i}\mapsto\partial_{x^i}-\Gamma_i\partial_{x^0},\qquad
\Sigma^*:\left\{
\begin{aligned}
\n{d} x^0&\mapsto -\Gamma_i\n{d}x^i\\
\n{d} x_i&\mapsto\n{d} x_i
\end{aligned}
\right.,
\end{gather*}
where $\Gamma_i$ are the components of the Christoffel form, $\df{\Gamma}=\Gamma_i\n{d}x^i$. %\par\medskip\noindent
We establish anholonomic dual frames by
\[
\vec{e}_0=\partial_{x^0},\quad \vec{e}_i=\Sigma\partial_{x^i}=\partial_{x^i}-\Gamma_i\partial_{x^0},
\]
and $\df{\varepsilon}^\nu(\vec{e}_\mu)=\delta^\nu_\mu$, which yields
\[
\df{\varepsilon}^0=\n{d}x^0+\Gamma_j\n{d}x^j,\quad\df{\varepsilon}^i=\n{d}x^i.
\]
The frame is adapted to the Ehresmann connection, since $\vec{e}_0$ spans the vertical subspace $V_pP$ and $(\vec{e}_i)$ spans the horizontal subspace $H_pP$ for all $p\in P$. With respect to this (co-)frame, we find
\begin{gather*}
\vec{w}=\vec{e}_0\otimes\n{d}t,\qquad
\df{\omega}=\df{\varepsilon}^0\otimes\partial_t,\\
\Sigma^*:\left\{
\begin{aligned}
\df{\varepsilon}^0&\mapsto 0\\
\df{\varepsilon}^i&\mapsto\df{\varepsilon}^i
\end{aligned}
\right..
\end{gather*}
Since the splitting map is composed of these three building blocks, its action on the components attains a very simple form, as we will see below.
\ep
\bph{The object of anholonomity}
We use the notation $\partial_\mu=\vec{e}_\mu$ to indicate that basis vector $\vec{e}_\mu$ acts as directional derivative operator on a multivariate differentiable function. For exterior products of coframe basis elements we use the shorthand
$
\df{\varepsilon}^{\mu_1\cdots \mu_k}=\df{\varepsilon}^{\mu_1}\wedge\cdots\wedge\df{\varepsilon}^{\mu_k}.
$
We compute the object of anholonomity $C_{\mu\nu}{}^\kappa$, which is defined by \cite[Eq.~(A.2.35)]{hehlbook}
\[
\n{d}\df{\varepsilon}^\kappa=\frac{1}{2}C_{\mu\nu}{}^\kappa\,\df{\varepsilon}^{\mu\nu}.
\]
As a result, $C_{\mu\nu}{}^i=0$, since $\n{d}\df{\varepsilon}^i=0$, while
\[
\n{d}\df{\varepsilon}^0=C_{0j}{}^0\,\df{\varepsilon}^{0j}+\frac{1}{2}C_{ij}{}^0\,\df{\varepsilon}^{ij},\quad
C_{0j}{}^0=\partial_0\Gamma_j,\quad
\frac{1}{2}C_{ij}{}^0=\partial_{\{i}\Gamma_{j\}}.
\]
\ep
\bph{The covariant exterior derivative, variance and curvature forms}
We receive from \ref{coordtime} the expressions for a differential $k$-form $\df{\gamma}$ and its covariant exterior derivative,
\begin{align*}
\df{\gamma}&=\frac{1}{k!}\gamma_{j_1\cdots j_k}\,\df{\varepsilon}^{j_1\cdots j_k},\quad\gamma_{[j_1\cdots j_k]}=0,\\
\n{D}\,\df{\gamma}&=\frac{1}{k!}\partial_{\{i}\gamma_{j_1\cdots j_k\}}\,\df{\varepsilon}^{ij_1\cdots j_k}.
\end{align*}
For the variance and curvature forms we obtain from \ref{variance} and \ref{curvature}
\begin{alignat*}{3}
\df{\chi}&=\groupderiv\df{\Gamma}&&=\chi_j\,\df{\varepsilon}^j\otimes(\partial_t\otimes\n{d}t),\quad&\chi_j&=C_{0j}{}^0,\\
\df{\Omega}&=\n{D}\,\df{\Gamma}&&=\frac{1}{2}\Omega_{ij}\,\df{\varepsilon}^{ij}\otimes\partial_t,\quad&\Omega_{ij}&=C_{ij}{}^0.
\end{alignat*}
For the chosen frame, anholonomic in general, the components of the variance and curvature forms coincide with the nonzero components of the object of anholonomity. Delphenich \cite[Sec.~3]{Delphenich2007} calls such frame fields which contain the minimum amount of information that it takes to describe the geometrical structure {\em semi-holonomic}
\ep
\bph{Electromagnetic fields and Maxwell's equations}
The electromagnetic fields admit coordinate expressions
\[
F=\frac{1}{2}F_{\mu\nu}\,\df{\varepsilon}^{\mu\nu},\quad
H=\bigl(\frac{1}{2}H_{\mu\nu}\,\df{\varepsilon}^{\mu\nu},\df{\varepsilon}^{0123}\bigr),\quad F_{[\mu\nu]}=H_{[\mu\nu]}=0,
\]
where $\df{\kappa}=\df{\varepsilon}^{0123}$ is the volume form in the representation of the twisted differential form $H$; compare \ref{twisted}. It holds that $(b,-\tilde{e})=\n{S}^{-*}F$, as well as $(d,\tilde{h})=\n{S}^{-*}H$, hence
\begin{alignat*}{5}
b&=\Sigma^*F&&=\frac{1}{2}b_{ij}\,\df{\varepsilon}^{ij},&&\quad &&b_{ij}&&=F_{ij},\\
\tilde{e}&=-\Sigma^*\cont{\vec{w}}F&&=\tilde{e}_i\,\df{\varepsilon}^i\otimes\n{d}t,&&\quad &&\tilde{e}_i&&=F_{i0},\\
d&=\Sigma^*H&&=\bigl(\frac{1}{2}d_{ij}\,\df{\varepsilon}^{ij},&&\,\df{\varepsilon}^{123}\otimes\n{d}t\bigr),\quad &&d_{ij}&&=H_{ij},\\
\tilde{h}&=\Sigma^*\cont{\vec{w}}H&&=\bigl(\tilde{h}_i\,\df{\varepsilon}^i\otimes\n{d}t,&&\,\df{\varepsilon}^{123}\otimes\n{d}t\bigr),\quad &&\tilde{h}_i&&=H_{0i}.
\end{alignat*}
We discuss the Faraday fields, and receive from \ref{max} the magnetic Gauss law and the Faraday law in components,
\[
\begin{aligned}
\partial_{\{i}b_{jk\}}&=\Omega_{\{ij}\tilde{e}_{k\}},\\
\partial_{\{i}\tilde{e}_{j\}}&=-\frac{1}{2}\partial_0b_{ij}+\chi_{\{i}\tilde{e}_{j\}}.
\end{aligned}
\]
We may rewrite this in terms of the electromagnetic field and the object of anholonomity,
\begin{align*}
\partial_{\{i}F_{jk\}}-C_{\{ij}{}^0F_{k\}0}&=0,\\
\partial_{\{i}F_{j\}0}+\frac{1}{2}\partial_0F_{ij}-C_{0\{i}{}^0F_{j\}0}&=0.
\end{align*}
These equations combine into
\[
\partial_{\{\mu}F_{\nu\kappa\}}-C_{\{\mu\nu}{}^0F_{\kappa\}0}=0,
\]
which is nothing but $\n{d}F=0$ in anholonomic components \cite[Eq.~(B.4.31)]{hehlbook}. Further components of the object of anholonomity would appear if we had chosen a noncoordinate basis for the horizontal subspace, too. The same line of reasoning is valid for the Maxwell-Amp\`ere fields.
\ep
\ifdefined\INCLUDEMETRIC
%%%%%%%%%%%%%%%%%%%%%%%%%%%%%%%%%%%%%%%%%%%%%%
%%%%%%%%%%%%%%%%%%%%%%%%%%%%%%%%%%%%%%%%%%%%%%
\newchapter{Metric Setting}\label{sec:metric}
In the metric setting we can require the spatial platforms of the Ehresmann connection to be orthogonal to the world-lines, in accordance with the hypothesis of locality. Only a splitting with an orthogonal connection produces measurable fields and physical laws. The observer metric turns the observer's space into a parametric Riemannian manifold. In this setting we can discuss constitutive relations and the splitting of the energy-momentum balance equation. Splittings which are not orthogonal are discussed next. Such splittings do not produce measurable fields. Still, they are useful for the solution of initial value problems. Examples can be found in the canonical formulation of general relativity \cite{Arnowitt1962,Fodor1994}. The kinematic parameters of observers are introduced and used to distinguish classes of observers. The observer classification is juxtaposed to a classification of splitting structures. Finally, we introduce ordinary fields and forms as proxies for their Lie-(co)algebra valued counterparts.
\section{Regular relativistic splittings}\label{sec:refobsmod}\labcount{metric}
An observer's relative space is the base manifold of the principal $\G$-bundle. It is equipped with the pullback metric under the horizontal lift of the Ehresmann connection. The lapse function is the pointwise norm of the fundamental field, pulled back to the base manifold. It induces in each event an inner product on the Lie-algebra that represents the metric in the time-direction. We decompose the space-time metric in terms of these building blocks. Lastly, Hodge operators in four and three dimensions are defined, and we show how they are related by the splitting map.
\bph{Preliminaries}\label{metricprelim}
Let $\mathbf{g}$ be the Lorentzian metric tensor field on space-time, $\pd(\mathbf{g})=\pdim{L}^2$,\footnote{See Annex~\ref{mpost} for more on the physical dimension of the metric tensor.} and let $\n{g}$ be the metric Riesz operator, $\n{g}:\ves{1}(P)\xrightarrow{\,\,\sim\,\,}\dfs{1}(P)$, $\n{g}\,\vec{v}=\mathbf{g}(\vec{v},\cdot)$.\footnote{In the literature, the Riesz operator $\n{g}$ and its inverse $\n{g}^{-1}$ are sometimes denoted flat operator $\flat$ and sharp operator $\sharp$, respectively.} The Riesz operator is extended to multi-vector fields by exterior compound, $\n{g}(\vec{v}_1\wedge\vec{v}_2)= \n{g}\,\vec{v}_1\wedge\n{g}\,\vec{v}_2$, where $\vec{v}_{1}, \vec{v}_{2} \in \ves{1}(P)$. Its physical dimension is $\pd(\n{g})=\pdim{L}^{2 k}$, with $k$ the degree of the multi-vector that is being acted upon. A metric tensor $\mathbf{g}^{-1}$ on 1-forms is induced by 
\[
\mathbf{g}^{-1}(\df{\gamma},\df{\gamma}^\prime)=\mathbf{g}(\vec{v},\vec{v}^\prime)\quad\mathrm{for}\quad\df{\gamma}=\n{g}(\vec{v}),\df{\gamma}^\prime=\n{g}(\vec{v}^\prime),\quad
\pd(\mathbf{g}^{-1})=\pdim{L}^{-2}.
\]
Finally, we introduce the notation
\[
|\,\cdot\,|:\ves{k}(P)\to C^\infty(P),\quad |\vec{v}|=\sqrt{|\n{g}(\vec{v})(\vec{v})|},\quad\pd{\,|\,\cdot\,|}=\pdim{L}^k\,\pd{(\cdot)}
\]
for the pointwise norm.\footnote{$|\,\cdot\,|$ is not a norm in the usual sense, because it fails to be subadditive.} A similar notation is used for $k$-forms, and for parametric fields and forms on $X$, respectively.
Denote spaces of smooth functions that are twisted with respect to $\group$ (short: $\group$-twisted) by $C^\infty_\timetwist$; compare \ref{gtwisted}. The pointwise norm may be extended to Lie-(co-)algebra valued fields and forms. In this case, the norm is Lie-(co)algebra valued and $\group$-twisted.\footnote{By introducing the norm as $\group$-twisted it becomes independent of the orientation of the Lie-(co)algebra; this would not be possible otherwise.} For example,
\[
|\,\cdot\,|:\ves{1}(P,\mathfrak{g}^*)\to C^\infty_\timetwist(P,\mathfrak{g}^*),\quad |\vec{w}|=\bigl(|\vec{w}(\vec{e})|\otimes\df{\varepsilon},\df{\varepsilon}\bigr),
\]
where $(\vec{e},\df{\varepsilon})\in\mathfrak{g}\times\mathfrak{g}^*$ denotes dual bases of the Lie algebra, $\df{\varepsilon}(\vec{e})=1$. Obviously, the norm is independent of the choice of bases.
\ep
\bph{Regular relativistic splitting structure}\label{regular} 
A relativistic splitting structure on a space-time $(P,\mathbf{g})$ is regular \cite[II.6.2.4]{Matolcsi}, \cite[Sec.~2.4]{Matolcsi1998} if the horizontal subspaces of the Ehresmann connection are orthogonal to the world-lines, $H_pP = (V_pP)_\perp$ for all $p\in P$. This condition depends only on the conformal class of the metric. The {\em hypothesis of locality} states that splittings that fail to be regular do not produce measurable quantities or physical laws that govern measurable quantities. %Only regular splittings are eligible for the modeling of observable quantities. 
For the rest of this section we assume the splitting structure to be regular.
\ep
\bph{Einstein synchronization}
A time synchronization (see \ref{timesynch}) of a natural regular splitting structure is an Einstein synchronization. 
\ep
\bph{Relative space}\label{relspace}
The {\em observer metric} on $X$ is obtained from a regular splitting as the parametric $\binom{0}{2}$-tensor field\footnote{Equivalently, it holds that $\mathbf{h}^{-1}=-\Pi\,\mathbf{g}^{-1}$. Note that for {\em nonregular} splittings this relation in general gives a different metric in the base manifold \cite[p.~45]{gourg}. We will further exploit this in Section~\ref{sec:nonreg}.}
\begin{alignat*}{2} \mathbf{h}&=-\Sigma^*\mathbf{g},&\quad\quad \pd(\mathbf{h})&=\pdim{L}^2,\\
\n{h}&=\Sigma^*\circ\n{g}\circ\Sigma\circ\signop,&\quad\quad \pd(\n{h})&=\pdim{L}^{2k};\end{alignat*}
see \ref{signop} for the definition of the sign operator $\signop$. The Riemannian manifold $(X,\mathbf{h})$ models an observer's relative space. To be precise, relative space is a {\em parametric manifold} \cite{Boersma1995,Boersma1995a,Boersma1995b}, that is, a family of Riemannian manifolds $\bigl\{\bigl(X,\mathbf{h}(\g)\bigr)\bigr\}_{\g\in \G}$.
\ep
\bph{Time (disambiguation)} \label{timedis}
The presented framework suggests the following distinctions in the use of the word `time':
\begin{itemize}
\item a {\em manifold}, as in ``somewhere in time" or ``a time axis".
\item an {\em instant} or a synchronization, as in ``at what time?".
\item a {\em group action}, as in ``time passes".
\item a {\em measure} of arclength along world-lines (divided by $c_0$, the speed of light), as in ``the roundtrip time is 5 seconds".
\end{itemize}
In a coordinate-based approach, all of the above concepts are usually amalgamated in the time coordinate. In the present framework, the manifold is the Lie group; the instant is an element of it and the synchronization a section; the group action is the principal right action. Only the measure carries the physical dimension of time.
\ep
\bph{Lapse function}\label{lapsefun}
The Lie-coalgebra valued $\group$-twisted function
\begin{alignat*}{2}
\lapse&=\Phi^*|\vec{w}|\in C^\infty_\timetwist(X,\group;\mathfrak{\g}^*),&&\quad\pd(\lapse)=\pdim{L},
\intertext{is called lapse function. It has a Lie-algebra valued $\group$-twisted inverse}
\lapse^{-1}&=\Phi^*|\df{\omega}|\in C^\infty_\timetwist(X,\group;\mathfrak{\g}),&&\quad\pd(\lapse^{-1})=\pdim{L}^{-1}.
\end{alignat*}
The lapse function has several interpretations:
\begin{enumerate}
\item For each $(x,\g)$, the lapse function is a twisted unit volume form on the Lie algebra.
\item For each $(x,\g)$, the lapse function induces an inner product on the Lie-algebra, whose metric tensor is given by $(\lapse\lapse)^\otimes$.
\item For each $(x,\g)$, the lapse function can be seen as a Hodge operator on the algebra $\mathsf{\Lambda}\,\mathfrak{g}^*$. Hence, the lapse function $\lapse=(\tilde{\lapse},\df{\varepsilon})$ provides isomorphisms\label{case3}
\begin{align*}
\lapse\;:\;\left\{
\begin{alignedat}{2}
\mathcal{F}^k(X,\group)&\xrightarrow{\,\,\sim\,\,}\mathcal{F}^k_\timetwist(X,\group;\mathfrak{g}^*)&&\;:\;\df{\beta}\mapsto(\tilde{N}\df{\beta},\df{\varepsilon}),\\
\mathcal{F}^k_\timetwist(X,\group)&\xrightarrow{\,\,\sim\,\,}\mathcal{F}^k(X,\group;\mathfrak{g}^*)&&\;:\;(\df{\beta},\df{\varepsilon})\mapsto\tilde{N}\df{\beta},\\
\mathcal{F}^k_\twist(X,\group)&\xrightarrow{\,\,\sim\,\,}\mathcal{F}^k_\doubletwist(X,\group;\mathfrak{g}^*)&&\;:\;(\df{\beta},\df{\kappa})\mapsto(\tilde{N}\df{\beta},\df{\kappa}\otimes\df{\varepsilon}),\\
\mathcal{F}^k_\doubletwist(X,\group)&\xrightarrow{\,\,\sim\,\,}\mathcal{F}^k_\twist(X,\group;\mathfrak{g}^*)&&\;:\;(\df{\beta},\df{\kappa}\otimes\df{\varepsilon})\mapsto(\tilde{N}\df{\beta},\df{\kappa}).
\end{alignedat}\right.
\end{align*}
\item The lapse function relates the fundamental field to the four-velocity, see \ref{4velo}.
\end{enumerate}
\ep
\bph{Regular splitting of the metric}\label{gdecomp}
%The decomposition of the metric tensor yields
%\footnote{The notation $\bigl(-\mathbf{h},(\lapse\lapse)^\otimes\bigr)$ is a shorthand for $-\n{pr}_1^*\mathbf{h}+\n{pr}_2^*(\lapse\lapse)^\otimes$.}
%%%\begin{alignat*}{5}
%%%\mathbf{g}(\vec{v},\vec{v}^\prime)\Big|_p&=-\mathbf{h}(\vec{\atest},\vec{\atest}^\prime)&&+
%%%(\lapse\lapse)^\otimes(\tilde{\btest},\tilde{\btest}^\prime)&&\Big|_{\varphi(p)},\quad
%%%&&(\vec{\atest},\tilde{\btest})&&=\n{S}\vec{v},\\
%%%\mathbf{g}^{-1}(\df{\gamma},\df{\gamma}^\prime)\Big|_p&=-\mathbf{h}^{-1}(\df{\alpha},\df{\alpha}^\prime)&&+
%%%(\lapse^{-1}\lapse^{-1})^\otimes(\tilde{\beta},\tilde{\beta}^\prime)&&\Big|_{\varphi(p)},\quad
%%%&&(\df{\alpha},\tilde{\beta})&&=\n{S}^{-*}\df{\gamma}.
%%%\end{alignat*}
%\begin{align*}
%\n{S}^{-*}\mathbf{g}&=\bigl(-\mathbf{h},(\lapse\lapse)^\otimes\bigr),\\
%\n{S}^{**}\mathbf{g}^{-1}&=\bigl(-\mathbf{h}^{-1},(\lapse^{-1}\lapse^{-1})^\otimes\bigr).
%\end{align*}
The splitting of the Riesz operator reads
\begin{align*}
\n{S}^{-*}\circ\n{g}\circ\n{S}^{-1}&=\begin{pmatrix}1&0\\0&
\,(\lapse\lapse)^\otimes\end{pmatrix}\n{h}\circ\signop,\quad\quad\\
\n{S}\circ\n{g}^{-1}\circ\n{S}^{*}&=\begin{pmatrix}1&0\\0&
(\lapse^{-1}\lapse^{-1})^\otimes\end{pmatrix}\n{h}^{-1}\circ\signop;
\end{align*}
see \ref{algebra} item \ref{algebra3a} for the product of (co)vector-valued scalar fields. The splitting of the Riesz operator encompasses the splitting of the metric tensor. The diagonal structure of the matrices is due to the regularity of the splitting.
\ep
\bph{Regular splitting of the Hodge operator}\label{dhodge}
Let $\df{\kappa}_4$ denote the twisted unit space-time volume form. That is, $\df{\kappa_4}(B_4)=1$, where $B_4$ is an orthonormal frame field in $P$, compare with \cite[Eq.~(C.2.17)]{hehlbook}. We define the twisted unit spatial volume form $\df{\kappa}_3$ by
\[
\n{S}^{-*}\df{\kappa}_4=(0,\lapse\df{\kappa}_3).
\]
It can be shown that $\df{\kappa_3}(B_3)=1$, where $B_3$ is an orthonormal frame field in $X$. 
% Proof can be found 22.05.13, p.3.
\begin{center}
\begin{tabular}{|c|c|c|}
\hline
&$\in$&$\pd(\,\cdot\,)$\\
\hline
$\df{\kappa}_4$&$\mathcal{F}^4_\twist(P)$&$\pdim{L}^4$\\
$\df{\kappa}_3$&$\mathcal{F}^3_\twist(X,\G)$&$\pdim{L}^3$\\
\hline
\end{tabular}
\end{center}
The Hodge operator in four dimensions can be written as\footnote{This definition fulfills the requirements i) $*1=\df{\kappa}_4$; ii) $*(\df{\beta}\wedge\df{\alpha})=\n{i}\bigl(\n{g}^{-1}(\df{\alpha})\bigr)*\df{\beta}$, where $\df{\alpha}$ is a 1-form, and $\df{\beta}$ is a form of arbitrary degree, compare with \cite[Eq.~(C.2.78)]{hehlbook} and \cite[p.~160]{burke}.}
\begin{align*} *_4\;:\;
\left\{
\begin{alignedat}{3}
\dfs{k}(P)&\xrightarrow{\,\,\sim\,\,}\mathcal{F}^{4-k}_\twist(P)\\
\mathcal{F}^{k}_\twist(P)&\xrightarrow{\,\,\sim\,\,}\dfs{4-k}(P)
\end{alignedat}\;\right\}\;:\;
\df{\gamma}\mapsto\lcont{\n{g}^{-1}\,\df{\gamma}}\,\df{\kappa}_4\,.
\end{align*}
Likewise, the parametric Hodge operator in three dimensions is defined by
\begin{align*} *_3\;:\;
\left\{
\begin{alignedat}{3}
\dfs{k}(X,G)&\xrightarrow{\,\,\sim\,\,}\mathcal{F}^{3-k}_\twist(X,G)\\
\mathcal{F}^{k}_\twist(X,G)&\xrightarrow{\,\,\sim\,\,}\dfs{3-k}(X,G)
\end{alignedat}\;\right\}\;:\;
\df{\gamma}\mapsto\lcont{\n{h}^{-1}\,\df{\gamma}}\,\df{\kappa}_3\,.
\end{align*}
It extends to covector-valued forms, and forms that are twisted with respect to $X$ and $\G$. We find that $\pd(*_n)=\pdim{L}^{n-2k}$. With \ref{splitop}, \ref{gdecomp}, and the above splitting of the volume form, the regular splitting of the Hodge operator reads
\[
{\n{S}}^{-*}\circ*_4\circ{\n{S}}^{*}=\begin{pmatrix}0&\lapse^{-1}*_3\circ\;\signop\\ \lapse\,*_3&0\end{pmatrix}.
\]
\ep
\labcount{metricchange}
\bph{Change of fiber chart}\label{lambdapull}
The Riesz and Hodge operators are local operators, acting on the base manifold $X$. They, therefore, commute with a change of the fiber chart,
\begin{align*}
\Phi_{ij}^*\circ\n{h}_{i}=\n{h}_{j}\circ\Phi_{ij}^*,\quad\quad
\Phi_{ij}^*\circ *_{i}=*_{j}\circ\Phi_{ij}^*,
\end{align*}
where $*=*_3$. The lapse function changes its scale under a change of fiber charts, $\lapse_{j}=\Phi_{ij}^*\lapse_{i}$.
% To prove this, introduce a fixed orientation of G. Next, write the lapse function as image of the four-velocity under the splitting map. Then use the invariance of the four-velocity, and the transformation properties of the splitting map.
\ep
\bph{Metric and Hodge operator in their component representation}\label{componenthodge}%We proceed to the metric setting, while restricting ourselves to regular relativistic splitting structures, that is, $H_pP=(V_pP)_\perp$ for all $p\in P$; compare \ref{regular}. It follows that
Referring to the concepts introduced in Section~\ref{amoreadapt}, the space-time metric tensor $\mathbf{g}$ and the observer metric tensor $\mathbf{h}=-\Sigma^*\mathbf{g}$ (see \ref{relspace}) can be written\footnote{We remark that greek indices are lowered and raised by the space-time metric tensor, while latin indices are lowered and raised by the observer metric tensor.}
\begin{alignat*}{3}
\mathbf{g}&=g_{\mu\nu}\,&\df{\varepsilon}^\mu&\otimes\df{\varepsilon}^\nu,\quad &&g_{\{\mu\nu\}}=0,\quad g_{0i}=0,\\
\mathbf{h}&=h_{ij}\,&\df{\varepsilon}^i&\otimes\df{\varepsilon}^j,\quad &&h_{ij}=-g_{ij}.
\end{alignat*}
The lapse function according to \ref{lapsefun} reads
$
N=\bigl(\sqrt{g_{00}}\otimes\n{d}t,\n{d}t\bigr).
$
For a coordinate representation of the Hodge operator for a Pseudo-Rie\-mannian manifold $(M,g)$, in $n$ dimensions, consider
\begin{alignat*}{3}
\df{\gamma}&=&&\frac{1}{k!}\gamma_{\nu_1\cdots\nu_k}\,&&\df{\varepsilon}^{\nu_1\cdots\nu_k},\quad \gamma_{[\nu_1\cdots\nu_k]}=0,\\
*\df{\gamma}&=\big(&&\frac{1}{\ell!}\gamma^*_{\mu_1\cdots\mu_\ell}\,&&\df{\varepsilon}^{\mu_1\cdots\mu_\ell},\df{\varepsilon}^{12\cdots n}\big),\quad k+\ell=n.
\end{alignat*}
Then it holds that \cite[Sec.~14.1.a]{frankelbook}
\[
\gamma^*_{\mu_1\cdots \mu_\ell}=\frac{1}{k!}\gamma^{\nu_1\cdots \nu_k}\sqrt{|g|}\,\hat{\epsilon}_{\nu_1\cdots\nu_k\mu_1\cdots\mu_\ell},
\]
where $\sqrt{|g|}=\sqrt{|\det(g_{\mu\nu})|}$, and $\hat{\epsilon}_{\kappa_1\cdots\kappa_n}$ is the Levi-Civita permutation symbol. Their product is the Levi-Civita tensor, a twisted covariant antisymmetric tensor \cite[Eq.~(C.2.24)]{hehlbook}.\par\medskip\noindent
\ep
%
%%%%%%%%%%%%%%%%%%%%%%%%%%%%%%%%%%%%%%%%%%%%%%
\section{Vacuum constitutive relations and energy-momentum balance}\labcount{energymomentum}
We discuss the splitting of the vacuum constitutive relations and the energy-momentum tensor. General splittings are assumed, with the exceptions of \ref{componentconst} and \ref{const}, which will be generalized to nonregular splittings in Section~\ref{sec:nonreg}. A treatment of more general constitutive relations in the presented framework is beyond the scope of this paper.
\labcount{hodge}
\bph{Hodge operator and constitutive relations in regular splitting}\label{const}
The vacuum constitutive relations in four dimensions read
\begin{alignat*}{2} H={Z_0}^{-1}*_4 F,\quad F&=-Z_0*_4H,\quad Z_0=\sqrt{\mu_0\varepsilon_0^{-1}}.
\end{alignat*}
The vacuum constitutive relations in three dimensions follow from \ref{dhodge},
\begin{align*}
d=Z_0^{-1} \lapse^{-1}*_3 \tilde{e},\quad\tilde{h}=Z_0^{-1} \lapse*_3b.
\end{align*}
\begin{center}
\begin{tabular}{|c|c|l|}
\hline
&$\pd(\,\cdot\,)$&name\\
\hline
$Z_0$&$\pdim{UI}^{-1}$&vacuum impedance\\
$\mu_0$&$\pdim{U}\pdim{I}^{-1}\pdim{T}\pdim{L}^{-1}$&magnetic constant, permeability of vacuum\\
$\varepsilon_0$&$\pdim{U}^{-1}\pdim{IT}\pdim{L}^{-1}$&electric constant, permittivity of vacuum\\
\hline
\end{tabular}
\end{center}
\ep
\bph{Vacuum constitutive relations in their component representation}\label{componentconst}
Referring to the concepts introduced in Section~\ref{amoreadapt}, we receive from \ref{const} after few manipulations the vacuum constitutive relations in three dimensions, 
\[
\begin{alignedat}{3}
d_{ij}&=&&Z_0^{-1}\sqrt{g_{00}}^{-1}&&\sqrt{|h|}\,\hat{\epsilon}_{ij}{}^k\tilde{e}_k,\\
\tilde{h}_i&=\frac{1}{2}&&Z_0^{-1}\sqrt{g_{00}}&&\sqrt{|h|}\,\hat{\epsilon}_i{}^{kl}b_{kl}.
\end{alignedat}
\]
We may rewrite this in terms of the electromagnetic field and excitation, respectively, and the space-time metric tensor. By taking into account $\sqrt{|g|}=\sqrt{g_{00}}\sqrt{|h|}$ and $g^{00}=g_{00}^{-1}$ we obtain
\begin{alignat*}{3}
H_{ij}&=&&Z_0^{-1}\sqrt{|g|}\,\hat{\epsilon}_{ijk0}g^{km}g^{00}F_{m0},\\
H_{i0}&=\frac{1}{2}&&Z_0^{-1}\sqrt{|g|}\,\hat{\epsilon}_{i0kl}g^{km}g^{ln}F_{mn}.
\end{alignat*}
These equations combine into
\[
H_{\mu\nu}=\frac{1}{2}Z_0^{-1}\sqrt{|g|}\,\hat{\epsilon}_{\mu\nu}{}^{\kappa\lambda}F_{\kappa\lambda},
\]
which is the component representation of $H=Z_0^{-1}*_4F$ in the chosen frame, \cite[Eq.~(D.6.14)]{hehlbook}.
\ep
\bph{Energy-Momentum density}\label{EM1}
The electromagnetic energy-momentum tensor is a covector-valued twisted 3-form $T$, $\pd(T)=\pdim{A}$, that we define with \cite{hehlbook}%\footnote{The opposite sign is used in \cite{Prechtl2007}.} 
\begin{alignat*}{2} T:\ves{1}(P)&\to\mathcal{F}^3_\twist(P)\\
\vec{n}&\mapsto T_\vec{n}=\frac{1}{2}(\cont{\vec{n}}H\wedge F - \cont{\vec{n}}F\wedge H).\end{alignat*} % Consistent with H&O, (B.5.7)
The vector field $\vec{n}$ is split into $(\vec{\atest},\tilde{\btest})$, and we obtain with \ref{splitop} and \ref{max} the four tensor quantities
\begin{alignat*}{2} % Consistent with H&O, (B.5.59) (B.5.60)
\n{S}^{-*}\circ{T}\circ\n{S}^{-1}=\begin{pmatrix}-p&\tilde{w}\\
-\tilde{m}&-\tilde{s}\end{pmatrix}
\end{alignat*}
which are characterized as follows:
\begin{center}
\begin{tabular}{|c|rcl|c|l|}
\hline
&$\cdot $&$\!\!\!\!\!\to\!\!\!\!\!$&$\cdot$&$\pd(\,\cdot\,)$&name\\
\hline
$p$&$\ves{1}(X,\G)$&$\!\!\!\!\!\to\!\!\!\!\!$&$\mathcal{F}^3_\doubletwist(X,\G)$&$\pdim{A}$&momentum density\\
$\tilde{w}$&$\ves{0}(X,\G;\mathfrak{\g})$&$\!\!\!\!\!\to\!\!\!\!\!$&$\mathcal{F}^3_\doubletwist(X,\G)$&$\pdim{A}$&energy density\\
$\tilde{m}$&$\ves{1}(X,\G)$&$\!\!\!\!\!\to\!\!\!\!\!$&$\mathcal{F}^2_\doubletwist(X,\G;\mathfrak{\g}^*)$&$\pdim{A}$&momentum flux density\\
$\tilde{s}$&$\ves{0}(X,\G;\mathfrak{\g})$&$\!\!\!\!\!\to\!\!\!\!\!$&$\mathcal{F}^2_\doubletwist(X,\G;\mathfrak{\g}^*)$&$\pdim{A}$&energy flux density\\
\hline
\end{tabular}
\end{center}
with
\begin{alignat*}{2}
p(\vec{\atest})&=\cont{\vec{\atest}}b\wedge d,\\ % Consistent with H&O, (B.5.63) p
\tilde{w}(\tilde{\btest})&=\frac{1}{2}\,\cont{\tilde{\btest}}(\tilde{e}\wedge d+\tilde{h}\wedge b),\\ % Consistent with H&O, (B.5.61) u
\tilde{m}(\vec{\atest})&=-\cont{\vec{\atest}}d\wedge \tilde{e}-\cont{\vec{\atest}}b\wedge \tilde{h}+\frac{1}{2}\,\cont{\vec{\atest}}(d\wedge\tilde{e}+b\wedge\tilde{h}),\\ % Consistent with H&O, (B.5.64) S
\tilde{s}(\tilde{\btest})&=\frac{1}{2}(\cont{\tilde{\btest}}\tilde{e}\wedge\tilde{h}+\tilde{e}\wedge\cont{\tilde{\btest}}\tilde{h}). % Consistent with H&O, (B.5.62) s
\end{alignat*}
Quantities $\tilde{m}$ and $\tilde{s}$ are also called Maxwell stress tensor and (covariant) Poynting vector, respectively. Note that this paragraph remains valid for constitutive relations beyond the vacuum relations.
\ep
\bph{\texorpdfstring{The $\df{\Theta}$ tensor}{The Theta tensor}}\label{genexp}
In view of the next paragraph and following \cite{Trautman1985}, we introduce the $\binom{1}{2}$-tensor field
\begin{align*}\df{\Theta}:\ves{1}(P)&\to\n{End}\bigl(\ves{1}(P)\bigr),\\
\vec{n}&\mapsto\df{\Theta}_\vec{n}= \mathbf{g}^{-1}(\lie{\vec{n}}\mathbf{g}), \end{align*} 
where $\Theta$ is dimensionless, that is, $\pd(\df{\Theta})=\oned$. Moreover, denote $\bar{\df{\Theta}}:\vec{n}\mapsto -\mathbf{g}\circ(\lie{\vec{n}}\,\mathbf{g}^{-1})$ the dual operation, acting on differential forms as a derivation of degree 0, for example, for a 2-form $\df{\gamma}=
\df{\gamma}^1\wedge\df{\gamma}^2$ 
\begin{align*} \bar{\df{\Theta}}_\vec{n}\,\df{\gamma}=
\bar{\df{\Theta}}_\vec{n}\,\df{\gamma}^1\wedge\df{\gamma}^2+\df{\gamma}^1\wedge\bar{\df{\Theta}}_\vec{n}\,\df{\gamma}^2.
\end{align*}
The trace $\Theta$ of $\df{\Theta}$ is given by
\begin{align*}\Theta:\ves{1}(P)\to C^\infty(P):\vec{n}\mapsto\n{Tr}(\df{\Theta}_\vec{n})=(\bar{\df{\Theta}}_\vec{n}\df{\kappa})(\vec{k}),\end{align*}
where $\df{\kappa}$ is a volume form on $P$, not necessarily unit, and $\df{\kappa}(\vec{k})=1$ everywhere.
\ep
\bph{Four-force density and energy-momentum balance}\label{momentumbalance}
The electromagnetic four-force density is defined in \cite{hehlbook} %\footnote{The opposite sign is used in \cite{Prechtl2007}.}
as
\begin{alignat*}{2} R:\ves{1}(P)&\to\mathcal{F}^4_\twist(P)\\ % Consistent with H&O, (B.4.3)
\vec{n}&\mapsto R_\vec{n}=\cont{\vec{n}}F\wedge J,\end{alignat*}
$\pd(R)=\pdim{A}$. The electromagnetic energy-momentum tensor fulfills 
\begin{align*} R_\vec{\vec{n}}&=\n{d}\, T_\vec{n} + X_\vec{n},\end{align*} % Consistent with H&O, (B.5.10)
{where} 
\begin{align*} X_\vec{n}& = -\frac{1}{2}(F\wedge\lie{\vec{n}} H - H\wedge\lie{\vec{n}} F)\end{align*} % Consistent with H&O, (B.5.8)
is a body-force term. Using the vacuum constitutive relation we obtain
\begin{align*} X_\vec{n}=-\frac{1}{2Z_0} F\wedge[\lie{\vec{n}},*_4] F.\end{align*}
The commutator is derived in \cite[Lemma 1]{Trautman1985} to be
\begin{align*} [\lie{\vec{n}},*_4]=(\bar{\df{\Theta}}_\vec{n}-\frac{1}{2}\Theta_\vec{n})*_4.\end{align*}
It measures the change of the Hodge map along the flow of the vector field $\vec{n}$. We see that a sufficient condition for $X_\vec{n}$ to vanish is that $\vec{n}$ is a Killing vector field of the space-time metric, $\lie{\vec{n}}\mathbf{g}=0$. For a comprehensive discussion of the energy-momentum balance see \cite[p.~169]{hehlbook}.
The four-force density splits into tensor quantities,
\begin{align*}
\n{S}^{-*}\circ{R}\circ\n{S}^{-1}=\begin{pmatrix}0&0\\
\tilde{f}&\tilde{r}\end{pmatrix},
\end{align*}
which are characterized as follows:
\begin{center}
\begin{tabular}{|c|rcl|c|l|}
\hline
&$\cdot $&$\!\!\!\!\!\to\!\!\!\!\!$&$\cdot$&$\pd(\,\cdot\,)$&name\\
\hline
$\tilde{f}$&$\ves{1}(X,\G)$&$\!\!\!\!\!\to\!\!\!\!\!$&$\mathcal{F}^3_\doubletwist(X,\G;\mathfrak{\g}^*)$&$\pdim{A}$&force density\\
$\tilde{r}$&$\ves{0}(X,\G;\mathfrak{\g})$&$\!\!\!\!\!\to\!\!\!\!\!$&$\mathcal{F}^3_\doubletwist(X,\G;\mathfrak{\g}^*)$&$\pdim{A}$&power density\\
\hline
\end{tabular}
\end{center}
with
\begin{alignat*}{2} % Consistent with H&O, (B.2.13) (B.2.14)
\tilde{f}(\vec{\atest})&=\cont{\vec{\atest}}\tilde{e}\wedge\rho+\cont{\vec{\atest}}b\wedge\tilde{\jmath},\\
\tilde{r}(\tilde{\btest})&=-\frac{1}{2}(\cont{\tilde{\btest}}\tilde{e}\wedge\tilde{\jmath}+\tilde{e}\wedge\cont{\tilde{\btest}}\tilde{\jmath}).
\end{alignat*}
Assume that $\vec{n}$ is a Killing vector field. The momentum and energy balance equations then read
\begin{align*} % Consistent with H&O, (B.5.71) (B.5.72)
\tilde{f}(\vec{\atest})&=-\timederiv p(\vec{\atest})-(\mult{\df{\chi}}-\n{D})\tilde{m}(\vec{\atest}),\\
\tilde{r}(\tilde{\btest})&=\timederiv\tilde{w}(\tilde{\btest})-(\mult{\df{\chi}}-\n{D})\tilde{s}(\tilde{\btest}).
\end{align*}
\ep
%
%%%%%%%%%%%%%%%%%%%%%%%%%%%%%%%%%%%%%%%%%%%%%%
\section[Metric in nonregular splittings]{Metric in nonregular splittings}\labcount{slicing}\label{sec:nonreg}
In general, a splitting structure cannot be both integrable and regular (that is, flat and orthogonal to an observer's world-lines). So far we have focused on regular splittings whose Ehresmann connection is determined by the conformal class of the space-time metric. This approach yields the standard form of the constitutive relations; see \ref{const}. However, Maxwell's equations feature extra terms reflecting the curvature and variance of the connection; see \ref{max}.
For the formulation and solution of initial-value problems, an alternative approach is often useful. It is centered around a natural splitting structure, whose Ehresmann connection is holonomic; see \ref{canonconnec}. Maxwell's equations, therefore, assume their standard -- and most
convenient -- form; see \ref{premetricEmagDiscussion}. Being integrable, a natural splitting structure is easily amenable to coordinate-based calculus. The convenience needs to be paid for when metric enters the stage. Orthogonality, that makes for a straight-forward splitting of the Hodge operator and, hence, the constitutive relations, is lost. What is also lost is the interpretation of parametric fields in the base manifold as measurable quantities. To obtain measurable predictions any results need to be transformed into a regular splitting.
In the sequel, we consider the splitting structure according to \ref{pmobserver} in full generality, with the sole restriction that horizontal subspaces be space-like. In the splitting of the metric operators, we get to make a choice for the metric in the base manifold. It might be either induced from the connection or from the fibration. Both cases are discussed and set in contrast with each other. Finally, we touch on the relationship to the canonical formulation of general relativity~\cite{Arnowitt1962,Fodor1994}.
%
%\bph{Admissible splitting structures}
%An Ehresmann connection is called admissible if its horizontal subspaces are space-like. Equivalently, its connection 1-form is time-like, $\mathbf{g}^{-1}\bigl(\df{\omega}(\df{\varepsilon}),\df{\omega}(\df{\varepsilon})\bigr)>0$, $\mathfrak{g}\ni\df{\varepsilon}\ne0$. A splitting structure is admissible if it is equipped with an admissible Ehresmann connection. Every regular splitting structure is admissible. In the metric setting we restrict ourselves to admissible splitting structures, as a rule.
%\ep
%
\bph{Reciprocal fundamental field and connection 1-form}\label{reciproc} 
We define lapse func\-tions\footnote{We use $\lapse^{-\dagger}$ as shorthand for $(\lapse^\dagger)^{-1}$.}
\[
\lapse=|\vec{w}|,\quad\lapse^{-\dagger}=|\df{\omega}|.
\]
It holds that $1\le\lapseprod=\lapse\lapse^{-\dagger}\in C^\infty(X,\group)$.\footnote{From the condition $\df{\omega}(\vec{w})^\otimes = 1\otimes\mathbf{t}$ (see \ref{ehresconnonprinc}) and Cauchy-Schwarz inequality it follows for time-like $\df{\omega}$ that $1\le|\vec{w}||\df{\omega}|$. The equal sign holds if and only if the splitting is regular.} The reciprocal\footnote{The term ``reciprocal" is borrowed from \cite[2.2]{dhaeseleer}.} fundamental field $\vec{w}^\dagger$ and the reciprocal connection 1-form $\df{\omega}^\dagger$ are given by%

\[
\vec{w}^\dagger=(\lapse^\dagger\lapse^\dagger)^\otimes\n{g}^{-1}\df{\omega},\qquad
\df{\omega}^\dagger=(\lapse^{-1}\lapse^{-1})^\otimes\n{g}\vec{w}.
\]
It holds that $\df{\omega}(\vec{w}^\dagger)^\otimes=\df{\omega}^\dagger(\vec{w})^\otimes=1\otimes\mathbf{t}$, and $\lapse^\dagger=|\vec{w}^\dagger|$, $\lapse^{-1}=|\df{\omega}^\dagger|$. For a regular splitting, we conclude from Cauchy-Schwarz inequality that $\vec{w}^\dagger=\vec{w}$, $\df{\omega}^\dagger=\df{\omega}$.
\ep
\bph{Shift vector field and shift 1-form}\label{shift}
The shift vector field $\shift\in\ves{1}(X,$ $\G;\mathfrak{\g}^*)$ \cite[3.2.2]{gourg} is defined by
\[
\shift=-\Pi\,\vec{w}^\dagger,\qquad\pd(\shift)=\oned,
\]
so that % Proof can be found 31.07.13
\[
\vec{w}=\Sigma\,\shift + \vec{w}^\dagger;
\]
see Fig.~\ref{fig:shift}.
% Internal: the splitting map has not been defined for covector-valued fields. The result would be $\n{S}\,\vec{w}^\dagger = (-\shift,1\otimes\mathbf{t})$.
%
\begin{figure}
\centering
\setlength{\unitlength}{0.055cm}
\begin{picture}(100,96)
\includegraphics[width=5.5cm]{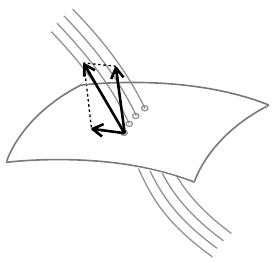}
\put(-76,72){$\vec{w}$}
\put(-54,68){$\vec{w}^\dagger$}
\put(-73,41){$\Sigma\,\shift$}
\end{picture}
\caption{Illustration of the shift vector field $\shift$.}\label{fig:shift}
\end{figure}
Likewise, the shift 1-form $\shiftform\in\dfs{1}(X,\G;\mathfrak{\g})$ is defined by
\[
\shiftform=-\Sigma^*\df{\omega}^\dagger,\qquad\pd(\shiftform)=\oned,
\]
so that % Proof runs along exactly the same lines as with the shift vector field.
\[
\df{\omega}=\Pi^*\shiftform + \df{\omega}^\dagger.
\]
\ep
\bph{Connection- and fiber-induced metric in the base manifold}\label{twometrics}
The connection-induced metric tensor in the base manifold is given by 
\[
\mathbf{h}_\Sigma=-\Sigma^*\mathbf{g}.
\]
The fiber-induced metric tensor\footnote{In adapted frames, the connection-induced metric amounts to taking the 3x3 submatrix $g_{ij}$ of $g_{\mu\nu}$ as spatial metric tensor. Conversely, the inverse of the fiber-induced metric tensor is given by the 3x3 submatrix $g^{ij}$ of $g^{\mu\nu}$.} is given by%
\footnote{$(\Pi\,\mathbf{g}^{-1})(\df{\alpha},\df{\alpha}^\prime)\stackrel{\n{def}}{=}\mathbf{g}^{-1}(\Pi^*\df{\alpha},\Pi^*\df{\alpha}^\prime)$.}

\[
\mathbf{h}_\Pi^{-1}=-\Pi\mathbf{g}^{-1}.
\]
Denote the horizontal lift corresponding to $\df{\omega}^\dagger$ as $\sigma_p^\dagger$; it is used to define the parametric map $\Sigma^\dagger$ in analogy to $\Sigma$ in \ref{paramaps}. One can also define the fiber-induced metric tensor by
\[
\mathbf{h}_\Pi=\mathbf{h}_{\Sigma^\dagger}=-(\Sigma^\dagger)^*\mathbf{g}.
\]
% Proof can be found 02.08.13, p.2
For regular splittings, $\Sigma^\dagger=\Sigma$, and the metric tensors agree, $\mathbf{h}_\Pi=\mathbf{h}_\Sigma$. Lastly, it can be shown that
\begin{alignat*}{2}
\shiftform&=(\lapse^{-1}\lapse^{-1})^\otimes&&\n{h}_\Sigma(\shift)\\
&=(\lapse^{-\dagger}\lapse^{-\dagger})^\otimes\,&&\n{h}_\Pi(\shift).
\end{alignat*}
% Proof can be found 05.08.13, p.2-3
\ep
\bph{Metric operators in nonregular splittings}\label{metopnonreg} % Proof can be found 20.08.13, pp.2-4
%Lengthy but trivial calculations yield the decomposition of the space-time metric tensor in terms of $\mathbf{h}_\Sigma$ and $\lapse^\dagger$,
%%%with $(\vec{\atest},\tilde{\btest})=\n{S}\vec{v}$, $(\df{\alpha},\tilde{\beta})=\n{S}^{-*}\df{\gamma}$,
%%%\begin{alignat*}{5}
%%%\mathbf{g}(\vec{v},\vec{v}^\prime)\Big|_p&=-\mathbf{h}_\Sigma(\vec{\atest}+\tilde{\btest}\shift,\vec{\atest}^\prime+\tilde{\btest}^\prime\shift)+
%%%(\lapse^\dagger\lapse^\dagger)^\otimes(\tilde{\btest},\tilde{\btest}^\prime)&&\Big|_{\varphi(p)},\\
%%%\mathbf{g}^{-1}(\df{\gamma},\df{\gamma}^\prime)\Big|_p&=-\mathbf{h}_\Sigma^{-1}(\df{\alpha},\df{\alpha}^\prime)+
%%%(\lapse^{-\dagger}\lapse^{-\dagger})^\otimes(\tilde{\beta}-\cont{\shift}\df{\alpha},\tilde{\beta}^\prime-\cont{\shift}\df{\alpha}^\prime)&&\Big|_{\varphi(p)};
%%%\end{alignat*}
%\begin{align*}
%\n{S}^{-*}\mathbf{g}&=\begin{pmatrix}1&\mult{\shift}\\0&1\end{pmatrix}^*\bigl(-\mathbf{h}_\Sigma,(\lapse^\dagger\lapse^\dagger)^\otimes\bigr),\\[3mm]
%\n{S}^{**}\mathbf{g}^{-1}&=\begin{pmatrix}1&0\\-\cont{\shift}&1\end{pmatrix}^*\bigl(-\mathbf{h}_\Sigma^{-1},(\lapse^{-\dagger}\lapse^{-\dagger})^\otimes\bigr);
%\end{align*}%
%compare with \cite[Eq.~(3.11)]{Arnowitt1962}\footnote{Note that the space-time signature in \cite{Arnowitt1962} is $(-,+,+,+)$. Moreover, their lapse function $\lapse$ according to (3.9a) corresponds to $\lapse^\dagger$ in our account.} For $\shift=0$ we recover the regular case, and the expression reduces to the one given in \ref{gdecomp}. The splitting of the Riesz operator is obtained in the same way,
Lengthy but trivial calculations yield the splitting of the space-time Riesz operator in terms of the connection-induced metric,%
\footnote{Compare with \cite[Eq.~(3.11)]{Arnowitt1962}. Note that the space-time signature in \cite{Arnowitt1962} is $(-,+,+,+)$. Moreover, their lapse function $\lapse$ according to (3.9a) corresponds to $\lapse^\dagger$ in our account.}
\begin{align*}
\n{S}^{-*}\circ\n{g}\circ\n{S}^{-1}&=
\begin{pmatrix}\n{Id}&-\mult{\shiftform}\\\cont{\shift}&\lapseprod^{-2}\n{Id}-\cont{\shift}\circ\mult{\shiftform}\end{pmatrix}\circ
\begin{pmatrix}1&0\\0&
\,(\lapse\lapse)^\otimes\end{pmatrix}\n{h}_\Sigma\circ\signop,\\[3mm]
\n{S}\circ\n{g}^{-1}\circ\n{S}^{*}&=\begin{pmatrix}1&0\\0&
(\lapse^{-1}\lapse^{-1})^\otimes\end{pmatrix}\n{h}_\Sigma^{-1}\circ\signop
\circ\lapseprod^2\begin{pmatrix}\lapseprod^{-2}\n{Id}-\mult{\shiftform}\circ\cont{\shift}&\mult{\shiftform}\\-\cont{\shift}&\n{Id}\end{pmatrix},
\end{align*}
where $\lapseprod=\lapse\lapse^{-\dagger}$; see \ref{reciproc}.
%Note the extra matrices compared to \ref{gdecomp}.
 In the regular case we have $\shift=\shiftform=0$, $\lapseprod=1$, and the expressions reduce to those given in \ref{gdecomp}.
Along the same lines as in \ref{dhodge} and with \ref{atools}, we obtain the splitting of the Hodge operator acting on differential forms,
\[
{\n{S}}^{-*}\circ*_4\circ{\n{S}}^{*}=\lapseprod
\begin{pmatrix}\n{Id}&-\mult{\shiftform}\\\cont{\shift}&\lapseprod^{-2}\n{Id}-\cont{\shift}\circ\mult{\shiftform}\end{pmatrix}\circ
\begin{pmatrix}0&\lapse^{-1}*_\Sigma\circ\;\signop\\ \lapse\,*_\Sigma&0\end{pmatrix},
\]
where $*_\Sigma$ is the Hodge operator induced by the metric tensor $\mathbf{h}_\Sigma$. Applying this Hodge operator to $(0,1)$ yields 
\[\n{S}^{-*}\df{\kappa}_4=\bigl(0,\lapse^\dagger\df{\kappa}_\Sigma\bigr);\]
this extends the definition of the twisted unit spatial volume form $\df{\kappa}_3$ in \ref{dhodge}. Similar expressions for the metric operators can be obtained in terms of $\mathbf{h}_\Pi$; these are collected in Appendix~\ref{ametopthread}.
\ep
\bph{Lagrangian and Eulerian observer}\label{lageulobs} 
In the theory of relativity, natural splitting structures and their fields are suitable for the formulation and solution of initial-value problems. The natural splitting is also called the {\em hypersurface approach}, where the space-like hypersurfaces are the leaves of the horizontal foliation (see \ref{frobenius}). It is suggested in \cite[2.3]{gourg} and \cite[p.~2530]{Smarr1978} that the natural splitting encompasses features belonging to regular splittings associated with so-called Lagrangian and Eulerian observers; see Fig.~\ref{fig:lageul}:
\begin{enumerate}
\item The Lagrangian observer and its regular splitting are defined on the world-lines of the hypersurface approach. The space platforms of the regular splitting, associated with the Lagrangian observer, are orthogonal to the world-lines, and the Lagrangian observer metric agrees with the fiber-induced metric in the hypersurface approach. 
\item The Eulerian observer's world-lines are orthogonal to the space platforms in the hypersurface approach. The world-lines of the Eulerian observer can be obtained from the natural splitting in the hypersurface approach via integration of the vector field $\vec{w}^\dagger$.\footnote{The vector field $\vec{w}^\dagger$ has to be complete, and the case of integral curves that come arbitrarily close to each other has to be ruled out \cite[16.10.3]{Dieudonne1972}.} The space platforms and the time synchronization are shared between the hypersurface approach and a regular splitting associated with the Eulerian observer. The Eulerian observer's metric is isometrically related to the connection-induced metric in the hypersurface approach. 
\end{enumerate}
% See scan 20.08.13, p.1 and pp.6-7
In the literature, the connection-induced metric is more prominent, most notably in the ADM Hamiltonian formulation of general relativity \cite{Arnowitt1962}. The fiber-induced metric is advocated in \cite{Fodor1994}.
\ep
\begin{figure}
\centering
\setlength{\unitlength}{0.04cm}
\begin{picture}(100,110)
\includegraphics[width=3.5cm]{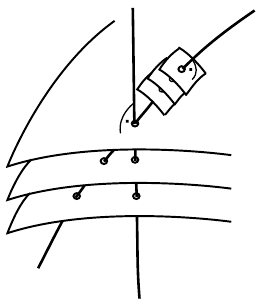}
\put(-82,8){L}
\put(-17,70){L}
\put(-39,8){E}
\put(-94,43){E}
\end{picture}
\caption{Illustration of space platforms and world-lines of the Lagrangian (L) and Eulerian (E) observer. Note that the space platforms of the Eulerian observer are given by the leaves of the foliation.}\label{fig:lageul}
\end{figure}
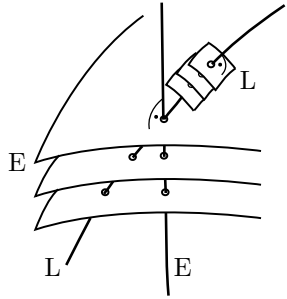
%
%%%%%%%%%%%%%%%%%%%%%%%%%%%%%%%%%%%%%%%%%%%%%%
\section{Kinematic parameters of observers}\label{sec:kinec}
The four-velocity in space-time encodes the kinematics of an observer. In this section we define the kinematic parameters in the observer's relative space, closely following \cite[2.2.3]{Jantzen2012}. The use of a regular splitting structure is mandatory to obtain physical kinematic parameters. In \ref{4velo} we introduce a time orientation, which will be used from there on to eliminate any twist with respect to $\group$.
\labcount{kinematics}
\bph{Four-velocity}\label{4velo}
An observer's four-velocity $\vec{u}$ is the vertical future-direc\-ted vector field with $|\vec{u}|=c_0$.
Here, $c_0$ denotes the vacuum speed of light, $\pd(c_0)=\pdim{L}\pdim{T}^{-1}$.\footnote{Note that the statement $\pd(c_0)=\pdim{L}\pdim{T}^{-1}$ does not imply a certain choice of units to express numerical values; for example, the speed of light might still be set to 1 and time be measured in meters.}
The four-velocity is independent of the choice of a principal $\G$-bundle. The metric Riesz dual of the four-velocity is denoted $\df{\mu}=\n{g}\,\vec{u}$.\label{urelations}
\ep
\bph{Time orientation, elimination of time twist}\label{timeorient}
It is through the four-velocity that space-time acquires time orientation. Denote by $\vec{e}^\uparrow$ a basis of the Lie algebra, such that $\zeta(\vec{e}^\uparrow)$ is future-directed. From now on, we will work with fixed time orientation and remove any twist with respect to $\group$ using $\vec{e}^\uparrow$. For instance, consider 
$\df{\alpha}_\doubletwist\in\mathcal{F}^k_\doubletwist(X,\G)$, which can be represented by $(\df{\alpha},\tilde{\df{\kappa}})\in\dfs{k}(X,\G)\times\dfs{n-1}(X,\G;\mathfrak{g}^*)$; compare with \ref{twistsplit}. Define $\df{\alpha}_\twist=\bigl(\df{\alpha},\tilde{\df{\kappa}}(\vec{e}^\uparrow)\bigr)\in\mathcal{F}^k_\twist(X,\G)$. We identify $\df{\alpha}_\doubletwist$ with $\df{\alpha}_\twist$, thus rendering the spaces $\mathcal{F}^k_\doubletwist(X,\G)$ and $\mathcal{F}^k_\twist(X,\G)$ isomorphic. Then, it holds that
\begin{align*}
\vec{u}=c_0(\Pi^*\lapse^{-1})\vec{w},\qquad \df{\mu}=c_0(\Pi^*\lapse)\df{\omega}; 
\end{align*}
% Internal: second relation only for regular splitting
see Fig.~\ref{fig:lapse}, and compare with \ref{lapsefun}.
\begin{center}
\begin{tabular}{|c|c|c|}
\hline
&$\in$&$\pd(\,\cdot\,)$\\
\hline
$\vec{u}$&$\mathcal{X}_V^1(P)$&$\pdim{T}^{-1}$\\
$\df{\mu}$&$\mathcal{F}_V^1(P)$&$\pdim{L}^{2}\pdim{T}^{-1}$\\\hline
$\vec{w}$&$\mathcal{X}_V^1(P;\mathfrak{\g}^*)$&$\oned$\\
$\df{\omega}$&$\mathcal{F}_V^1(P;\mathfrak{\g})$&$\oned$\\
\hline
\end{tabular}
\end{center}
\ep
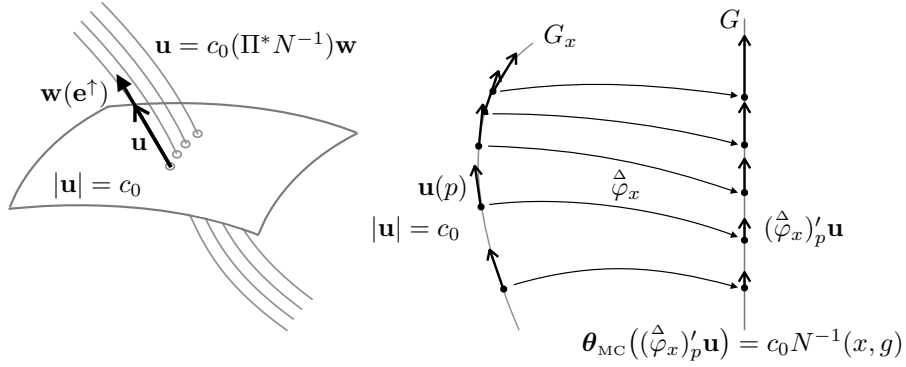
\begin{figure}
\centering
\hskip-20mm
\setlength{\unitlength}{0.8mm}
\begin{picture}(100,57)
\includegraphics[width=4.8cm,clip]{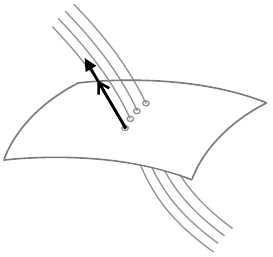}
\put(-39,31){$\vec{u}$}
\put(-52,24){$|\vec{u}|=c_0$}
\put(-54,39){$\vec{w}(\vec{e}^\uparrow)$}
\put(-35,47){$\vec{u}=c_0(\Pi^*\lapse^{-1})\vec{w}$}
\end{picture}\hskip-20mm
\setlength{\unitlength}{0.39mm}
\begin{picture}(100,110)
\includegraphics[width=3.9cm]{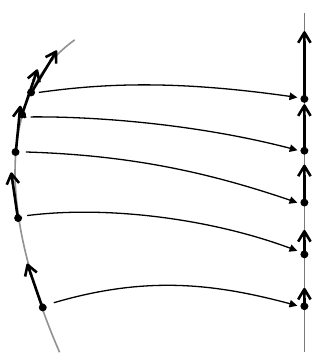}
\put(-73,100){$\G_x$}
\put(-13,105){$\G$}
\put(-50,48){$\hhat{\varphi}_{x}$}
\put(2,34){$(\hhat{\varphi}_x)'_p\vec{u}$}
\put(-60,-5){$\MC\bigl((\hhat{\varphi}_x)'_p\vec{u}\bigr)=c_0\lapse^{-1}(x,\g)$} 
\put(-116,48){$\vec{u}(p)$}
\put(-132,34){$|\vec{u}|=c_0$}
\end{picture} % Proof can be found 09.08.13, p.1
\caption{Illustration of elements in the relation between fundamental field $\vec{w}$, lapse function $\lapse$, and four-velocity $\vec{u}$, $\vec{e}^\uparrow\in\mathfrak{\g}$.}\label{fig:lapse}
\end{figure}
\bph{Proper time, time derivative}\label{propertime}
Denote $\tau$ the observer's proper time, that is, the arc-length along world-lines scaled by $c_0^{-1}$, $\pd(\tau)=\pdim{T}$. The time derivative on the observer's relative space is denoted $\partial_\tau$. It is related to the group derivative \ref{groupderiv} and the four-velocity by
\begin{alignat*}{2}\partial_\tau&=c_0\lapse^{-1}\timederiv=\Sigma^*\circ\lie{\vec{u}}\circ\Pi^*,&\quad\quad \pd(\partial_\tau)&=\pdim{T}^{-1}.
\end{alignat*}
Like the group derivative, the time derivative is a derivation of degree 0.
% Internal: The factor $c_0\lapse^{-1}$ can be included in the vector field argument of the Lie derivative without add'l extra term due to orthogonality reasons.
\ep
\bph{Kinematic parameters}\label{kine}
Denote $\nabla$ the covariant derivative of the Levi-Civita connection on space-time. $\nabla\df{\mu}$ is a $\binom{0}{2}$-tensor field that is decomposed as%
\footnote{The second row of the matrix is zero, because for any vector $\vec{v}$ it holds that $0=\nabla_\vec{v}\mathbf{g}^{-1}(\df{\mu},\df{\mu})=2\,\mathbf{g}^{-1}(\df{\mu},\nabla_\vec{v}\df{\mu})=2\,\cont{\vec{u}}\nabla_\vec{v}\df{\mu}\sim\cont{\vec{w}}\nabla_\vec{v}\df{\mu}$.}
\[
\n{S}^{-*}\circ \nabla\df{\mu}\circ\n{S}^{-1} = \begin{pmatrix}\df{\gamma}&\tilde{\df{\delta}}\\0&0\end{pmatrix}.
\]
We obtain the $\binom{0}{2}$-tensor field $\df{\gamma}$ and the Lie coalgebra-valued 1-form $\tilde{\df{\delta}}$,
\begin{alignat*}{3}
\df{\gamma}&\stackrel{\n{def}}{=}\Sigma^*\nabla\df{\mu}:\quad&&\ves{1}(X,\G)&&\to\dfs{1}(X,\G),\\
\tilde{\df{\delta}}&\stackrel{\n{def}}{=}\Sigma^*\nabla_\vec{w}\,\df{\mu}:\quad&&\ves{0}(X,\G;\mathfrak{g})&&\to\dfs{1}(X,\G).
\end{alignat*}
The tensor field $\df{\gamma}$ is further decomposed into its antisymmetric and symmetric parts,\footnote{Our decomposition reads $\df{\gamma}=-\df{\eta}-\df{\lambda}$, while in literature we also find $\df{\gamma}=-\df{\eta}+\df{\lambda}$ \cite[Eq.~(2.93)]{Jantzen2012} and $\df{\gamma}=\df{\eta}+\df{\lambda}$ \cite[Eq.~(2.8)]{ellis}, \cite[Eq.~(4.17)]{hawking}. The different sign of $\df{\eta}$ is conventional, while the different sign of $\df{\lambda}$ is due to the fact that space-time metric has the signature $(-,+,+,+)$ in the references.}
\[
\df{\eta}\stackrel{\n{def}}{=}-\n{Asy}(\df{\gamma}),\qquad
\df{\lambda}\stackrel{\n{def}}{=}-\n{Sym}(\df{\gamma}),
\]
where the antisymmetric part should be read as a two-form. The quantities $\tilde{\df{\delta}}$, $\df{\eta}$, and $\df{\lambda}$ are the {\em kinematic parameters}. They are characterized as follows:\footnote{\label{pdimkine}The unusual physical dimension of the quantities is a direct consequence of their being derived from the metric dual $\df{\mu}$ of the four-velocity. In this way, the quantities are well adjusted to fit into the electromagnetic theory discussed here. For a purely kinematic treatise, a derivation solely based on the four-velocity may be preferable.}
\begin{center}
\begin{tabular}{|c|c|c|l|}
\hline
& $\in$&$\pd(\,\cdot\,)$&name\\
\hline
$\tilde{\df{\delta}}$&$\mathcal{F}^1(X,\G;\mathfrak{g}^*)$&$\pdim{L}^{2}\pdim{T}^{-1}$&acceleration form\\
$\df{\eta}$&$\mathcal{F}^2(X,\G)$&$\pdim{L}^{2}\pdim{T}^{-1}$&vorticity form\\
\hline
\end{tabular}
\end{center}
\begin{center}
\begin{tabular}{|c|rcl|c|l|}
\hline
&$\cdot $&$\!\!\!\!\!\to\!\!\!\!\!$&$\cdot$&$\pd(\,\cdot\,)$&name\\
\hline
$\df{\lambda}$&$\ves{1}(X,\G)$&$\!\!\!\!\!\to\!\!\!\!\!$&$\mathcal{F}^1(X,\G)$&$\pdim{L}^2\pdim{T}^{-1}$&expansion tensor\\
\hline
\end{tabular}
\end{center}
The Levi-Civita connection is not strictly necessary for the definition of the kinematic parameters. They can be equivalently defined in terms of characteristics of the splitting structure,
\begin{align*}
\tilde{\df{\delta}}&=c_0^{-1}\lapse\,\Sigma^*\lie{\vec{u}}\df{\mu},\\
2\,\df{\eta}&=\Sigma^*\n{d}\,\df{\mu},\\
2\,\df{\lambda}&=-\Sigma^*\lie{\vec{u}}\mathbf{g};
\end{align*}
see \ref{akine}. Note that $(2\,\df{\eta},\tilde{\df{\delta}})=\n{S}^{-*}\n{d}\,\df{\mu}.$ % Proof can be found 28.06.13
The acceleration and vorticity forms are related to the curvature and variance forms via
\begin{align*}
\tilde{\df{\delta}}&=c_0\lapse(\df{\chi}-\lapse^{-1}\n{D}\lapse),\\
2\,\df{\eta}&=c_0\lapse\,\df{\Omega},
\end{align*}
% Proof can be found 28.06.13
% Internal: These relations hold only for regular splittings.
which is seen by setting $\df{\mu}=c_0(\Pi^*\lapse)\,\df{\omega}$; compare with \ref{4velo}.\footnote{In literature, $\lapse^{-1}\n{D}\,\lapse$ is frequently written as $\n{D}\,\n{ln}\lapse$. We cannot do this here since i) the lapse function is Lie-coalgebra valued, and ii) the lapse function has physical dimension~$\pdim{T}$.}
Finally, according to \ref{aderobsmed}, twice the expansion tensor is the time derivative of the observer metric tensor,
\[
2\,\df{\lambda}=\partial_\tau\n{h}.
\]
\ep
\bph{Expansion scalar and shear tensor}\label{exptensor}
For completeness, we note that the expansion tensor is usually decomposed \cite[Eq.~(2.96)]{Jantzen2012} into
\[
\df{\lambda}=\df{\sigma}+\frac{1}{3}\lambda\n{h},
\]
where $\df{\sigma}$ is the trace free shear tensor, and $\lambda$ the expansion scalar.%
\footnote{It can be shown that this definition is consistent with the standard definiton $\lambda=\n{div}\,\vec{u}$ \cite[Eq.~(152)]{ellisnotes}, \cite[Appendix~E]{Fecko1997}.}
% Proof can be found 18.09.13, p.2
It is defined by 
\[
\lambda=\n{Tr}(\n{h}^{-1}\df{\lambda}),\qquad\pd(\lambda)=\pdim{T}^{-1}.
\]
% Internal: Trace in index notation is a self contraction. To do this, we need a (1,1)- rather than a (0,2)-tensor. We have to raise an index, and that is why $h^{ij} = \n{h}^{-1}$ is required. Moreover, for (1,1)-tensors the framework of paragraph \ref{genexp} applies. The second equation follows from the first one by applying $\n{h}^{-1}$, taking the trace, and considering $\n{Tr}\n{Id}=3$.
It is shown in \ref{aexpansion} that 
\[
\partial_\tau\df{\kappa}_3=\lambda\,\df{\kappa}_3,
\]
% Internal: this holds only for regular splittings. 
where $\df{\kappa}_3$ is the twisted unit volume form.
% \cite[Appendix~E]{Fecko1997}
\ep
%
%%%%%%%%%%%%%%%%%%%%%%%%%%%%%%%%%%%%%%%%%%%%%%
\section{Classification of observers and of splitting structures}\label{sec:classification}\labcount{classification}
Recall the distinction between observers and splitting structures. Observers are persons or entities experiencing electromagnetic phenomena. Splitting structures are used to predict or explain measurements by the formulation and solution of initial-value problems. Observers are classified with respect to their kinematic parameters. Splitting structures are classified with respect to properties of the time-translation, the Ehresmann connection, and the time synchronization, as well as the compatibility of time-translation and Ehresmann connection with the space-time metric; see Fig.~\ref{fig:class}.
\addtocounter{footnote}{-1}
\begin{figure}
\centering
\setlength{\unitlength}{0.1cm}
\begin{picture}(100,40)
\includegraphics[width=10cm]{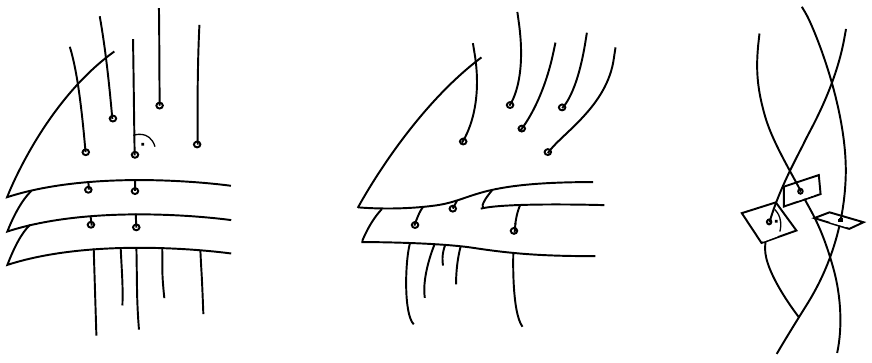}
\end{picture}
\caption{Illustrations of different splittings: holonomic and regular (left); integrable, but neither principal nor regular (middle)\protect\footnotemark; {regular, but not integrable} (right). The sheets and tiles represent space platforms, the lines depict world-lines.}\label{fig:class}
\end{figure}
\footnotetext{From an integrability point of view, the middle surface does not ``end". This rather has to do with discretization. If we write $\df{\omega}=f\n{d}\nu$ in some basis of the Lie algebra, we might normalize $f$ in the considered region such that $0<f\le 1$. We display parts of level surfaces of $\nu$ subject to the condition $f>0.5$. This approach emphazises that integrals of $\df{\omega}$ are path-dependent. Further discussion on how (not) to visualize 1-forms can be found in \cite[5.7]{bachmann}.}
\bph{Classification of observers} 
\begin{itemize}
\item A {\em nonrotating} observer has vanishing vorticity, $\df{\eta}=0$. 
\item A {\em geodesic} observer experiences no acceleration, ${\tilde{\df\delta}}=0$.
\item An observer is called {\em Born rigid} \cite[p.~15]{Born1909} if there is no expansion, $\df{\lambda}=0$.
\item An {\em inertial} observer is nonrotating, geodesic, and Born rigid, $\nabla\vec{u}=0$.
\end{itemize}
Note that except for flat Minkowski space-time there might be no inertial observers, since covariantly constant vector fields may not exist.
\ep
\bph{Classification of splitting structures continued} 
Recall the classification of splitting structures from \ref{classobs} and \ref{regular}. %
We introduce a further classification by studying the interplay between elements of the splitting and the space-time metric.
\begin{itemize}
\item A {\em metric} splitting\footnote{The qualifier `metric' is borrowed from the theory of linear connections \cite[15.3]{Fecko}. In many ways, the Ehresmann connection in our application plays a similar role as a linear connection, albeit with the sole focus on the time direction vs.\ the spatial subspace.} is regular, with constant lapse function, $\n{D}\,\lapse^{-1}=0$ and $\timederiv N^{-1}=0$.
\item A {\em standard} splitting is natural and metric.
\item A splitting is {\em stationary} if the fundamental field $\vec{w}$ is a Killing field of the space-time metric, $\lie{\vec{w}}\mathbf{g}=0$ \cite[Sec.~12]{Jantzen1992}. A space-time that admits time-like Killing vector fields is called a {\em stationary space-time}.
\end{itemize}
The results are summarized in the table below, where a cross means that the classification in the respective column imposes a condition on the mathematical structure in the respective row.
\begin{center}\small
\renewcommand{\tabcolsep}{1mm}\renewcommand{\arraystretch}{1}
\newcommand{\cmacro}[1]{\multicolumn{1}{c|}{#1}}
\newcommand{\rmacro}[1]{\cmacro{\raisebox{-2mm}[0mm][0mm]{#1}}}
\newcommand{\xmacro}{\rmacro{\sf X}}
\begin{tabular}{|p{2.2cm}|p{1.2cm}|p{1.4cm}|p{1.2cm}|p{1.2cm}|p{1.6cm}|p{1.6cm}|}\hline
& \multicolumn{5}{c|}{\it Standard} & \\ \cline{2-6}
\raisebox{-2mm}[0mm][0mm]{\bf Classification}& \multicolumn{3}{c|}{\it Natural} & \multicolumn{2}{c|}{\raisebox{-2mm}[0mm][0mm]{\it Metric}} & \rmacro{\it Stationary} \\ \cline{2-4}
& \multicolumn{2}{c|}{\it Holonomic} && \multicolumn{2}{c|}{} & \\ \cline{2-3}\cline{5-6}
& \cmacro{\it Flat} & \cmacro{\it Principal} & & \cmacro{\it Regular} & &\\\hline\hline
{\bf Condition/} &\rmacro{$\df{\Omega}=0$} & \rmacro{$\df{\chi}=0$} & \rmacro{$\df{\Gamma}=0$} & \rmacro{$\shift=0$} & \multicolumn{1}{r|}{\rule[0mm]{0mm}{3.5mm}$\n{D}\,\lapse^{-1}=0$} & \rmacro{$\lie{\vec{w}}\mathbf{g}=0$}\\ 
{\bf Structure}&&&&&\multicolumn{1}{r|}{$\timederiv\lapse^{-1}=0$}&\\\hline
Ehresmann\newline connection & \xmacro & \xmacro & \xmacro & \xmacro & & \\ \hline
Principal\newline action & & \xmacro & \xmacro & & \xmacro & \xmacro \\ \hline
Global\newline section & & & \xmacro & & & \\ \hline
Conformal \newline class & & & & \xmacro & & \\ \hline
Space-time \newline metric & & & & & \xmacro & \xmacro\\ \hline
\end{tabular}
\end{center}
\ep
\bph{Observers and splitting structures}\label{implic}Some implications of the above classi\-fications are given below:
\begin{itemize}
\item For metric splittings, the characteristics of the splitting structure are directly linked to the observer's kinematic parameters, 
\begin{align*}(\tilde{\df{\delta}},2\,\df{\eta},2\,\df{\lambda})=c_0\lapse\,(\df{\chi},\df{\Omega},\df{\vartheta}),\end{align*} 
where $\df{\vartheta}=(\lapse^{-1}\lapse^{-1})^\otimes\timederiv\mathbf{h}$.
\item For regular splittings, the following conditions are equivalent (see \ref{aderobsmed}):
\begin{enumerate}
\item The splitting is stationary, $\lie{\vec{w}}\mathbf{g}=0$.
\item The splitting is principal, $\df{\chi} = 0$, with time-independent observer metric and lapse function, $\timederiv\mathbf{h}=0$ and $\timederiv\lapse^{-1}=0$.
\end{enumerate}
Regular stationary splittings may, therefore, only be constructed on world-lines of Born-rigid observers. 
\item For regular principal splittings it follows from $\df{\chi} = 0$ that $\tilde{\df{\delta}}=c_0\n{D}\lapse$. Hence, $\lapse$ takes the role of an acceleration potential.
\item For regular splittings, a rotation of the observer, $\df{\eta}\neq0$, implies nonvanishing curvature of the splitting, $\df{\Omega}\neq0$, and vice versa; see \ref{kine}. It follows that for rotating observers the regular splitting cannot be natural; see \ref{classobs}. As a consequence, a global Einstein synchronization (see \ref{regular}) cannot exist for rotating observers.
\end{itemize}
\ep
%
%%%%%%%%%%%%%%%%%%%%%%%%%%%%%%%%%%%%%%%%%%%%%%
\section{Proxies for Lie-(co)algebra valued fields}\label{sec:proxy}
In this section we recover the customary parametric fields, that is to say, we split fields in such a way that the vertical fields are not Lie-(co)algebra valued. The quantities thus found are proxies for their Lie-(co)algebra valued counterparts, in the same sense as vector fields in three dimensions can act as proxies for their differential-form counterparts.\footnote{The term ``vector proxy" was coined by Bossavit \cite{bossavit1998a}.} The proxy map is established with the help of the lapse function, which encodes the vertical part of the space-time metric. The construction of the proxy map is analogous to that of the so-called translation isomorphisms \cite[10.2]{jaenich_va_eng}, which leverage the horizontal metric to map differential forms to their vector proxy representation.
\labcount{normalization}
\bph{The proxy map}\label{proxymap}
Recall \ref{lapsefun}, Point \ref{case3}, where we stated that for each $(x,\g)$ the lapse function can be seen as a Hodge operator on the algebra $\mathsf{\Lambda}\,\mathfrak{g}^*$. Since we introduced a fixed time orientation in \ref{4velo}, the lapse function provides isomorphisms
\begin{align*}
\lapse\;:\;\left\{
\begin{alignedat}{2}
\mathcal{F}^k(X,\group)&\xrightarrow{\,\,\sim\,\,}\mathcal{F}^k(X,\group;\mathfrak{g}^*)&&\;:\;\df{\beta}\mapsto\lapse\df{\beta},\\
\mathcal{F}^k_\twist(X,\group)&\xrightarrow{\,\,\sim\,\,}\mathcal{F}^k_\twist(X,\group;\mathfrak{g}^*)&&\;:\;(\df{\beta},\df{\kappa})\mapsto(\lapse\df{\beta},\df{\kappa}).
\end{alignedat}\right.
\end{align*}
Hence, the lapse function induces the proxy map
\begin{alignat*}{2}
 \proxy^{-*}:\dfs{k}(X,\G)\times\dfs{k-1}(X,\G;\mathfrak{\g}^*)&\to\dfs{k}(X,\G)\times\dfs{k-1}(X,\G)\\
(\df{\alpha},\tilde{\df{\beta}})&\mapsto(\df{\alpha},\df{\beta})=(\df{\alpha},c_0\lapse^{-1}\,\tilde{\df{\beta}}),
\end{alignat*}
$\pd(\proxy^{-*})=(\oned,\pdim{T}^{-1})$. For vector fields, the proxy map reads
\begin{alignat*}{2}
 \proxy:\ves{k}(X,\G)\times\ves{k-1}(X,\G;\mathfrak{\g})&\to\ves{k}(X,\G)\times\ves{k-1}(X,\G)\\
(\vec{\atest},\tilde{\df{\btest}})&\mapsto(\vec{\atest},\df{\btest})=(\vec{\atest},c_0^{-1}\lapse\,\tilde{\df{\btest}}),
\end{alignat*}
$\pd(\proxy)=(\oned,\pdim{T})$.\footnote{The splitting and proxy maps can be concatenated for regular splittings into $(\proxy\circ\n{S})^{-*}:\dfs{k}(P)\xrightarrow{\,\,\sim\,\,}\dfs{k}(X,\G)\times\dfs{k-1}(X,\G):
\df{\gamma}\mapsto(\df{\alpha},{\df{\beta}})=(\Sigma^*\df{\gamma},\Sigma^*\cont{\vec{u}}\df{\gamma})$. The inverse is given by $(\proxy\circ\n{S})^{*}:(\df{\alpha},{\df{\beta}})\mapsto\df{\gamma}=\Pi^*\df{\alpha}+c_0^{-2}\df{\mu} \wedge\Pi^*\tilde{\df{\beta}}$.
} We call the subsequent application of splitting and proxy maps $\proxy^{-*}\circ\n{S}^{-*}=(\proxy\circ\n{S})^{-*}$ proxy-splitting, and the fields in the image of the proxy-splitting we call proxies. 
\ep
\bph{Exterior derivative}\label{obsderiv}
The proxy-splitting of the exterior derivative reads 
\begin{align*}(\proxy\circ \n{S})^{-*}\circ\n{d}\circ({\proxy\circ \n{S}})^*=\begin{pmatrix}{{\n{D}}}&c_0^{-2}\,\mult{2\bar{\df{\eta}}}\\
\;\partial_\tau\;&c_0^{-2}\,\mult{\bar{\df{\delta}}} - {\n{D}}\end{pmatrix},\end{align*}
where
\begin{alignat*}{2}
\bar{\df{\delta}}&\stackrel{\n{def}}{=}c_0^2\lapse^{-1}(\df{\chi}-\n{D})\lapse,&\quad\quad \pd(\bar{\df{\delta}})=\pdim{L}^2\pdim{T}^{-2},\\
2\bar{\df{\eta}}&\stackrel{\n{def}}{=}c_0\lapse\,\df{\Omega},&\quad\quad \pd(\bar{\df{\eta}})=\pdim{L}^2\pdim{T}^{-1}.
\end{alignat*}
For regular splittings, these are the proxies of the acceleration form and the vorticity form, $(2\bar{\df{\eta}},\bar{\df{\delta}})=(2\df{\eta},{\df{\delta}}){=}(\proxy\,\n{S})^{-*}\n{d}\,\df{\mu}$; see \ref{kine}. The covariant exterior derivative can be written
\begin{align*} \n{D} = \n{d}-(\df{\gamma}\specialwedge\partial_\tau),\end{align*}
where we set $\df{\gamma}=c_0^{-1}\lapse\,\df{\Gamma}$, so that $(\df{\gamma}\specialwedge\partial_\tau)=(\df{\Gamma}\specialwedge\timederiv)$.
The splittings of contraction, exterior product, and Lie derivative in terms of proxies can be found in Appendix~\ref{normsplitop}.
\ep % Proof can be found 11.08.13, p.2
\bph{Maxwell's equations}
The splitting of the space-time fields $A$, $F$, $H$, and $J$ yields proxy forms
\begin{alignat*}{4}
&(\proxy\,\n{S})^{-*}\,A&&=(a,-\varphi),&&\quad\quad(\proxy\,\n{S})^{-*}\,F&&=(b,-e),\\
&(\proxy\,\n{S})^{-*}\,H&&=(d,h),&&\quad\quad(\proxy\,\n{S})^{-*}\,J&&=(\rho,-j),
\end{alignat*}
with
\begin{center}
\begin{tabular}{|c|c|c|c|}
\hline
&$\in$&$\pd(\,\cdot\,)$&$\pd(|\cdot|)$\\
\hline
$\varphi$&$\dfs{0}(X,\G)$&$\pdim{U}$&$\pdim{U}$\\
$e$&$\dfs{1}(X,\G)$&$\pdim{U}$&$\pdim{UL}^{-1}$\\
$h$&$\mathcal{F}^1_\twist(X,\G)$&$\pdim{I}$&$\pdim{IL}^{-1}$\\
$j$&$\mathcal{F}^2_\twist(X,\G)$&$\pdim{I}$&$\pdim{IL}^{-2}$\\
\hline
\end{tabular}
\end{center}
Maxwell's equations read in their general form
\begin{alignat*}{2}
\n{D}\,&b&&=c_0^{-2}2\bar{\df{\eta}}\wedge e,\\
\n{D}\,&e&&=-\partial_\tau b+c_0^{-2}\bar{\df{\delta}}\wedge e,\\
\n{D}\,&d&&=\rho-c_0^{-2}2\bar{\df{\eta}}\wedge h,\\
\n{D}\,&h&&= j+\partial_\tau d+c_0^{-2}\bar{\df{\delta}}\wedge h.
\end{alignat*}
The potential equations and charge continuity read
\begin{alignat*}{2}
\n{D}\,&a&&=b+c_0^{-2}2\bar{\df{\eta}}\wedge\varphi,\\
-\n{D}\,&\varphi&&=e+\partial_\tau a-c_0^{-2}\bar{\df{\delta}}\wedge\varphi,\\
\n{D}\,&j&&=-\partial_\tau\rho+c_0^{-2}\bar{\df{\delta}}\wedge j.
\end{alignat*}
\ep
\bph{Riesz operator} % Proof can be found 11.08.13, p. 4
The regular splitting of the Riesz operator reads for proxies
\begin{align*} (\proxy\circ \n{S})^{-*}\circ\n{g}\circ(\proxy\circ\n{S})^{-1}=
\begin{pmatrix}1&0\\0&c_0^2\end{pmatrix}\n{h}\circ\signop.
\end{align*}
\ep
\bph{Hodge operator} % Proof can be found 11.08.13, p.4
With the regular splitting of the volume form, $(\proxy\,\n{S})^{-*}\df{\kappa}_4=(0,c_0\,\df{\kappa}_3)$, the regular splitting of the Hodge operator in terms of proxies reads
\begin{alignat*}{2}(\proxy\circ \n{S})^{-*}\circ*_4\circ(\proxy\circ \n{S})^{*}&=\begin{pmatrix}0&c_0^{-1}*_3\circ\;\signop\\ c_0\,*_3&0\end{pmatrix}.
\end{alignat*}
Nonregular splittings of Riesz and Hodge operators are collected in Appendix \ref{anormetop}.
\ep
\bph{Constitutive relations}
The vacuum constitutive relations in three dimensions, resulting from a regular splitting, read for proxy forms
\[
d=\varepsilon_0 *_3 e,\quad h=\mu_0^{-1} *_3b.
\]
\ep
\bph{Energy-Momentum density}\label{EM1proxy} % Proof can be found 12.08.13, p.1
We refer to \ref{EM1} and find for the proxies
\begin{alignat*}{2}
(\proxy\circ \n{S})^{-*}\circ T\circ (\proxy\circ \n{S})^{-1}=\begin{pmatrix}-p&w\\
-m&-s\end{pmatrix},
\end{alignat*}
where
\begin{center}
\begin{tabular}{|c|rcl|c|l|}
\hline
& $\cdot\!\!\!\!\!$&$\to$&$\!\!\!\!\!\cdot$&$\pd(\,\cdot\,)$&name\\
\hline
$p$&$\ves{1}(X,\G)\!\!\!\!\!$&$\to$&$\!\!\!\!\!\mathcal{F}^3_\twist(X,\G)$&$\pdim{A}$&momentum density\\
$m$&$\ves{1}(X,\G)\!\!\!\!\!$&$\to$&$\!\!\!\!\!\mathcal{F}^2_\twist(X,\G)$&$\pdim{A}\pdim{T}^{-1}$&momentum flux density\\
\hline
\end{tabular}\end{center}\begin{center}
\begin{tabular}{|c|c|c|l|}
\hline
& $\in$&$\pd(\,\cdot\,)$&name\\
\hline
$w$&$\mathcal{F}^3_\twist(X,\G)$&$\pdim{AT}^{-1}$&energy density\\
$s$&$\mathcal{F}^2_\twist(X,\G)$&$\pdim{AT}^{-2}$&energy flux density\\
\hline
\end{tabular}
\end{center}
and from 
\begin{align*}
w&\stackrel{\n{def}}{=}c_0\tilde{w}(\lapse^{-1}),\\
m&\stackrel{\n{def}}{=}c_0\lapse^{-1}\tilde{m},\\
s&\stackrel{\n{def}}{=}c_0^2\lapse^{-1}\tilde{s}(\lapse^{-1}),
\end{align*}
follows
\begin{align*}
p(\vec{\atest})& =\cont{\vec{\atest}}b\wedge d,\\
w&=\frac{1}{2}\,(e\wedge d+h\wedge b),\\
m(\vec{\atest})&=-\cont{\vec{\atest}}d\wedge e-\cont{\vec{\atest}}b\wedge h+\frac{1}{2}\,\cont{\vec{\atest}}(d\wedge e+b\wedge h),\\
s&=e\wedge h.
\end{align*}
\ep
\bph{Four-Force density and energy-momentum balance}\label{momentumbalanceproxy} % Proof can be found 12.08.13, p.1
Starting from \ref{momentumbalance} we obtain
\begin{align*}
(\proxy\circ \n{S})^{-*}\circ{R}\circ(\proxy\circ\n{S})^{-1}=\begin{pmatrix}0&0\\
f&r\end{pmatrix},
\end{align*}
where $f$ and $r$ are characterized as follows:
\begin{center}
\begin{tabular}{|c|rcl|c|l|}
\hline
& $\cdot\!\!\!\!\!$&$\to$&$\!\!\!\!\!\cdot$&$\pd(\,\cdot\,)$&name\\
\hline
$f$&$\ves{1}(X,\G)$&$\!\!\!\!\!\to\!\!\!\!\!$&$\mathcal{F}^3_\twist(X,\G)$&$\pdim{AT}^{-1}$&force density\\
\hline
\end{tabular}\end{center}\begin{center}
\begin{tabular}{|c|c|c|l|}
\hline
& $\in$&$\pd(\,\cdot\,)$&name\\
\hline
$r$&$\mathcal{F}^3_\twist(X,\G)$&$\pdim{AT}^{-2}$&power density\\
\hline
\end{tabular}
\end{center}
where from 
\begin{align*}
f&\stackrel{\n{def}}{=}c_0\lapse^{-1}\tilde{f},\\
r&\stackrel{\n{def}}{=}c_0^2\lapse^{-1}\tilde{r}(\lapse^{-1})
\intertext{follows}
f(\vec{\atest})&=\cont{\vec{\atest}}e\wedge\rho+\cont{\vec{\atest}}b\wedge j,\\
r&=-e\wedge j.
\end{align*}
Assume there exists a Killing vector field $\vec{n}$ of the space-time metric such that $(\proxy\circ \n{S})\vec{n}=(\vec{\atest},\cdot)$. The momentum and energy balance equations then read
\begin{align*}
f(\vec{\atest})&=-\partial_\tau p(\vec{\atest})-(c_0^{-2}\mult{\bar{\df{\delta}}}-\n{D})m(\vec{\atest}),\\
r&=\phantom{-}\partial_\tau w-(c_0^{-2}\mult{\bar{\df{\delta}}}-\n{D})s.
\end{align*}
The last equation is also known as Poynting's theorem.
\ep
\fi
\ifdefined\INCLUDEAPPS
%%%%%%%%%%%%%%%%%%%%%%%%%%%%%%%%%%%%%%%%%%%%%%
%%%%%%%%%%%%%%%%%%%%%%%%%%%%%%%%%%%%%%%%%%%%%%
\newchapter{Applications}\label{sec:applications}
In the final chapter we make a short remark on the Ehrenfest paradox, and apply the splitting framework to the study of Schiff's answer to his 1939 ``Question in General Relativity'' \cite{Schiff1939}.
%
%%%%%%%%%%%%%%%%%%%%%%%%%%%%%%%%%%%%%%%%%%%%%%
\section{Ehrenfest paradox}\labcount{ehrenfest}
\bph{Original reference}We cite Paul Ehrenfest's 1909 formulation \cite{Ehrenfest1909} of the paradox from \cite{gron}:
\begin{quote}
Consider a relativistically rigid cylinder with radius $R$ and height $H$. It is given a rotating motion about its axis, which finally becomes constant. As measured by an observer at rest, the radius of the rotating cylinder is $R'$. Then $R'$ has to fulfill the following two contradictory requirements:
\begin{enumerate}
\item[a)]The circumference of the cylinder must obtain a contraction $2\pi R' < 2\pi R$ relative to its rest
length, since each of its elements move with an instantaneous velocity $R' \omega$.
\item[b)]If one considers each element along a radius, then the instantaneous velocity of each element is directed perpendicular to the radius. Hence, the elements of a radius cannot show any
contraction relative to their rest length. This means that: $R' = R$.
\end{enumerate}
\end{quote}
In our own words, the paradox comes down to the following question: If the cylinder is rigid in the relativistic sense of Born, then how can the ratio of its circumference and diameter not be equal to $\pi$ when it rotates? An extensive survey of the decades-long discussion that ensued is found in \cite{gron}, as well as a summary of the resolution of the paradox. We do not intend to make a contribution to the debate; rather, we show how the above observation and the resolution of the paradox come about in the regular splitting from a purely kinematic point of view, omitting practical questions of elasticity.
\ep
\bph{Kinematic setting}\label{kinsetting}
\begin{figure}
\centering
\includegraphics[width=0.9\linewidth]{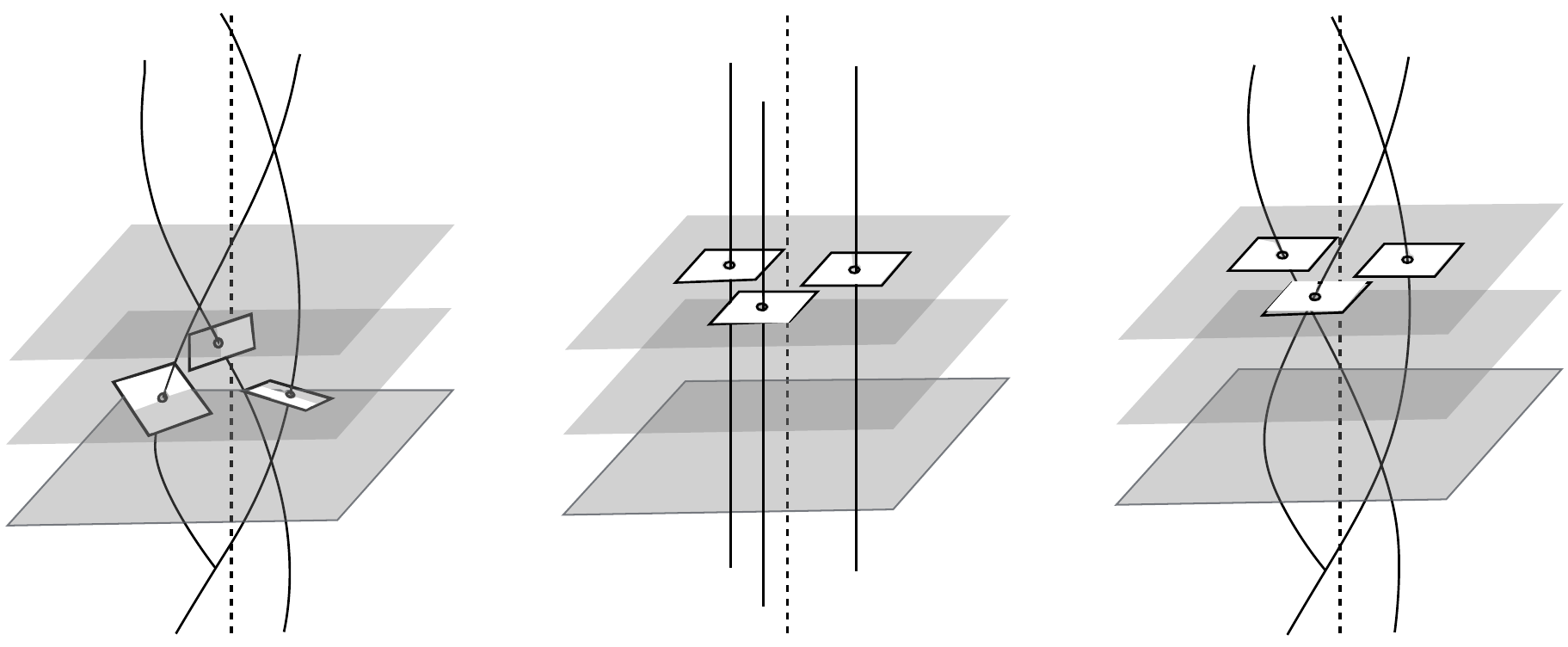}\\[-\baselineskip]
\vspace*{4mm}
\hspace*{\fill}(a)\hspace*{14mm}\hspace*{\fill}(b)\hspace*{14mm}\hspace*{\fill}(c)\hspace*{\fill}
\caption{World-lines of a (a) rotating observer, with a regular stationary splitting; (b) observer at rest, with a standard stationary splitting; (c) rotating observer, with a nonregular natural stationary splitting. Solid lines are world-lines (fibers), the dashed line is the axis of rotation, tiles depict local space platforms (Ehresmann connection), the translucent plane with border depicts a time synchronization (section), and the set of translucent planes depicts the simultaneity structure (foliation).}\label{fig:rotating}
\end{figure}%
Ehrenfests's observer at rest is equipped with a standard stationary splitting, see Fig.~\ref{fig:rotating}~(b). We introduce a rotating observer, co-moving with the cylinder. The splitting associated with the rotating observer is based on a congruence of helical world-lines. A regular splitting on helical world-lines is not integrable and a global Einstein synchronization cannot exist;\footnote{The synchronization defect is related to the Sagnac effect \cite{Sagnac1914,Minguzzi2003}, see also the enlightening explanations about synchronization on a rotating disk in \cite{Matolcsi1998}.} see \ref{implic}. In the absence of accelerated rotation, there exist regular splittings that are stationary, and, hence, principal, see Fig.~\ref{fig:rotating}~(a). The splittings associated with the observer at rest and with the rotating observer share their simultaneity structure. The subsequent analysis is based on this setting. The regular splitting associated with the rotating observer is not metric since $\n{D}\,\lapse\neq 0$. The lapse function is the acceleration potential of the centrifugal acceleration, $\tilde{\df{\delta}}=c_0\n{D}\lapse$; see \ref{implic}. The curvature form follows from $2\,\df{\eta}=c_0\lapse\,\df{\Omega}$; compare \ref{kine}.  Coordinate expressions for various quantities related to the rotating observer can be found in \ref{arotating}. 
\ep
\bph{Resolution in a regular splitting}
As shown in \cite[Eq.~(32)]{Rizzi2002}, the observer metric $\mathbf{h}=-\Sigma^*\mathbf{g}$ in the regular splitting on helical world-lines is not flat but hyperbolic; $\mathbf{g}$ is the Minkowski metric. The observer metric depends on the frequency of revolution, approaching a flat Euclidean metric as the frequency of revolution tends to zero. A cylinder can, therefore, not be Born-rigid in a state of accelerated rotation, $\df{\lambda}=\frac{1}{2}\partial_\tau\n{h}\neq 0$. The very premise that led to the apparent contradiction above is invalidated: The cylinder is Born-rigid and flat at rest and Born-rigid and curved at constant frequency of revolution; to go from one state to the other, it has to undergo a nonstationary and non-Born-rigid phase of accelerated rotation. Note that the curvature of the relative space is in no way due to a curvature of space-time, for the discussion is based on a Minkowski space-time.
\ep
\ifdefined\LONGVERSION
\bph{Pre-metric and metric points of view}Assume now that the rotating cylinder is contained in a solid block at rest with a cylindrical hole just large enough to house the cylinder. Again omitting practical questions of elasticity, that is, using only kinematic arguments, we assume the block's inner wall to be infinitely close to the cylinder's outer walls. As the cylinder starts to rotate in the hole, one could naively assume that, due to Lorentz contraction, the outer wall of the cylinder would have to shrink away from the inner wall of the block. {\st3fan From which observer's point of view?}\footnote{Of course, as in the original Ehrenfest paradox, this would be in contradiction to the constancy of the radius as measured by a rotating observer from the cylinder axis outwards.} To model the situation, let the world-lines be given by the evolution of material points, so that there is a discontinuity between the rotating cylinder and the block at rest. {\st3fan Can this be a fiber bundle? In \cite[9.1]{kolar} $\pi$ is required to be smooth, which is not the case here.}\footnote{For the sake of the argument, we omit the technical discussion of discontinuities in the splitting structure.} Note that in a pre-metric setting an opening of a gap is impossible by definition. Lorentz-contraction is not a pre-metric concept. The world-lines are densely packed in the chosen topology of space-time.\footnote{We may employ the topology induced by the Minkowski metric.} Adding the metric structure, we find that the regular observer metric has a jump discontinuity on the cylinder wall. The measures of circumference do not agree inside and outside of the cylinder. As the relative space's topology is inherited from the principal bundle, rather than being induced by the observer metric, we do not end up in any conceptual difficulty. {\st3fan I am not sure if the congruence of world-lines has the structure of differentiable manifold in this case.}
\ep
\fi
\bph{Discussion}
As pointed out in \cite{Rizzi2002} and more recently in \cite{Kassner2012}, in the resolution of the paradox it is important to employ a congruence of world-lines (the principal bundle), an orthogonal decomposition of space-time (the orthogonal Ehresmann connection), and a model of relative space that is a quotient space $P/\!\!\sim$ under the equivalence relation that identifies all points in a world-line (the base space of the principal bundle). Moreover it is important to realize that a time synchronization is an element of convention and that a global Einstein synchronization is not an option. Defining the observer metric as the pullback metric under the horizontal lift, the above observations are evident and not contradictory. In a framework where topology, metric, time synchronization, and relative space are all amalgamated into a coordinate chart, such deliberations are harder to come by.
\ep
%
%%%%%%%%%%%%%%%%%%%%%%%%%%%%%%%%%%%%%%%%%%%%%%
\section{Schiff's ``Question in General Relativity''}\label{sec:schiff}\labcount{schiff}
\bph{Original reference}
In his 1939 paper \cite{Schiff1939} Leonard Schiff writes:
\begin{quote}
Consider two concentric spheres with equal and opposite total charges uniformly distributed over their surfaces. When the spheres are at rest, the electric and magnetic fields outside the spheres vanish. [...]
Then an observer traveling in a circular orbit around the spheres should find no field, for since all of the components of the electromagnetic field tensor vanish in one coordinate system, they must vanish in all coordinate systems. On the other hand, the spheres are rotating with respect to this observer, and so he should experience a magnetic field. [...] It is clear in the above arrangement that an observer~A at rest with respect to the spheres does not obtain the same results from physical experiments as an observer~B who is rotating about the spheres. [...] We know experimentally that the fields outside the charged spheres vanish in system~A [...] and so the covariance of Maxwell's equations guarantees that the fields will also vanish outside the spheres in system~B. It is of interest, however, to see by direct calculation how it is that the spheres, which are rotating with respect to system~B, do not give rise to a magnetic field outside. [...] To see this, we must of course know the expression for the metric in system~B, and we shall obtain this by transformation from the (approximately) Galilean metric of system~A [in which the spheres are at rest].  
\end{quote}
\ep
\bph{Schiff's treatment}\label{schifftreatment}
Retracing Schiff's steps in our framework, we find that he employs the fibre bundle and section as defined in \ref{kinsetting}, but the chart-associated Ehresmann connection (see \ref{canonconnec}) rather than the orthogonal connection. This yields the natural, nonregular, stationary splitting depicted in Fig.~\ref{fig:rotating}~(c). To see this, first notice that Schiff transforms the Minkowski metric tensor from system~A (related to the observer at rest), into the metric tensor of system~B (related to the rotating observer), by means of a coordinate transformation.
\begin{quote}
The metric tensor of observer~B assumes the form:
\begin{alignat*}{4}
g_{11}&=g_{22}&=g_{33}&=-1,&\quad&&g_{44}&=1-\omega^2(x^2+y^2),\\
g_{14}&=g_{41}&=\omega y,&&&&g_{24}&=g_{42}=-\omega x,
\end{alignat*}
and all other $g_{\mu\nu}$ vanish. 
\end{quote}
This is not the regular splitting of the space-time Riesz operator of \ref{gdecomp}; it corresponds to the nonregular splitting given in \ref{metopnonreg}. The connection-induced metric in the base manifold is Euclidean. Similarly, Schiff obtains the field quantities in system~B by coordinate transformation. Again, this corresponds to a natural splitting, where Maxwell's equations take their usual form \ref{premetricEmagDiscussion}.
\begin{quote}
Since the determinant $g= -1$, the electromagnetic field equations [...] are unaltered by the transformation. But the connections between the covariant and contravariant components of the field tensor depend on the metrical tensor.
\end{quote}
The rotation enters the field problem of system~B in the constitutive relation.
\begin{quote}
It is of interest to note that the vanishing of the fields [in system~B] in this calculation is due to the cancellation of the actual current $J$ with other terms [...] that behave in this respect like a current.
\end{quote}
To reproduce Schiff's results, we split the constitutive relations $H=Z_0^{-1}*_4F$ and $F=-Z_0*_4H$, and obtain from \ref{metopnonreg}
\begin{alignat*}{5}
d&=Z_0^{-1}&&\lapse^{-\dagger}*_\Sigma(\tilde{e}\,+\,&&\cont{\shift}b)&\;\stackrel{\n{def}}{=}\;&
Z_0^{-1}&&\lapse^{-\dagger}*_\Sigma\tilde{e}+ p_\n{\scriptscriptstyle S},\\
b&=Z_0&&\!\lapse^{-\dagger}*_\Sigma(\tilde{h}\,-\,&&\cont{\shift}d)&\;\stackrel{\n{def}}{=}\;&Z_0&&\!\lapse^{-\dagger}*_\Sigma(\tilde{h}+\tilde{m}_\n{\scriptscriptstyle S}).
\end{alignat*}
% Proof can be found 27.10.13, p.1
Introducing $({d}^*,{h}^*)$ 
\begin{alignat*}{3}
{d}^*&=Z_0^{-1}\lapse^{-\dagger}*_\Sigma&&\tilde{e}&\;=\;&d-p_\n{\scriptscriptstyle S},\\
{h}^*&=Z_0^{-1}\lapse^\dagger*_\Sigma &&b&\;=\;&\tilde{h}+\tilde{m}_\n{\scriptscriptstyle S},
\end{alignat*}
we find
\begin{alignat*}{2}
\n{d}\,{h}^*&=\tilde{\jmath}+\tilde{\jmath}_\n{\scriptscriptstyle S}+\groupderiv\bar{d},&\quad\quad\quad \tilde{\jmath}_\n{\scriptscriptstyle S}&=\n{d}\,\tilde{m}_\n{\scriptscriptstyle S}+\groupderiv p_\n{\scriptscriptstyle S},\\
\n{d}\,{d}^*&=\rho+\rho_\n{\scriptscriptstyle S},&\quad\quad\quad\rho_\n{\scriptscriptstyle S}&=-\n{d}\,p_\n{\scriptscriptstyle S}.
\end{alignat*}
$(\rho_\n{\scriptscriptstyle S},\tilde{\jmath}_\n{\scriptscriptstyle S})$ are widely known as {\em Schiff charges and currents}.
We have shown in \cite[5.3]{Kurz2009a} that in the system~B there holds that $\tilde{\jmath}=-\tilde{\jmath}_\n{\scriptscriptstyle S}$ and $\rho_\n{\scriptscriptstyle S}=0$, in agreement with Schiff's paper. A classical treatment is found, for example, in \cite[11.2]{bladelbook}.\par
\ep
\bph{Discussion of Schiff's treatment}
In the natural nonregular splitting, vacuum generally has nonzero polarization $p_\n{\scriptscriptstyle S}(b,\lapse^\dagger,\shift)$ and magnetization $\tilde{m}_\n{\scriptscriptstyle S}(d,\lapse^\dagger,\shift)$. This is, of course, not a statement on the vacuum constitutive relation, but a consequence of the nonregular splitting that yields fields that are not measurable, and relations between the fields that are not physical laws; as {\sc T. Mo} writes in \cite[p.~2596]{Mo1970}: {\em Mistaken results, such as $D$ depends constitutively on $B$ even for vacuum, will occur if one does not treat the constitutive physics and the constitutive tensor properly}; and by ``properly'' we understand ``in a regular splitting''. The question we might want to raise is, therefore, whether the above reasoning does indeed represent a valid resolution of the paradox. What is demonstrated by means of Schiff charges and currents is that the field problem that is formulated in a natural nonregular splitting gives consistent results. Schiff's system~B does not make statements on measurable fields seen by observer~B. It would, consequently, appear that a proper resolution of the paradox should be derived in a regular splitting that involves the standard constitutive relations of vacuum, and Maxwell's equations that feature terms due to the rotation. 
\ep
\bph{Static equations in the regular splitting}
We return to the regular splitting structure, according to \ref{kinsetting} and Fig.~\ref{fig:rotating}~(a). The static Maxwell's equations in the regular splitting are given by
\begin{alignat*}{2}
\n{d}\,{b} &= \df{\Omega}\wedge\tilde{e},&\quad\n{d}\,d&=\rho-\df{\Omega}\wedge \tilde{h},\\
\n{d}\,\tilde{e} &=0,&\quad\n{d}\,\tilde{h}&= \tilde{\jmath};
\end{alignat*}
compare \ref{max} and \ref{premetricEmagDiscussion}. The corresponding constitutive relations \ref{const} read
\[
d=Z_0^{-1} \lapse^{-1}*_3 \tilde{e},\quad b=Z_0\lapse^{-1}*_3\tilde{h},
\]
with the non-Euclidean Hodge operator $*_3$. The electromagnetic fields in these equations are observable quantities of observer~B.
\ep
\bph{Axial splitting}\label{axialsplit}
The rotating observer, the spheres, and the electromagnetic field exhibit axisymmetry with respect to the rotation axis. The above fields and equations can be split further in order to benefit from this symmetry and make the problem more easily accessible analytically. The axial splitting structure for dimensional reduction consists of
\begin{enumerate}
\item The principal $U(1)$-bundle $\bigl(X,\mathring{\pi},Y,U(1)\bigr)$\footnote{On the axis of rotation, the group action of $U(1)$ is not free. Therefore, to be precise, the bundle space is $X$ without the axis of rotation.}, the fibers of which are circles around the axis. $Y$ is a two-dimensional open half-space, $\mathring{\pi}$ is the projection map. $U(1)$ or short $U$ is the unitary group with Lie algebra $\mathfrak{u}$.
\item The $\mathbf{h}$-orthogonal Ehresmann connection.
\item A global section that maps $Y$ to an (arbitrary) meridian plane.
\end{enumerate}
We use the notation of the previous chapters and the symbol $\mathring{}$ to indicate that objects are defined with respect to the axial splitting. The azimuthal fundamental field $\mathring{\vec{w}}\in\ves{1}(X,\mathfrak{u}^*)$ is a Killing field of the observer metric. We orient the Lie algebra $\mathfrak{u}$ in such a way that the fundamental field points in the direction of rotation.\footnote{Fixed orientation eliminates twist with respect to $U$. This is the same policy as in \ref{timeorient}, where twist with respect to $\group$ was eliminated.} The splitting structure is natural and regular. It is also stationary, that is, $\partial_U\mathbf{h}=0$. Axisymmetry is characterized by $\partial_U=0$ for fields, metric, and material parameters.  Axial splitting yields two decoupled systems of fields and equations \cite{Auchmann2010,Raumonen2011}. Since the currents are in azimuthal direction, $\cont{\mathring{\vec{w}}}^\otimes\tilde{\jmath}=0$, one of the systems is trivial. Taking this into account, the splitting map yields 
\begin{alignat*}{8}
&\mathring{\n{S}}^{-*}\tilde{e} &&= (\bar{e},0),\quad&&\bar{e}&&\in\dfs{1}(Y;\mathfrak{g}^*),\quad\quad
&&\mathring{\n{S}}^{-*}{b} &&= (0,\bar{b}),\quad&&\bar{b}&&\in\dfs{1}(Y;\mathfrak{u}^*),\\
&\mathring{\n{S}}^{-*}\tilde{h} &&= (\bar{h},0),\quad&&\bar{h}&&\in\mathcal{F}^1_\twist(Y;\mathfrak{g}^*),\quad\quad
&&\mathring{\n{S}}^{-*}{d} &&= (0,\bar{d}),\quad&&\bar{d}&&\in\mathcal{F}^1_\twist(Y;\mathfrak{u}^*),\\
&\mathring{\n{S}}^{-*}\tilde{\jmath} &&= (\bar{\jmath},0),\quad&&\bar{\jmath}&&\in\mathcal{F}^2_\twist(Y;\mathfrak{g}^*),\quad\quad
&&\mathring{\n{S}}^{-*}{\rho} &&= (0,\bar{\rho}),\quad&&\bar{\rho}&&\in\mathcal{F}^2_\twist(Y;\mathfrak{u}^*).
\end{alignat*}
The curvature form and lapse function of the relativistic splitting are split axially into
\begin{alignat*}{3}
\mathring{\n{S}}^{-*}\df{\Omega} &\,=\,& (0,\bar{\df{\Omega}}),&\quad\quad\bar{\df{\Omega}}&\,\in\;&\dfs{1}(Y;\mathfrak{u}^*\otimes\mathfrak{g}),\\
\mathring{\n{S}}^{-*}\lapse &\,=\,& (\bar{\lapse},0),&\quad\quad\bar{\lapse}&\,\in\;& C^\infty(Y;\mathfrak{g}^*).
\end{alignat*}
\ep
\bph{Dimensionally reduced equations}\label{dimredmax}
Axial splitting yields static Max\-well's equations in $Y$,
\begin{alignat*}{4}
-\n{d}\,\bar{b} &= \bar{\df{\Omega}}\wedge\bar{e},&\quad-\n{d}\,\bar{d}&=\bar{\rho}-\bar{\df{\Omega}}\wedge \bar{h},\\
\n{d}\,\bar{e} &=0,&\quad\n{d}\,\bar{h}&= \bar{\jmath}.
\end{alignat*}
In analogy to the lapse function in \ref{lapsefun}, we introduce 
\[\mathring{\lapse}=\mathring{\Phi}^*|\mathring{\vec{w}}|.\] 
The pullback metric, $\bar{\mathbf{h}}=\mathring{\Sigma}^*\mathbf{h}$, is Euclidean, with induced Hodge operator $*_2$.\footnote{The non-Euclidean nature of the 3-dimensional metric is relegated to $\mathring{\lapse}$.} The axial splitting of the Hodge operator yields  the constitutive relations
\[
\bar{d}=Z_0^{-1}(\mathring{\lapse}\bar{\lapse}^{-1})^\otimes*_2 \bar{e},\quad\bar{b}=Z_0(\mathring{\lapse}\bar{\lapse}^{-1})^\otimes*_2\bar{h}.
\]
Denote by $\omega$ the angular frequency of the rotating observer, $\pd(\omega)=\pdim{T}^{-1}$. Define 
$\beta=\omega r/c_0,$
where $r$ is the radial distance from the rotation axis measured by $\bar{\mathbf{h}}$. The Lorentz factor 
$\gamma=(1-\beta^2)^{-1/2}$
ensues. It is convenient to rewrite the constitutive relations in the form
\[
\bar{d}=Z_0^{-1}\gamma^2\Lambda*_2\bar{e},\quad\bar{b}=Z_0\gamma^2\Lambda*_2\bar{h},
\]
where 
\[\Lambda=\gamma^{-2}(\mathring{\lapse}\bar{\lapse}^{-1})^\otimes.\] 
It holds that $\Lambda\sim r$, independent of $\omega$, see Appendix~\ref{arotating}.
\begin{center}
\begin{tabular}{|c|c|c|}
\hline
&$\in$&$\pd(\,\cdot\,)$\\
\hline
$r$&$ C^\infty(Y)$&$\pdim{L}$\\
$\beta$&$ C^\infty(Y)$&$\oned$\\
$\gamma$&$ C^\infty(Y)$&$\oned$\\\hline
$\mathring{\lapse}$&$C^\infty(Y;\mathfrak{u}^*)$&$\pdim{L}$\\
$\Lambda$&$C^\infty(Y;\mathfrak{u}^*\otimes\mathfrak{g})$&$\oned$\\
\hline
\end{tabular}
\end{center}
\ep
\bph{Solving Schiff's paradox}\label{schiffsolv}
The excitation in the regular splitting associated with the rotating observer can be written as
\[
(\bar{\rho},\bar{\jmath})=(\gamma^2,-\beta\Lambda^{-1})\bar{\rho}_0,
\]
where $\bar{\rho}_0$ represents the limit for $\omega \to 0$ of the charge distribution (see \ref{aschiffsolv}), that is, equal but opposite total charges, uniformly distributed over the respective spheres. We define the problem domain as a sphere of radius $R$, fulfilling $\omega R < c_0$. The equations of \ref{dimredmax}, with the above sources and with homogeneous Dirichlet boundary conditions on the domain boundary, constitute a boundary-value problem, that can be solved by a perturbation method in terms of the paramter $\hat{\beta}=\frac{\omega R}{c_0}$. There holds $\beta \leq \hat{\beta}$ in the problem domain. The zeroth order electric field has $\beta=0$ and $\gamma=1$, so that
\begin{align*}
\n{d}\,\bar{e}_0 &=0,\\
-\n{d}\,\bar{d}_0 &=\bar{\rho}_0,\\
\bar{d}_0&=Z_0^{-1}\Lambda*_2\bar{e}_0.
\intertext{This is an axisymmetric electrostatic problem in the meridian plane, with Euclidean vacuum constitutive relation. Its solution is a radial electric field $(\bar{e}_0,\bar{d}_0)$, confined to the interior of the spheres. From this we obtain the first order magnetic field, with $|\beta|\ll 1$ and $\gamma=1$, as the solution of}
-\n{d}\,\bar{b}_1 &= \bar{\df{\Omega}}_1\wedge\bar{e}_0,\\
\n{d}\,\bar{h}_1&=-\beta\Lambda^{-1}\bar{\rho}_0,\\
\bar{b}_1&=Z_0\Lambda*_2\bar{h}_1.
\intertext{It holds that $\bar{\df{\Omega}}=\n{d}(\beta\gamma^2\Lambda)$, $\bar{\df{\Omega}}_1=\n{d}(\beta\Lambda)$, and $\n{d}(\beta\Lambda^{-1})=0$; see \ref{arotating}. It can be easily checked that}
(\bar{b}_1,\bar{h}_1)&=\beta(-\Lambda\bar{e}_0,\Lambda^{-1}\bar{d}_0).
\intertext{Since the coupling term $\bar{\df{\Omega}}_1\wedge\bar{h}_1$ is of second order, the first order electric field is given by}
(\bar{e}_1,\bar{d}_1)&=(\bar{e}_0,\bar{d}_0).
\intertext{It can be verified that the exact solution for $|\beta|<1$ is given by }
(\bar{e},\bar{d})&=(\bar{e}_1,\gamma^2\bar{d}_1),\\
(\bar{b},\bar{h})&=(\gamma^2\bar{b}_1,\bar{h}_1). \end{align*}
Outside the spheres, all electromagnetic fields vanish. The solution of the paradox is essentially found in the first-order step, undergoing only a scaling in the exact-solution step.
\ep
\bph{Discussion}
We have, citing from \cite{Schiff1939} again, shown {\em ``by direct calculation how it is that the spheres, which are rotating with respect to observer~B, do not give rise to a magnetic field outside.''} In the demonstration we have used a regular splitting, ensuring that fields and equations are those that are experienced by the observer~B. This is in contrast to Schiff's treatment in the coordinates of system~B, which represents a nonregular splitting with fields and laws that are mainly of mathematical interest. We, therefore, find that a conceptual disentanglement of the mathematical structures and the coordinate systems used to represent them is conducive to enhanced insight.
\ep

%
%%%%%%%%%%%%%%%%%%%%%%%%%%%%%%%%%%%%%%%%%%%%%%
%%%%%%%%%%%%%%%%%%%%%%%%%%%%%%%%%%%%%%%%%%%%%%
\newchapter{Conclusion}\textit{Nature uses as little as possible of anything (J. Kepler).}\par
$\!$\\
We have introduced the relativistic splitting structure in a geometric, that is, frame- and coordinate-free way. It allows to relate four-dimensional fields and equations of electromagnetism to their three-dimensional manifestations in observer's space. We aimed for a minimal set of mathematical structures that are directly motivated by the physical theory: space-time, world-lines, time translation, space platforms, and time synchronization are all modeled by distinct structural elements. A concise and insightful classification of splitting structures ensues that can be studied in its relation to a classification of observers. The application of the framework to the Ehrenfest paradox and Schiff's ``Question in General Relativity'' illustrates the advantages of the framework: the resolution of the Ehrenfest paradox becomes straight-forward; and we can discuss Schiff's paper in a new light, detect limitations in the original treatment, and offer a more complete resolution of the paradox. The increased conceptual clarity has to be paid for with the necessity to familiarize with the relevant mathematical structures. We believe, however, that, in the absence of at least a reliable mental image of the implied concepts, a coordinate-based approach is bound to lead into the same pit-falls that have been encountered many times over. In our own experience, the study of relativistic splittings lends itself to render those powerful mathematical structures more accessible. As we were writing this article, we aimed to ensure the maturity of the presented content by proving that all concepts can be generalized in a straight-forward way to manifolds of arbitrary dimensions and to arbitrary Lie groups. As a consequence, the presented model is but an instance of a wider framework. We are confident that the splitting structure can be of use in physical theories other than electromagnetism, and in contexts other than relativity; for an example, we mention the axial splitting in the preceding section.
\chapter*{Acknowledgements}
The authors wish to express their gratitude to Alberto Favaro for his critical reading of the document and helpful comments. Bernd Flemisch was a part of this research several years back when we used earlier versions of the splitting for the dimensional reduction of relativistic electrodynamics and 3-D boundary-value problems with symmetries. We are grateful to Professors Friedrich W. Hehl, Lauri Kettunen, Michael Kiessling, and Peter W. Michor for their encouragement and suggestions. This work was partially funded by Tekes, the Finnish Funding Agency for Technology and Innovation, FiDiPro project 1481/31/09, Advanced Electromagnetic Modeling and Simulation for Engineering.

\fi
\ifdefined\LONGVERSION
\part{Multiple Observers}
\part{Constitutive Relations of Simple Materials}
\part{Jump Conditions}
\fi
%
%%%%%%%%%%%%%%%%%%%%%%%%%%%%%%%%%%%%%%%%%%%%%%
\ifdefined\INCLUDEMETRIC
\else\label{sec:metric}%\label{classificationation}
\fi %\INCLUDEMETRIC
%%%%%%%%%%%%%%%%%%%%%%%%%%%%%%%%%%%%%%%%%%%%%%
%%%%%%%%%%%%%%%%%%%%%%%%%%%%%%%%%%%%%%%%%%%%%%
%\part*{}
\appendix
\renewcommand{\thechapter}{\Alph{chapter}}
\renewcommand{\thesection}{\thechapter\arabic{section}}
\newcommand{\refcount}[1]{\setcounter{para}{0}\setcounter{section}{\value{#1}}}
%%%%%%%%%%%%%%%%%%%%%%%%%%%%%%%%%%%%%%%%%%%%%
\chapter{Some Proofs and Details}
\ifdefined\LONGVERSION
\refcount{notation}
\bp
We show that the adjoint representation of a Lie group on its Lie algebra is a well defined operation $\n{Ad}:\group\to L(\mathfrak{g},\mathfrak{g})$ not only via its operation on the tangent space of the identity element, $\n{Ad}:\group\to L(T_e\group,T_e\group)$ (see \cite[p.~166]{amp}, \cite[4.24]{kolar}), but also when it operates in every point of $\group$ on left-invariant vector fields. To this end we have $\n{Ad}(h^{-1})=L_{h^{-1}}'\circ R_h'$ act on an element $\vec{v}\in\mathfrak{g}$ of the Lie algebra in point $g$. Since $\vec{v}_g=L_g'\vec{v}_e$ we find
\begin{align*}
\n{Ad}(h^{-1})_g\circ(L_g')_e&=(L_{h^{-1}}')_{gh}\circ(R_h')_g\circ(L_g')_e\\
&=(L_{h^{-1}}')_{gh}\circ(L_g')_h\circ(R_h')_e\\
&=(L_{h^{-1}}')_{gh}\circ(L_g')_h\circ(L_h')_e\circ(L_{h^{-1}}')_h\circ(R_h')_e\\
&=(L_{h^{-1}gh}')_e\circ\n{Ad}(h^{-1})_e.
\end{align*}
It follows that both definitions are equivalent.
\ep
\refcount{lie}
\bph{Principal right action and the fundamental field map}\label{afundam}
Referring to \ref{canonisomorphism}, we show that $r_h'\circ\zeta = \zeta\circ \n{Ad}(h^{-1})$ for $\varphi(p)=(x,g)$:
\begin{align*}
(r_h')_p\circ\zeta 
&= (\hhat{\varphi}_x^{-1})'_{gh}\circ (R_h')_g\circ(\hhat{\varphi}_x)'_p\circ (\hhat{\varphi}_x^{-1})'_g\circ (L_g')_e\\
&= (\hhat{\varphi}_x^{-1})'_{gh}\circ (R_h')_g\circ (L_g')_e\\
&= (\hhat{\varphi}_x^{-1})'_{gh}\circ (L_g')_h\circ (R_h')_e\\
&= (\hhat{\varphi}_x^{-1})'_{gh}\circ (L_g')_h\circ(L_h')_e\circ(L_{h^{-1}}')_h\circ (R_h')_e\\
&= (\hhat{\varphi}_x^{-1})'_{gh}\circ (L_{gh}')_e\circ \bigl(\n{Ad}(h^{-1})\bigr)_e\\
&=\zeta\circ \bigl(\n{Ad}(h^{-1})\bigr)_e.
\end{align*}
\ep
\refcount{parametric}

\refcount{splitting}
\bph{Algebra of the unit tensor}\label{unittens}
Let $(V, V^*)$ denote a pair of $q$-dimensio\-nal dual vector spaces. The mixed exterior algebra $\Wedge(V,V^*)=\Wedge V\otimes\Wedge V^*$ is equipped with exterior and interior multiplication via
\begin{align*}\lmult{\vec{m}\otimes\df{\mu}}\,\vec{n}\otimes\df{\nu}&=\mult{\vec{m}}\vec{n}\otimes\mult{\df{\mu}}\df{\nu}\\
\lcont{\vec{m}\otimes\df{\mu}}\,\vec{n}\otimes\df{\nu}&=\cont{\df{\mu}}\vec{n}\otimes\cont{\vec{m}}\df{\nu}.
\end{align*}
Let $(\vec{e}_i)$ and $(\df{\varepsilon}^i)$, $1\le i\le q$ be their respective dual bases. The unit tensor reads
\begin{align*}w=\sum_{i=1}^q\vec{e}_i\otimes\df{\varepsilon}^i.\end{align*}
The $k$-th exterior power of the unit tensor 
\begin{align*}\Wedge^kw=\frac{1}{k!}\underbrace{w\wedge\dots\wedge w}_{\times k},\end{align*}
is the unit tensor of $\Wedge^k (V,V^*)$, and
\begin{align*}\Wedge w = \bigoplus_{k=0}^q \Wedge^kw,\end{align*}
with $w^0=1\otimes 1$, is the unit tensor of $\Wedge(V,V^*)$. $\Wedge w$ is an element of the diagonal subalgebra $\mathsf{\Delta}(V,V^*)=\bigoplus_{k=0}^q\Wedge^k V\otimes\Wedge^k V^*$; compare with \cite[6.4, 6.7]{greub2}. It follows that
\begin{align*}
w(w)&=q,\\
\Wedge^kw(\Wedge^kw)&={\binom{q}{k}},\\
\Wedge w(\Wedge w)&=\sum_{k=0}^q{\binom{q}{k}}=2^q.
\end{align*}
\ep
\bph{The operator $Q$ and the unit tensor}\label{Qop} 
The operator $Q$ (see \cite[6.3]{greub2}) maps $\Wedge(V,V^*)\to\Wedge(V^*,V):\vec{m}\otimes\df{\mu}\mapsto\df{\mu}\otimes\vec{m}$. 
We define the contraction $\cont{}:\mathsf{\Delta}(V,V^*)\times\mathsf{\Delta}(V^*,V)\to\mathsf{\Delta}(V^*,V)$ such that for 
\begin{align*}\lcont{\vec{m}\otimes\df{\mu}}\,\df{\nu}\otimes\vec{n}&=\bigoplus_{k=0}^q\bigl(\lcont{\vec{m}^k}\df{\nu}^k\bigr){\df{\mu}^k}\otimes \vec{n}^k.\end{align*}
It follows with $Q\,\Wedge w\in\Wedge(V^*,V)$ that
\begin{align*}\lcont{\Wedge w}\,Q\,\Wedge w = Q\,\Wedge w.\end{align*}
Vice versa we define $\cont{}:\mathsf{\Delta}(V^*,V)\times\mathsf{\Delta}(V,V^*)\to\mathsf{\Delta}(V,V^*)$ such that
\begin{align*}\lcont{\df{\mu}\otimes\vec{m}}\,\vec{n}\otimes\df{\nu}&=\bigoplus_{k=0}^q\bigl(\lcont{\df{\mu}^k}\vec{n}^k\bigr){\vec{m}^k}\otimes \df{\nu}^k,\end{align*}
and, therefore,
\begin{align*}\lcont{Q\,\Wedge w}\,\Wedge w = \Wedge w.\end{align*}
\ep
\fi
\refcount{variance}
%\section{Derivatives in the base manifold: Christoffel form, curvature, and variance}
\bph{Principal connection}\label{avariance}
We prove the assertion of \ref{variance}, that is, $\df{\chi}=0$ iff the Ehresmann connection is principal with respect to the given $\group$-bundle. For Abelian Lie groups, a principal connection is characterized by vanishing Lie derivative of the connection 1-form with respect to the fundamental field, $\lie{\vec{w}}\df{\omega}=0$; see \cite[11.1(3)]{kolar}.
We first show that $\df{\chi}=0$ implies $\lie{\vec{w}}\df{\omega}=0$. Consider the chart-associated connection and use \ref{par:verhor},
\begin{alignat*}{2}\vec{v}&=\n{hor\,}\vec{v}&&+\n{ver}\,\vec{v}\\
&=\Phi\,\Pi\,\vec{v}&&+(\vec{w}\otimes\df{\theta})\vec{v},\quad\vec{v}\in\ves{1}(P).
\end{alignat*}
By testing the equation with $\df{\omega}\in\dfs{1}(P;\mathfrak{g})$ and invoking duality we obtain
\begin{align*}
\df{\omega}=(\Phi\,\Pi)^*\df{\omega}+\df{\theta}.
\end{align*}
From the definitions of the variance of the Ehresmann connection, the group derivative and the Christoffel form we have
\begin{align*}
0=\df{\chi}&=\groupderiv\df{\Gamma}=\Sigma^*\lie{\vec{w}}\Pi^*\Phi^*\df{\omega}\\
&=\Sigma^*\lie{\vec{w}}(\df{\omega}-\df{\theta})=\Sigma^*\lie{\vec{w}}\df{\omega},
\end{align*}
since $\lie{\vec{w}}\df{\theta}=\cont{\vec{w}}\n{d}\df{\theta}=0$. This means that $\lie{\vec{w}}\df{\omega}\in\n{Ker\,}\Sigma^*=\mathcal{F}_V^1(P,\mathfrak{g}^*\otimes\mathfrak{g})$. On the other hand, $\lie{\vec{w}}\df{\omega}=\cont{\vec{w}}\n{d}\df{\omega}\in\mathcal{F}_H^1(P,\mathfrak{g}^*\otimes\mathfrak{g})$. Hence $\lie{\vec{w}}\df{\omega}=0$, which proves the first part of the assertion.
The converse direction of the proof is obvious.
\ep
%
%\refcount{frobenius}
\bph{Frobenius integrability condition}\label{afrobenius}
To show that $\df{\Omega}=0$ implies that $(\n{d}\,\df{\omega}\wedge\df{\omega})^\otimes=0$ and vice versa (see \ref{frobenius}), we proceed along the same lines as in \ref{avariance}:
% Internal: The $\binom{2}{0}$-tensor vanishes iff it vanishes on a basis. Since the Lie algebra is 1-dimensional, a basis is given by $\df{\varepsilon}$. Therefore, it suffices to consider the expression $\n{d}\,\df{\omega}\wedge\df{\omega}(\df{\varepsilon})$.
\begin{align*}
0=\df{\Omega}&=\n{D}\df{\Gamma}=\Sigma^*\n{d}\,\Pi^*\Phi^*\df{\omega}\\
&=\Sigma^*\n{d}(\df{\omega}-\df{\theta})=\Sigma^*\n{d}\df{\omega}\\
&\Leftrightarrow\n{d}\df{\omega}\in\n{Ker\,}\Sigma^*=\mathcal{F}_V^2(P,\mathfrak{g})\\
&\Leftrightarrow\n{d}\df{\omega}=\n{ver}^*\n{d}\df{\omega}=\mult{\df{\omega}(\df{\varepsilon})}\cont{\vec{w}(\vec{e})}\n{d}\df{\omega},
\end{align*}
where we used \ref{asplit} for the algebraic representation of the map $\n{ver}^*$. Note that its natural extension to vector-valued arguments according to \ref{extend} demands for introduction of a pair of dual bases $(\vec{e},\df{\varepsilon})\in\mathfrak{g}\times\mathfrak{g}^*$. Lastly, we find the implication by applying $\mult{\df{\omega}(\df{\varepsilon})}$ from the left.
Conversely, we infer from $(\n{d}\,\df{\omega}\wedge\df{\omega})^\otimes=0$ that
\begin{align*}
0=\cont{\vec{w}(\vec{e})}\mult{\df{\omega}(\df{\varepsilon})}\n{d}\df{\omega}&=\n{hor}^*\n{d}\df{\omega}\\
&\Leftrightarrow\n{d}\df{\omega}\in\mathcal{F}_V^2(P,\mathfrak{g}),
\end{align*}
which yields the assertion.
\ep
\refcount{splitop}
%\section{The relativistic splitting structure}
\bph{Space-time splitting of operators}\label{splitop} % Proof can be found 17.03.13, pp.1-2
The splitting of fields readily leads to a splitting of operators. The splitting of the contraction reads for $\n{S}\,\vec{v}=(\vec{\atest},\tilde{\df{\btest}})$, $\vec{v}\in\ves{k}(P)$, 
\begin{align*}\n{S}^{-*}\circ\cont{\vec{v}}\circ\n{S}^*
=\begin{pmatrix}\cont{\vec{\atest}}\;&\cont{\tilde{\df{\btest}}}\\0\;&\cont{\signop(\vec{\atest})}\end{pmatrix}.\end{align*}
The splitting of the exterior product reads for $\n{S}^{-*}\,\df{\gamma}=(\df{\alpha},\tilde{\df{\beta}})$, $\df{\gamma}\in\dfs{k}(P)$, 
\begin{align*}\n{S}^{-*}\circ\mult{\df{\gamma}}\circ\n{S}^*
=\begin{pmatrix}\mult{\df{\alpha}}\;&0\\\mult{\tilde{\df{\beta}}}\;&\mult{\signop(\df{\alpha})}\end{pmatrix}.\end{align*}
The splitting of the Lie derivative reads for $\n{S}\,\vec{v}=(\vec{\atest},\tilde{\btest})$, $\vec{v}\in\ves{1}(P)$,
\begin{align*}\n{S}^{-*}\circ\lie{\vec{v}}\circ\n{S}^*
&=\begin{pmatrix}\n{L}_\vec{\atest}+\tilde{\btest}\,\timederiv&\lmult{\n{D}\tilde{\btest}}+\tilde{\btest}\,\mult{\df{\chi}}
+\lmult{\cont{\vec{\atest}}\df{\Omega}}\\
[\timederiv,\cont{\vec{\atest}}]&\n{L}_\vec{\atest}+\timederiv\circ{\cont{\tilde{\btest}}}-{\lmult{\cont{\vec{\atest}}\df{\chi}}}
\end{pmatrix},
\intertext{where} 
\n{L}_\vec{\atest}&=\n{D}\circ\cont{\vec{\atest}}+\cont{\vec{\atest}}\circ\n{D}.\end{align*}
\ep
\bph{Factorization of the splitting of the exterior derivative}
\label{ddecomp}The splitting of the exterior derivative of \ref{splitd} can be factorized as follows,
\begin{align*}
\begin{pmatrix}{{\n{D}}}&\mult{{\df{\Omega}}}\\\timederiv&\;\mult{\df{\chi}}- {\n{D}}\end{pmatrix}
=\begin{pmatrix}\n{Id}&-\mult{\df{\Gamma}}\\0&\n{Id}\end{pmatrix}\circ\begin{pmatrix}\n{d}&0\\\timederiv&-\n{d}\end{pmatrix}\circ\begin{pmatrix}\n{Id}&\mult{\df{\Gamma}}\\0&\n{Id}\end{pmatrix},
\end{align*}
which follows from change-of-connection to a chart-associated connection in \ref{changeconnec}, $\df{\omega}_\alpha=\df{\omega}$, $\df{\omega}_\beta=\df{\theta}$. % More proof can be found 18.03.13
\ep
\ifdefined\INCLUDEMETRIC
\refcount{metric}
%\section{Regular relativistic splittings}
%
\bph{Action, charge, and the physical dimension of the metric tensor}\label{mpost}
In \ref{max}, Footnote~\ref{aqfoot}, we have seen that the electromagnetic fields divide into Faraday fields with physical dimension $\pdim{AQ}^{-1}$, and Amp\`ere-Maxwell fields with physical dimension $\pdim{Q}$. Products between these two classes have the physical dimension of action. E.J.~Post in \cite[Ch.~II, \S 3]{post} states that any physical field $X$ should have a physical dimension that can be expressed in terms of action $\pdim{A}$ and charge $\pdim{Q}$, that is, $\pd(X)=\pdim{A}^p\pdim{Q}^q$. In \cite[Ch.~II, \S 4]{post} he applies this concept to the metric tensor, stating that only $\pd(\mathbf{g})=\oned$ ($p=q=0$) would be an appropriate choice under this premise. With regard to the role of metric and its physical dimension (called ``gauge factor"), E.J.~Post writes in \cite[Ch.~II, \S 4]{post} {\em ``For our present purpose we shall find [...] that the question whether or not a gauge factor for $g_{ab}$ should be used is not too urgent for electromagnetism. Either the gauge factor of the metric or the metric itself cancels out in all quantities and equations of importance.''} In our framework, we confirm that the choice for $\pd(\mathbf{g})$ leaves Maxwell's equations and the constitutive laws unaltered. The splitting map is of neutral dimension, and the lapse function ensures consistency of physical dimensions in the constitutive relations in three dimensions. A change of $\pd(\mathbf{g})$ would only affect Section~\ref{sec:kinec} on kinematic parameters; see Footnote~\ref{pdimkine}. Following \cite[p.~398]{schouten} we stick to the conventional $\pd(\mathbf{g})=\pdim{L}^2$.
\ep
\refcount{slicing}
%\section{Metric in nonregular splittings}
\bph{Nonregular splitting of the Hodge operator}\label{atools} 
We give useful formulas related to the nonregular splitting of the Hodge operator in \ref{metopnonreg} and \ref{ametopthread}:
\begin{alignat*}{2}
\lcont{\df{\gamma}}\circ\n{h}^{-1} &=\n{h}^{-1}\circ \lcont{\n{h}^{-1}\,\df{\gamma}},&\quad\quad
\lmult{\vec{v}}\circ\n{h}^{-1} &=\n{h}^{-1}\circ \lmult{\n{h}\,\vec{v}},\\[1.5mm]
\lcont{\vec{v}}\circ *_3&=*_3\circ\lmult{\n{h}\,\vec{v}}\circ \signop,&\quad\quad
\lmult{\df{\gamma}}\circ *_3&=-*_3\circ\;\lcont{\n{h}^{-1}\,\df{\gamma}}\circ \signop.
\end{alignat*}
% Internal: The second one is a simple consequence of exterior compound, while the first one is derived by dualizing the second one.
\ep
\bph{Nonregular splitting of metric operators in terms of the fiber-induced metric}\label{ametopthread} 
We leverage the relation $\mathbf{h}_\Pi=\mathbf{h}_{\Sigma^\dagger}$ in \ref{twometrics}. Denote the nonregular splitting under consideration by $\beta$. Replace its connection 1-form $\df{\omega}$ by $\df{\omega}^\dagger$. This yields a regular splitting $\alpha$, whose observer metric is $\mathbf{h}_\Pi$, by construction. Both splittings are related by a change of Ehresmann connection, with $\df{\Gamma}_\alpha-\df{\Gamma}_\beta=-\shiftform$; compare with \ref{changeconnec},
% Internal: Proof can be found 07.08.13, pp.1-4 and 20.08.13, p.5. This also demonstrates that the transformations are independent from the choice of fiber chart.
\[
\n{S}_\beta^{-*}\circ\n{S}_\alpha^*=\begin{pmatrix}\n{Id}&-\mult{\shiftform}\\0&\n{Id}\end{pmatrix}.
\]
The splitting $\n{S}_\beta^{-*}\circ\n{g}\circ\n{S}_\beta^{-1}$ follows from the regular splitting $\n{S}_\alpha^{-*}\circ\n{g}\circ\n{S}_\alpha^{-1}$ in \ref{gdecomp}. %The decomposition of the space-time metric tensor in terms of $\mathbf{h}_\Pi$ and $\lapse$ ensues,
%%%with $(\vec{\atest},\tilde{\btest})=\n{S}\vec{v}$, $(\df{\alpha},\tilde{\beta})=\n{S}^{-*}\df{\gamma}$,
%%%\begin{alignat*}{5}
%%%\mathbf{g}(\vec{v},\vec{v}^\prime)\Big|_p&=-\mathbf{h}_\Pi(\vec{\atest},\vec{\atest}^\prime)+
%%%(\lapse\lapse)^\otimes(\tilde{\btest}-\cont{\shiftform}\vec{\atest},\tilde{\btest}^\prime-\cont{\shiftform}\vec{\atest}^\prime)&&\Big|_{\varphi(p)},\\
%%%\mathbf{g}^{-1}(\df{\gamma},\df{\gamma}^\prime)\Big|_p&=-\mathbf{h}_\Pi^{-1}(\df{\alpha}+\tilde{\beta}\shiftform,\df{\alpha}^\prime+\tilde{\beta}^\prime\shiftform)+
%%%(\lapse^{-1}\lapse^{-1})^\otimes(\tilde{\beta},\tilde{\beta}^\prime)&&\Big|_{\varphi(p)};
%%\end{alignat*}
%\begin{align*}
%\n{S}^{-*}\mathbf{g}&=\begin{pmatrix}1&0\\-\cont{\shiftform}&1\end{pmatrix}^*\bigl(-\mathbf{h}_\Pi,(\lapse\lapse)^\otimes\bigr),\\
%\n{S}^{**}\mathbf{g}&=\begin{pmatrix}1&\mult{\shiftform}\\0&1\end{pmatrix}^*\bigl(-\mathbf{h}_\Pi^{-1},(\lapse^{-1}\lapse^{-1})^\otimes\bigr);
%\end{align*}%
%compare with \cite[Eq.~(22)]{Fodor1994}. 
Denote $\df{\kappa}_\Pi$ and $*_\Pi$ the twisted unit spatial volume form and the Hodge operator induced by the metric tensor $\mathbf{h}_\Pi$, respectively. The splittings of the Riesz\footnote{Compare with \cite[Eq.~(22)]{Fodor1994}.} and Hodge operator read
%with $\lapseprod=\lapse\lapse^{-\dagger}$
\begin{alignat*}{2}
\n{S}^{-*}\circ\n{g}\circ\n{S}^{-1}&=
\lapseprod^2&&\begin{pmatrix}\lapseprod^{-2}\n{Id}-\mult{\shiftform}\circ\cont{\shift}&-\mult{\shiftform}\\\cont{\shift}&\n{Id}\end{pmatrix}\circ
\begin{pmatrix}1&0\\0&
\,(\lapse^\dagger\lapse^\dagger)^\otimes\end{pmatrix}\n{h}_\Pi\circ\signop,\\[3mm]
\n{S}\circ\n{g}^{-1}\circ\n{S}^{*}&=&&\begin{pmatrix}1&0\\0&
(\lapse^{-\dagger}\lapse^{-\dagger})^\otimes\end{pmatrix}\n{h}_\Pi^{-1}\circ\signop\circ
\begin{pmatrix}\n{Id}&\mult{\shiftform}\\-\cont{\shift}&\lapseprod^{-2}\n{Id}-\cont{\shift}\circ\mult{\shiftform}\end{pmatrix},\\[3mm]
{\n{S}}^{-*}\circ*_4\circ{\n{S}}^{*}&=
\lapseprod&&\begin{pmatrix}\lapseprod^{-2}\n{Id}-\mult{\shiftform}\circ\cont{\shift}&-\mult{\shiftform}\\\cont{\shift}&\n{Id}\end{pmatrix}\circ
\begin{pmatrix}0&\lapse^{-\dagger}*_\Pi\circ\;\signop\\ \lapse^\dagger*_\Pi&0\end{pmatrix}.
\end{alignat*}
As a consequence of the last equation, 
\[\n{S}^{-*}\df{\kappa}_4=\bigl(0,\lapse\df{\kappa}_\Pi\bigr).\]
\ep
\refcount{kinematics}
%\section{Kinematic parameters of observers}
\bph{Kinematic parameters}\label{akine}
We study the definition of the kinematic parameters in \ref{kine}. The results are also stated in \cite[Eq.~(2.98)]{Jantzen2012}. The acceleration 1-form is given with $\df{\mu}=\n{g}(\vec{u})$ by
\[\tilde{\df{\delta}}\stackrel{\n{def}}{=}\Sigma^*\nabla_\vec{w}\df{\mu}=c_0^{-1}\lapse\Sigma^*\nabla_\vec{u}\df{\mu},\]
where we used \cite[p.~375, 5.]{Fecko}. From \cite[Eq.~(6.23)]{blau}
\begin{align*}(\lie{\vec{u}}\df{\mu})_i&=\vec{u}^j\nabla_j \df{\mu}_i+(\nabla_i \vec{u}^j) \df{\mu}_j\end{align*}
we find with \cite[Eq.~(2.92)]{Jantzen2012}
\begin{align*}\lie{\vec{u}}\df{\mu}&=\nabla_\vec{u}\df{\mu}+\frac{1}{2}\,\nabla\bigl(\mathbf{g}(\vec{u},\vec{u})\bigr)\\
&=\nabla_\vec{u}\df{\mu}.
\end{align*}
It follows that 
\[\tilde{\df{\delta}}=c_0^{-1}\lapse\Sigma^*\lie{\vec{u}}\df{\mu}.\]
To show that $\df{\eta}\stackrel{\n{def}}{=}-\n{Asy}(\Sigma^*\nabla\df{\mu})=\frac{1}{2}\Sigma^*\n{d}\df{\mu}$ we observe that \cite[6.2.5]{Fecko}
\[\n{d}\,\df{\mu}=-2\df{\mu}_{[i,j]},\]
which gives the assertion, with
\[\df{\mu}_{[i,j]}=\df{\mu}_{[i;j]}=\n{Asy}(\nabla\df{\mu}).\]
Lastly, we show that $\df{\lambda}\stackrel{\n{def}}{=}-\n{Sym}(\Sigma^*\nabla\df{\mu})=-\frac{1}{2}\Sigma^*\lie{\vec{u}}\mathbf{g}$. The assertion follows with \cite[Eq.~(6.26)]{blau},
\[\lie{\vec{u}}\mathbf{g}_{ij}=2\,\df{\mu}_{\{i;j\}}=2\,\n{Sym}(\nabla\df{\mu}).\]
\ep
\bph{Group- and time derivative of metric tensor}\label{aderobsmed}
For regular splittings it can be proven that
% Proof can be found 08.09.13, pp.2-6
\begin{align*}
(\lie{\vec{w}}\mathbf{g})(\vec{v},\vec{v}^\prime)\Big|_p=&-(\groupderiv\mathbf{h})(\vec{\atest},\vec{\atest}^\prime)
+\bigl(\groupderiv(\lapse\lapse)^\otimes\bigr)(\tilde{\btest},\tilde{\btest}^\prime)\\
&+\df{\chi}(\vec{\atest})(\lapse\lapse)^\otimes\tilde{\btest}^\prime
+\df{\chi}(\vec{\atest}^\prime)(\lapse\lapse)^\otimes\tilde{\btest}\,\Big|_{\varphi(p)},
\end{align*}
where $(\vec{\atest},\tilde{\btest})=\n{S}\,\vec{v}$, $\vec{v}\in\ves{1}(P)$. The third term in the first line is defined by
\[
\bigl(\groupderiv(\lapse\lapse)^\otimes\bigr)(\tilde{\btest},\tilde{\btest}^\prime)=\groupderiv\bigl(\lapse(\vec{e})\lapse(\vec{e})\bigr)\bigl(\tilde{\btest}(\df{\varepsilon}),\tilde{\btest}^\prime(\df{\varepsilon})\bigr),
\]
where $(\vec{e},\df{\varepsilon})\in\mathfrak{g}\times\mathfrak{g}^*$, $\df{\varepsilon}(\vec{e})=1$. Since vectors $\vec{v}$, $\vec{v}^\prime$ can be chosen arbitrarily, we conclude that for regular splittings the following conditions are equivalent:
\begin{enumerate}
\item The splitting is stationary, $\lie{\vec{w}}\mathbf{g}=0$.
\item The splitting is principal, $\df{\chi} = 0$, with time-independent observer metric and lapse function, $\timederiv\mathbf{h}=\timederiv\lapse^{-1}=0$.
\end{enumerate}
We give the product rule for the Lie derivative of the metric tensor, for $\nu\in C^\infty(P;\mathfrak{g})$,
% Proof can be found 06.09.13, p.1
\[
\lie{\nu\vec{w}}\mathbf{g}=\nu\lie{\vec{w}}\mathbf{g}+2\,\n{Sym}\bigl(\n{d}\nu\otimes\n{g}(\vec{w})\bigr).
\]
For regular splittings,
\[
\Sigma^*\lie{\nu\vec{w}}\mathbf{g}=\Sigma^*\nu\lie{\vec{w}}\mathbf{g}.
\]
Let $\nu=c_0\Pi^*\lapse^{-1}$, hence $\vec{u}=\nu\vec{w}$; compare with \ref{4velo}. With $\btest=\btest^\prime=0$ we obtain
\[
\Sigma^*\lie{\vec{u}}\mathbf{g}=c_0\lapse^{-1}\Sigma^*\lie{\vec{w}}\mathbf{g}=-c_0\lapse^{-1}\groupderiv\mathbf{h}.
\]
As a consequence,
\[
2\df{\lambda}\stackrel{\n{def}}{=}-\Sigma^*\lie{\vec{u}}\mathbf{g}=c_0\lapse^{-1}\groupderiv\mathbf{h}=\partial_\tau\mathbf{h}.
\]
\ep
\bph{Expansion scalar and volume form}\label{aexpansion} 
Given a regular splitting and a twisted unit volume form $\df{\kappa}_3$, we prove that $\partial_\tau\df{\kappa}_3=\lambda\df{\kappa}_3$; see \ref{exptensor}. The results are also stated in \cite[Eq.~(2.99)]{Jantzen2012}. For simplicity, introduce an orientation and regard $\df{\kappa}_3$ as ordinary unit volume form. From $\mathbf{h}^{-1}(\df{\kappa}_3,\df{\kappa}_3)=1$ it follows that
\[
0=\partial_\tau\mathbf{h}^{-1}(\df{\kappa}_3,\df{\kappa}_3)=
\bigl((\partial_\tau\n{h}^{-1})\df{\kappa}_3+2\n{h}^{-1}\partial_\tau\df{\kappa}_3\bigr)\df{\kappa}_3,
\]
hence
\[
\partial_\tau\df{\kappa}_3=-\frac{1}{2}\n{h}(\partial_\tau\n{h}^{-1})\df{\kappa}_3
=\frac{1}{2}(\partial_\tau\n{h})\n{h}^{-1}\df{\kappa}_3
=\frac{1}{2}\n{Tr}(\n{h}^{-1}\partial_\tau\n{h})\df{\kappa}_3.
\]
% Internal: The last equation leverages the symmetry of the tensors $\partial_\tau\n{h})$ and $\n{h}^{-1}$, respectively.
By \ref{aderobsmed} we have $2\df{\lambda}=\partial_\tau\n{h}$ for regular splittings. Therefore,
\[
\partial_\tau\df{\kappa}_3=\n{Tr}(\n{h}^{-1}\df{\lambda})\df{\kappa}_3=\lambda\df{\kappa}_3,
\]
where we used the definition of $\lambda$. Obviously, the result is independent of the chosen orientation, and therefore extends to the twisted unit volume form.
\ep
\refcount{normalization}
%\section{Proxies for Lie-(co)algebra valued fields}
\bph{Splitting of operators in terms of proxies}\label{normsplitop} % Proof can be found 11.08.13, p.3
The splitting of the contraction reads for $\proxy\,\n{S}\,\vec{v}=(\vec{\atest},\df{\btest})$, $\vec{v}\in\ves{k}(P)$,
\begin{align*}(\proxy\circ \n{S})^{-*}\circ\cont{\vec{v}}\circ(\proxy\circ \n{S})^*
=\begin{pmatrix}\cont{\vec{\atest}}\;&\cont{\df{\btest}}\\0\;&\cont{\signop(\vec{\atest})}\end{pmatrix}.\end{align*}
The splitting of the exterior product reads for $(\proxy\,\n{S})^{-*}\,\df{\gamma}=(\df{\alpha},\df{\beta})$, $\df{\gamma}\in\dfs{k}(P)$,
\begin{align*}(\proxy\circ \n{S})^{-*}\circ\mult{\df{\gamma}}\circ(\proxy\circ \n{S})^*
=\begin{pmatrix}\mult{\df{\alpha}}\;&0\\\mult{\df{\beta}}\;&\mult{\signop(\df{\alpha})}\end{pmatrix}.\end{align*}
The splitting of the Lie derivative reads for $\proxy\,\n{S}\,\vec{v}=(\vec{\atest},\btest)$, $\vec{v}\in\ves{1}(P)$,
\begin{align*}(\proxy\circ \n{S})^{-*}\circ\lie{\vec{v}}\circ(\proxy\circ \n{S})^*
=\begin{pmatrix}\n{L}_\vec{\atest}+\btest\,\partial_\tau&\lmult{\n{D}\btest}+
c_0^{-2}\bigl(\btest\,\mult{\df{\bar{\df{\delta}}}}+{\lmult{\cont{\vec{\atest}}2\bar{\df{\eta}}}}\bigr)\\
[\partial_\tau,\cont{\vec{\atest}}]&\n{L}_\vec{\atest}+\partial_\tau\circ\cont{\btest}-c_0^{-2}\lmult{\cont{\vec{\atest}}\bar{\df{\delta}}}
\end{pmatrix},
\end{align*}
{with $\n{L}_\vec{\atest}$ as defined in \ref{splitop} and $(2\bar{\df{\eta}},\bar{\df{\delta}})$ in \ref{obsderiv}, respectively.}
\ep
\bph{Nonregular splitting of metric operators in terms of proxies}\label{anormetop}
Based on \ref{metopnonreg} we find with $\bar{\vec{v}}=c_0\lapse^{-1}\shift$ and $\vdual=c_0^{-1}\lapse\shiftform=c_0^{-2}\n{h}_\Sigma\bar{\vec{v}}$
\begin{alignat*}{2}
(\proxy\circ \n{S})^{-*}\circ\n{g}\circ(\proxy\circ\n{S})^{-1}&=
&&\begin{pmatrix}\n{Id}&-\mult{\vdual}\\\cont{\bar{\vec{v}}}&\lapseprod^{-2}\n{Id}-\cont{\bar{\vec{v}}}\circ\mult{\vdual}\end{pmatrix}\circ
\begin{pmatrix}1&0\\0&
c_0^2\end{pmatrix}\n{h}_\Sigma\circ\signop,\\[3mm]
(\proxy\circ\n{S})\circ\n{g}^{-1}\circ(\proxy\circ \n{S})^{*}&=
\lapseprod^2&&\begin{pmatrix}1&0\\0&
c_0^{-2}\end{pmatrix}\n{h}_\Sigma^{-1}\circ\signop
\circ\begin{pmatrix}\lapseprod^{-2}\n{Id}-\mult{\vdual}\circ\cont{\bar{\vec{v}}}&\mult{\vdual}\\-\cont{\bar{\vec{v}}}&\n{Id}\end{pmatrix},\\[3mm]
(\proxy\circ \n{S})^{-*}\circ *_4\circ(\proxy\circ \n{S})^{*}&=\lapseprod
&&\begin{pmatrix}\n{Id}&-\mult{\vdual}\\\cont{\bar{\vec{v}}}&\lapseprod^{-2}\n{Id}-\cont{\bar{\vec{v}}}\circ\mult{\vdual}\end{pmatrix}\circ
\begin{pmatrix}0&\!\!\!\!\!c_0^{-1}\!*_\Sigma\circ\signop\\ c_0\,*_\Sigma&0\end{pmatrix}.
\end{alignat*} % Proof can be found 13.08.13, p.1, and 21.08.13, p.1
\begin{center}
\begin{tabular}{|c|c|c|c|l|}
\hline
&$\in$&$\pd(\,\cdot\,)$&$\pd(|\cdot|)$&name\\
\hline
$\vec{v},\bar{\vec{v}}$&$\ves{1}(X,\G)$&$\pdim{T}^{-1}$&$\pdim{L}\pdim{T}^{-1}$&proxies of shift vector field\\
$\vdual$&$\dfs{1}(X,\G)$&$\pdim{T}$& $\pdim{L}^{-1}\pdim{T}$&proxy of shift 1-form\\
\hline
\end{tabular}
\end{center}
Based on \ref{ametopthread} we find with $\vec{v}=\lapseprod^2c_0\lapse^{-1}\shift$ and $\vdual=c_0^{-1}\lapse\shiftform=c_0^{-2}\n{h}_\Pi\vec{v}$
\begin{align*}
(\proxy\circ \n{S})^{-*}\circ\n{g}\circ(\proxy\circ\n{S})^{-1}&=
\begin{pmatrix}\n{Id}-\mult{\vdual}\circ\cont{\vec{v}}&-\mult{\vdual}\\\cont{\vec{v}}&\n{Id}\end{pmatrix}\circ
\begin{pmatrix}1&0\\0&
c_0^2\end{pmatrix}\n{h}_\Pi\circ\signop,\\[3mm]
(\proxy\circ\n{S})\circ\n{g}^{-1}\circ(\proxy\circ \n{S})^{*}&=
\begin{pmatrix}1&0\\0&
c_0^{-2}\end{pmatrix}\n{h}_\Pi^{-1}\circ\signop
\circ\begin{pmatrix}\n{Id}&\mult{\vdual}\\-\cont{\vec{v}}&\n{Id}-\cont{\vec{v}}\circ\mult{\vdual}\end{pmatrix},\\[3mm]
(\proxy\circ \n{S})^{-*}\circ *_4\circ(\proxy\circ \n{S})^{*}&=
\begin{pmatrix}\n{Id}-\mult{\vdual}\circ\cont{\vec{v}}&-\mult{\vdual}\\\cont{\vec{v}}&\n{Id}\end{pmatrix}\circ
\begin{pmatrix}0&c_0^{-1}*_\Pi\circ\;\signop\\ c_0\,*_\Pi&0\end{pmatrix}.
\end{align*} % Proof can be found 21.08.13, pp.1-2
The proxy field $\vec{v}$ is the velocity of the Eulerian observer with respect to the Lagrangian observer, up to sign; compare \ref{lageulobs}.
% Proof can be found 19.08.13, p.2
\ep
%
%%%%%%%%%%%%%%%%%%%%%%%%%%%%%%%%%%%%%%%%%%%%%%
\fi %\INCLUDEMETRIC
%%%%%%%%%%%%%%%%%%%%%%%%%%%%%%%%%%%%%%%%%%%%%%
\ifdefined\INCLUDEAPPS
%%%%%%%%%%%%%%%%%%%%%%%%%%%%%%%%%%%%%%%%%%%%%%
\refcount{ehrenfest}
%\section{Ehrenfest paradox}
\bph{Rotating observer's splitting in coordinates}\label{arotating}
In \ref{kinsetting}, Fig.~\ref{fig:rotating}~(a) we described a splitting structure on the world-lines of a rotating observer. The splitting uses the simultaneity structure of a standard stationary splitting of the nonrotating observer; see Fig.~\ref{fig:rotating}~(b).
%Standard stationary splittings are a 1-parameter family.
We introduce coordinates $(t,r,\varphi,z)$ adapted to the splitting structure \ref{kinsetting}, where $\pd(t,r,\varphi,z)=(\oned,\pdim{L},\oned,\pdim{L})$, as follows. By choosing oriented dual bases $(\mathbf{e}^\uparrow,\df{\varepsilon}^\uparrow)$ $\sim$ $(\partial_t,\n{d}t)$ of the Lie algebra we obtain coordinate time $t$; compare with \ref{coordtime}. The helical world-lines are coordinate lines of $t$, the leaves of the foliation are the hyperplanes of constant $t$, see Fig.~\ref{fig:rotating}~(a). Coordinates $(r,\varphi,z)$ are cylindrical coordinates in the base manifold. Let $L=\lapse_0(\partial_t)$, where $\lapse_0$ is the value of $\lapse$ on the axis of rotation. For a dimensionless coordinate time, $L$ is the distance that light travels along the axis of rotation during $\Delta t=1$, and so $\pd(L)=\pdim{L}$.
%$L$ is the parameter by which we can distinguish different splittings in the standard stationary family.
The standard stationary splitting is unique up to a constant factor that scales coordinate time. The factor is fixed by the parameter $L$.\footnote{In the Born chart \cite[Footnote~13]{Rizzi2002}, $t$ measures proper time of the observer at rest along its straight world-lines, $\pd(t)=\mbox{\ssar T}$. We find $L = c_0$ and $\pd(L)=\mbox{\ssar LT}^{-1}$. The same concept is used in the Global Positioning System on the rotating earth, where it is called Earth-Centered Inertial Frame \cite[Sec.~VIII.]{Ashby1994}.} Define $\beta=\omega r/c_0$, where $\omega$ is the angular frequency of the rotating observer, and the Lorentz factor $\gamma=(1-\beta^2)^{-1/2}$. Then it holds that
% Proof can be found 13.11.13, p.1-4. 
\begin{alignat*}{2}
\bigl(g_{\mu\nu}\bigr)&=\begin{pmatrix}\gamma^{-2}L^2&0&-\beta rL&0\\0&-1&0&0\\-\beta rL&0&-r^2&0\\0&0&0&-1\end{pmatrix},\quad\quad
\bigl(h_{ij}\bigr)&=\begin{pmatrix}1&0&0\\0&(\gamma r)^2&0\\0&0&1\end{pmatrix},
\end{alignat*}
\begin{alignat*}{3}
\vec{w}&=\partial_t\otimes\n{d}t,&\quad\quad\df{\omega}&=\left(\n{d}t-\frac{\omega}{c_0L}(\gamma r)^2\,\n{d}\varphi\right)\otimes\partial_t,\\[2mm]
\lapse&=\gamma^{-1}L\otimes\n{d}t,&\quad\quad\df{\Gamma}&=-\frac{\omega}{c_0L}(\gamma r)^2\,\n{d}\varphi\otimes\partial_t,\\[2mm]
\df{\chi}&=0\otimes(\n{d}t\otimes \partial_t),&\quad\quad\tilde{\df{\delta}}&=\beta\gamma\omega L\n{d}r\otimes\n{d}t\\[1mm]
&&&=c_0^{-1}\lapse\gamma^2\omega^2r\,\n{d}r,\\[2mm]
% This result agrees with what can be found under "relativistic centripetal acceleration" in the web.
\df{\Omega}&=-2\beta{\gamma^4}L^{-1}\,\n{d}r\wedge\n{d}\varphi\otimes\partial_t&\quad\quad
\df{\eta}&=-\gamma^3\omega r\,\n{d}r\wedge\n{d}\varphi\\
&=-\frac{2}{c_0\lapse}\gamma^3\omega r\,\n{d}r\wedge\n{d}\varphi,&\quad\quad&=-\gamma^2\omega*_3\n{d}z;
\intertext{compare with \cite[Appendix~A]{Rizzi2002}. The quantities related to the axial splitting \ref{axialsplit} have the following representation in the chart:}
\mathring{\vec{w}}&=\partial_\varphi\otimes\n{d}\varphi,&\quad\quad\mathring{\lapse}&=\gamma r\otimes\n{d}\varphi,\\[2mm]
\bar{\lapse}&=\lapse,&\quad\quad\Lambda&=\frac{r}{L}\otimes(\n{d}\varphi\otimes\partial_t),\\[2mm]
\bar{\df{\Gamma}}&=-\frac{\omega}{c_0L}(\gamma r)^2(\n{d}\varphi\otimes\partial_t)
&\quad\quad\bar{\df{\Omega}}&=2\beta\gamma^4L^{-1}\n{d}r\otimes(\n{d}\varphi\otimes\partial_t)\\
&=-\beta\gamma^2\Lambda,
&\quad\quad&=\n{d}(\beta\gamma^2\Lambda)=-\n{d}\bar{\df{\Gamma}}.
\end{alignat*}
\ep
\refcount{schiff}
%\section{Schiff's ``Question in General Relativity"}
\bph{Excitation terms for Schiff's rotating observer}\label{aschiffsolv}
We calculate the excitation terms of Schiff's rotating observer, used in \ref{schiffsolv}. Denote by $(\underline{t},\underline{r},\underline{\varphi},\underline{z})$ coordinates adapted to the standard stationary splitting of Schiff's observer at rest; see Fig.~\ref{fig:rotating}~(b). The coordinates are related to the coordinates $(t,r,\varphi,z)$ introduced in \ref{arotating} by the following transformations:
\begin{alignat*}{3}
t&=\underline{t},&\quad\quad r&=\underline{r},\\
\varphi&=\underline{\varphi}-\frac{\omega L}{c_0}\underline{t},&\quad\quad z&=\underline{z},
\intertext{and}
\n{d}t&=\n{d}\underline{t},&\quad\quad\n{d}r&=\n{d}\underline{r},\\
\n{d}\varphi&=\n{d}\underline{\varphi}-\frac{\omega L}{c_0}\n{d}\underline{t},&\quad\quad\n{d}z&=\n{d}\underline{z}.
\end{alignat*}
A charge distribution $\underline{\rho}$ that is static with respect to Schiff's observer at rest gives rise to an electric charge-current in space-time of the form
\begin{align*}
J&=\rho_{r\varphi z}\,\n{d}\underline{r}\wedge\n{d}\underline{\varphi}\wedge\n{d}\underline{z}\\
&=\rho_{r\varphi z}\,\n{d}r\wedge\bigl(\n{d}\varphi+\frac{\omega L}{c_0}\n{d}t\bigr)\wedge\n{d}z.
\end{align*}
The connection 1-form $\df{\omega}$ in \ref{arotating} induces a parametric map $\Sigma$ and its dual,
\[
\Sigma:\left\{\begin{aligned}
\partial_r&\mapsto \partial_r\\
\partial_\varphi&\mapsto \partial_\varphi+\frac{r}{L}\beta\gamma^2\partial_t\\
\partial_z&\mapsto \partial_z
\end{aligned}\right.\,,\qquad
\Sigma^*:\left\{\begin{aligned}
\n{d}t&\mapsto\frac{r}{L}\beta\gamma^2\n{d}\varphi\\
\n{d}r&\mapsto \n{d}r\\
\n{d}\varphi&\mapsto \n{d}\varphi\\
\n{d}z&\mapsto \n{d}z
\end{aligned}\right.\,.
\]
For the relativistic splitting of the charge-current, $\n{S}^*J=(\Sigma^*J,\Sigma^*\cont{\vec{w}}J)=(\rho,-\tilde{\jmath})$ we obtain
\begin{align*}
\rho&=\Sigma^*J\\
&=\gamma^2\rho_{r\varphi z}\,\n{d}r\wedge\n{d}\varphi\wedge\n{d}z\\
&=\gamma^2\rho_0,
\intertext{where we set}
\rho_0&=\rho_{r\varphi z}\,\n{d}r\wedge\n{d}\varphi\wedge\n{d}z,
\intertext{independent of $\omega$, and}
\tilde{\jmath}&=-\Sigma^*\cont{\vec{w}}J\\
&=\beta\frac{L}{r}\rho_{r\varphi z}\,\n{d}r\wedge\n{d}z\otimes\n{d}t\\
&=-\beta\Lambda^{-1}\cont{\mathring{\vec{w}}}\rho_{r\varphi z}\,\n{d}r\wedge\n{d}\varphi\wedge\n{d}z\\
&=-\beta\Lambda^{-1}\cont{\mathring{\vec{w}}}\rho_0.
\end{align*}
After dimensional reduction by axial splitting this yields
\[
(\bar{\rho},\bar{\jmath})=(\gamma^2,-\beta\Lambda^{-1})\bar{\rho}_0.
\]
\ep
%

%
%%%%%%%%%%%%%%%%%%%%%%%%%%%%%%%%%%%%%%%%%%%%%%
\fi %\INCLUDEAPPS
%%%%%%%%%%%%%%%%%%%%%%%%%%%%%%%%%%%%%%%%%%%%%%
\clearpage
\setcounter{section}{1}
\chapter{Notations}
\begin{small}
\begin{longtable}[l]{@{}p{20mm}p{74mm}r@{}}
% Latin Letters
$a$ & magnetic vector potential & \ref{max}\\
$\vec{a},\vec{b},\vec{c}$ & generic (multi-)vector fields or differential forms\\
$\vec{a}$ & element of the Lie algebra, $\vec{a}\in\mathfrak{g}$ & \ref{MC}\\
$A$ & electromagnetic potential & \ref{max}\\
$\pdim{A},\pdim{B},\pdim{D}$ & generic physical dimensions & \ref{pdsyst}\\
$\pdim{A}$ & physical dimension action & \ref{pdsyst}, \ref{EM1}\\
$\Xspace,\Yspace,\Zspace$ & generic spaces&\ref{algebra}\\
$\surf$ & two-dimensional compact subdomain of the base manifold, $\surf\subset X$ & \ref{intquant}\\
$\n{Ad}$ & adjoint representation & \ref{principalconnect}\\
$\n{Asy}$ & antisymmetric part of a tensor & \ref{kine}\\
$b$ & magnetic flux density & \ref{max}\\
$B$ & basis of physical dimensions system & \ref{pdsyst}\\
$B_3$, $B_4$ & orthonormal frame fields in $X$, $P$ & \ref{dhodge}\\
$c_0$ & vacuum speed of light & \ref{4velo}\\
$C_{\mu\nu}{}^\kappa$ & object of anholonomity & \ref{amoreadapt}\\
$C^\infty(M)$ & smooth functions on $M$&\\
$\curv$ & one-dimensional compact subdomain of the base manifold, $\curv\subset X$ & \ref{intquant}\\
$d$ & electric flux density & \ref{max}\\
$\n{d}$ & exterior derivative&\\
$\n{div}$ & divergence (in four dimensions) & \ref{exptensor}\\
$D$ & system of physical dimensions & \ref{pdsyst}\\
$\n{D}$ & exterior covariant derivative in the base manifold & \ref{covderiv}\\
$\n{Diff}(\,\cdot\,)$ & diffeomorphism group & \ref{transfunc}\\
$e$ & (proxy of) electric field strength & \ref{max}\\
$e$ & group identity element & \ref{Gstructure}\\
$\vec{e}$ & basis element of the Lie algebra, $\vec{e}\in\mathfrak{g}$&\\
$(\vec{e}_\mu,\df{\varepsilon}^\nu)$ &  (anholonomic) dual frames / frame fields in $P$ & \ref{amoreadapt}\\
$\mult{}$ & exterior product / multiplication & \ref{intext}\\
$\n{End}(\,\cdot\,)$ & set of endomorphisms & \ref{genexp}\\
$f$ & (proxy of) force density & \ref{momentumbalance}, \ref{momentumbalanceproxy}\\
$f(\cdot,\cdot)$ & bilinear operation & \ref{algebra}\\
$f$ & $\mathfrak{g}^*\otimes\mathfrak{g}$-valued scalar function & \ref{frobenius}\\
$F$ & electromagnetic field & \ref{max}\\
$F,F_x$ & typical fiber, typical fiber at $x$ & \ref{fiberbundle}\\
$\dfs{k}(M)$ & smooth differential $k$-forms on $M$ & \ref{fieldsforms}\\
$\dfs{k}(M;V)$ & smooth differential $k$-forms on $M$ that take values in $V$ & \ref{fieldsforms}\\
$\dfs{}(M)$ & direct sum over $\dfs{k}(M)$ & \ref{fieldsforms}\\
$\dfs{k}(X,\G)$ & $\G$-parametric differential forms on $X$ & \ref{paramfields}\\
$g,h$ & group elements, $g,h\in\group$ / instants & \ref{Gstructure}\\
$g_{ij}$ & transition function of a $\group$-bundle & \ref{Gstructure}\\
$\mathbf{g}$ & Lorentzian metric tensor field & \ref{metricprelim}\\
$\n{g}$ & Riesz operator associated with $\mathbf{g}$ & \ref{metricprelim}\\
$g_{\mu\nu},g^{\mu\nu}$ & covariant, contravariant components of the metric tensor & \ref{twometrics}\\
$|g|$ & modulus of determinant $\det(g_{\mu\nu})$ & \ref{componenthodge}\\
$\mathfrak{g}$ & Lie algebra of $\group$ & \ref{liealgebra}\\
$\group,\group_x$ & Lie group / typical fiber, typical fiber at $x$ & \ref{Gstructure}\\
$h$ & (proxy of) magnetic field strength & \ref{max}\\
$\mathbf{h}$ & observer metric tensor field & \ref{relspace}\\
$\n{h}$ & Riesz operator associated with $\mathbf{h}$ & \ref{relspace}\\
$\n{hor},\n{ver}$ & horizontal, vertical map & \ref{par:verhor}\\
$H$ & electromagnetic excitation  & \ref{max}\\
$H_p$ & horizontal space at $p$ & \ref{ehresconnonprinc}\\
$H$ & index for spaces of horizontal (multi-)vector fields and differential forms & \ref{horfields}\\
$\interval$ & compact domain in the Lie group, $\interval\subset\group$ & \ref{integralgroup}, \ref{intquant}\\
$i,j,k ,\ldots$ & indices, latin tensor indices $\in\{1,2,3\}$\\
$\cont{}$ & interior product / contraction & \ref{intext}\\
$I$ & electric current & \ref{intquant}\\
$\pdim{I}$ & physical dimension electric current & \ref{pdsyst}\\
$\n{Id}$ & identity map\\
$\n{Im}$ & image space\\
$j$ & (proxy of) electric current density & \ref{max}\\
$J$ & electric charge current & \ref{max}\\
$k,\ell$ & degree of a multi-vector field or differential form\\
$(\vec{\atest},\tilde{\btest}),(\vec{\atest},\tilde{\df{\btest}})$ & (multi-)vector fields in the image of the splitting map $\n{S}$ & \ref{psipush}\\
$(\vec{\atest},\btest),(\vec{\atest},\df{\btest})$ & (multi-)vector fields in the image of the proxy map $\proxy$ & \ref{proxymap}\\
$\vec{k}$ & multivector, dual to a volume form $\df{\kappa}$ & \ref{genexp}\\
$\n{Ker}$ & kernel space\\
$\ell$ & left action of the Lie group on the typical fiber & \ref{Gstructure}\\
$L,\tilde{l}$ & Lagrangian of the electromagnetic field in four and three dimensions & \ref{max}\\
$L$ & characteristic distance along the axis of rotation & \ref{arotating}\\
$\pdim{L}$ & physical dimension length & \ref{pdsyst}\\
$L_g$ & left group action / translation & \ref{liealgebra}\\
$\n{L}_\vec{\atest}$ & operator that occurs in the splitting of the Lie derivative & \ref{splitop}\\
$\n{Lin}(\cdot,\cdot)$ & vector space of linear maps & \ref{translie}\\
$m$ & (proxy of) momentum flux density / Maxwell stress tensor & \ref{EM1}, \ref{EM1proxy}\\
\ifdefined\INCLUDEAPPS
$m$ & vacuum magnetization & \ref{schifftreatment}\\
\fi
$M$ & smooth differentiable manifold & \ref{fieldsforms}\\
$\pdim{M}$ & physical dimension mass & \ref{pdsyst}\\
$n$ & dimension of a manifold & \ref{fieldsforms}\\
$\vec{n}$ & vector field in energy-momentum balance & \ref{EM1}\\
$\signop$ & sign operator & \ref{signop}\\
$\lapse$ & lapse function & \ref{lapsefun}\\
$\shift$ & shift vector field & \ref{shift}\\
$p$ & momentum density & \ref{EM1}, \ref{EM1proxy}\\
\ifdefined\INCLUDEAPPS
$p$ & vacuum polarization & \ref{schifftreatment}\\
\fi
$p$ & point/event in $P$ & \ref{fiberbundle}\\
$\pd$ & physical-dimension map & \ref{pdsyst}\\
$\n{pr}$ & canonical projection & \ref{fiberbundle}\\
$P$ & bundle manifold & \ref{fiberbundle}\\
$\proxy$ & proxy map & \ref{proxymap}\\
$q$ & exponent in the physical dimension system & \ref{pdsyst}\\
$q$ & dimension of $F,G$ & \ref{fiberbundle}\\
$Q$ & electric charge & \ref{intquant}\\
$r$ & (proxy of) power density & \ref{momentumbalance}, \ref{momentumbalanceproxy}\\
$r$ & rank of the physical dimension system & \ref{pdsyst}\\
$r$ & principal right action & \ref{principaction}\\
$(r,\varphi,z)$ & cylindrical coordinates in the base manifold & \ref{axialsplit}, \ref{arotating}\\
$R$ & electromagnetic four-force density & \ref{momentumbalance}\\
$R$ & ring & \ref{pdsyst}\\
$R_g$ & right group action & \ref{principaction}\\
$R$ & radius of problem domain & \ref{schiffsolv}\\
$s$ & (proxy of) energy flux density / Poynting form & \ref{EM1}, \ref{EM1proxy}\\
$s$	& section & \ref{section}\\
$\vec{s},\vec{t}$ & tensors, $\vec{s},\vec{t}\in T$& \ref{algebra}\\
$S$ & sets of physical quantities & \ref{pdsyst}\\
$S_g$ & leaf of a foliation, labeled by $g$ & \ref{foliation}\\
$\n{S}$ & splitting map & \ref{psipush}\\
$\n{Sym}$ & symmetric part of a tensor & \ref{kine}\\
$t$ & coordinate time & \ref{coordtime}\\
$\vec{t}$ & unit tensor for the pair $(\mathfrak{g},\mathfrak{g}^*)$ & \ref{MC2}\\
$T$ & electromagnetic energy-momentum tensor & \ref{EM1}\\
$T$ & space of $\binom{1}{1}$-tensors\\
$T_pP$ & tangent space at $p\in P$\\
$\pdim{T}$ & physical dimension time & \ref{pdsyst}\\
$\n{Tr}$ & trace of a tensor\\
$\vec{u}$ & four-velocity & \ref{4velo}\\
$\vec{u}$ & (multi-)vector, (multi-)vector field\\
$\mathfrak{u}$ & Lie algebra of $U$ & \ref{axialsplit}\\
$U$ & electric voltage & \ref{intquant}\\
$U$ & open neighbourhood of $x\in X$ & \ref{fiberbundle}\\
$U$ & unitary group & \ref{axialsplit}\\
$\pdim{U}$ & physical dimension voltage & \ref{pdsyst}\\
$\vec{v}$ & (multi-)vector, (multi-)vector field\\
$\hat{\vec{v}}$ & element of the Lie algebra, $\hat{\vec{v}}\in\mathfrak{g}$ & \ref{canonisomorphism}\\
$\check{\vec{v}}$ & left invariant vector field associated with $\hat{\vec{v}}$  & \ref{canonisomorphism}\\
$\bar{\vec{v}}$ & parametric (multi-)vector field, related to $\vec{v}\in\ves{}(P)$ & \ref{paramaps}\\
$\vec{v},\bar{\vec{v}}$ & proxies of the shift vector field & \ref{anormetop}\\
$V$ & magnetic voltage & \ref{intquant}\\
$V,V^\prime$ & finite-dimensional vector spaces & \ref{algebra}\\
$V_p$ & vertical space at $p$ & \ref{verfields}\\
$V$ & index for spaces of vertical (multi-)vector fields and differential forms & \ref{verfields}\\
$\vol$ & three-dimensional compact subdomain of the base manifold, $\vol\subset X$ & \ref{intquant}\\
$w$ & (proxy of) energy density & \ref{EM1}, \ref{EM1proxy}\\
$\vec{w}$ & fundamental field & \ref{MC2b}\\
$\wMC$ & Maurer-Cartan vector field & \ref{MC2}\\
$x,y$ & generic physical quantities & \ref{algstrucphys}\\
$x$ & point in $X$ & \ref{fiberbundle}\\
$(x^i),(x^\mu)$ &  coordinates in $X,P$ & \ref{coordtime}, \ref{amoreadapt}\\
$X$ & base manifold & \ref{fiberbundle}\\
$X$ & body-force term in the energy-momentum balance & \ref{momentumbalance}\\
$Y$ & two-dimensional open half-space (meridian plane) & \ref{axialsplit}\\
$Z_0$ & vacuum impedance & \ref{const}\\
\\
% Greek Letters
$\alpha,\beta$ & indices\\
$\df{\alpha},\df{\beta}$ & differential forms\\
$\df{\alpha}$ & element of the Lie co-algebra, $\df{\alpha}\in\mathfrak{g}^*$ & \ref{MC2}\\
$(\df{\alpha},\tilde{\df{\beta}})$ & differential forms in the image of the splitting map $\n{S}^{-*}$ & \ref{psipush}\\
$(\df{\alpha},\df{\beta})$ & differential forms in the image of the proxy map $\proxy^{-*}$ & \ref{proxymap}\\
$\beta$, $\hat{\beta}$ & (maximum) normalized velocity & \makebox[0mm][r]{\ref{axialsplit}, \ref{schiffsolv}, \ref{arotating}}\\
$\gamma$ & Lorentz factor & \ref{axialsplit}, \ref{arotating}\\
$\df{\gamma}$ & covector, differential form\\
$\df{\gamma}$ & $\binom{0}{2}$-tensor field in the definition of kinematic para\-meters & \ref{kine}\\
$\df{\gamma}$ & proxy of the Christoffel form & \ref{obsderiv}\\
$\df{\Gamma}$ & Christoffel form & \ref{gaugepot}\\
$\Gamma_i$ & components of $\df{\Gamma}$ & \ref{coordtime}\\
$\df{\delta}$ & (proxy of) acceleration form & \ref{kine}, \ref{obsderiv}\\
$\hat{\epsilon}_{\kappa_1\cdots\kappa_n}$ & Levi-Civita permutation symbol, in $n$ dimensions & \ref{componenthodge}\\
$\varepsilon_0$ & electric constant (permittivity of vacuum) & \ref{const}\\
$\df{\varepsilon}$ & basis element of the Lie co-algebra, $\df{\varepsilon}\in\mathfrak{g}^*$\\
$\df{\varepsilon}^{\mu_1\cdots\mu_k}$ & basis element for differential $k$-forms & \ref{amoreadapt}\\
$\zeta$ & fundamental field map & \ref{canonisomorphism}\\
$\df{\eta}$ & vorticity form & \ref{kine}, \ref{obsderiv}\\
$\df{\vartheta}$ & twice the expansion tensor & \ref{implic}\\
$\df{\theta}$ & connection 1-form of chart-associated connection & \ref{canonconnec}\\
%$(\df{\theta}^i)$ & coframe basis & \ref{classobs}\\
$\df{\Theta}$& $\binom{1}{2}$-tensor field & \ref{genexp}\\
$\bar{\df{\Theta}}$, $\Theta$ & dual of $\df{\Theta}$, trace of $\df{\Theta}$ & \ref{genexp}\\
$\MC$ & Maurer-Cartan form & \ref{MC}\\
$\df{\kappa}$ & volume form & \ref{twisted}\\
$\df{\kappa}_3$ & twisted unit spatial volume form & \ref{dhodge}\\
$\df{\kappa}_4$ & twisted unit space-time volume form & \ref{dhodge}\\
$\lambda$ & scalar, smooth function\\
$\lambda$ & expansion scalar & \ref{exptensor}\\
$\df{\lambda}$ & expansion tensor & \ref{kine}\\
$\Lambda$ & combined lapse function in the axial splitting & \ref{axialsplit}, \ref{arotating}\\
$\mathsf{\Lambda}^k$ & $k$-th exterior power\\
$\mu,\nu,\ldots$ & greek tensor indices, $\in\{0,1,2,3\}$ & \ref{amoreadapt}, \ref{arotating}\\
$\mu_0$ & magnetic constant (permeability of vacuum) & \ref{const}\\
$\df{\mu}$ & metric Riesz dual of the four-velocity & \ref{4velo}\\
$\nu$ & $\mathfrak{g}$-valued scalar function & \ref{frobenius}\\
$\shiftform$ & shift 1-form & \ref{shift}\\
$\xi$ & scalar function to characterize nonregular splittings & \ref{reciproc}\\
$\pi$ & projection & \ref{fiberbundle}\\
$\Pi$ & parametric map, induced by $\pi$ and $\varphi$ & \ref{paramaps}\\
$\Pi$ & index for fiber-induced objects, like in $\mathbf{h}_\Pi$ & \ref{twometrics}\\
$\rho$ & electric charge density & \ref{max}\\
$\sigma_p$ & horizontal lift & \ref{ehresconnonprinc}\\
$\df{\sigma}$ & shear tensor & \ref{exptensor}\\
$\Sigma$ & parametric map, induced by $\sigma_p$ and $\varphi$ & \ref{paramaps}\\
$\Sigma$ & index for connection-induced objects, like in $\mathbf{h}_\Sigma$ & \ref{twometrics}\\
$\tau$ & proper time & \ref{propertime}\\
$\df{\upsilon}$ & proxy of the shift 1-form & \ref{anormetop}\\
$\Upsilon$ & connection form of an Ehresmann connection & \ref{ehresconnonprinc}\\
$\varphi$ & (proxy of) electric potential & \ref{max}\\
$\varphi$ & diffeomorphism & \ref{extend}\\
$\varphi=(\pi,\hhat{\varphi})$ & fiber chart & \ref{fiberbundle}\\
$\hhat{\varphi}_x$ & restriction of $\hhat{\varphi}$ to $F_x$ & \ref{fiberbundle}\\
$\varphi_{ij}$ & transition function & \ref{fiberbundle}\\
$\check{\varphi}_{ij}$ & field of invertible linear maps $\mathfrak{g}\to\mathfrak{g}$ induced by $\varphi_{ij}$ & \ref{translie}\\
$\phi$ & adapted chart on $\G$ & \ref{adaptchart}\\
$\tilde{\phi}$ & coordinate chart on $P$ & \ref{amoreadapt}\\
$\phi_p$ & horizontal lift of chart-associated connection & \ref{canonconnec}\\
$\Phi$ & magnetic flux & \ref{intquant}\\
$\Phi$ & parametric map, induced by $\phi_p$ and $\varphi$ & \ref{canonconnec}\\
$\Phi_{ij}^*$ & transition map for parametric fields in the base manifold, induced by $\varphi_{ij}$ & \ref{pulliecoalg}\\
$\ves{k}(M)$ & smooth $k$-vector fields on $M$ & \ref{fieldsforms}\\
$\ves{k}(M;V)$ & smooth $k$-vector fields on $M$ that take values in $V$ & \ref{fieldsforms}\\
$\ves{}(M)$ & direct sum over $\ves{k}(M)$ & \ref{fieldsforms}\\
$\ves{k}(X,\G)$ & $\G$-parametric $k$-vector fields on $X$ & \ref{paramfields}\\
$\df{\chi}$ & variance of the Ehresmann connection & \ref{variance}\\
$\psi$ & chart on $X$ & \ref{amoreadapt}\\
$\Psi$ & electric flux & \ref{intquant}\\
$\affineform$ & affine part in the transformation of the Christoffel form & \ref{canonshift}\\
$\omega$ & angular frequency of the rotating observer & \ref{axialsplit}, \ref{arotating}\\
$\df{\omega}$ & connection 1-form & \ref{ehresconnonprinc}\\
$\df{\Omega}$ & curvature of the Ehresmann connection & \ref{curvature}\\
\\
% Miscellaneous
$\oned$ & neutral element of the physical dimensions system & \ref{pdsyst}\\
$*$ & index for dual space, like in $V^*$\\
$\prime$ & index for push map, like in $\varphi^\prime$ & \ref{extend}\\
$*$ & index for pull map, like in $\varphi^*$ & \ref{extend}\\
$\circ$ & concatenation of operators\\
$[\,\cdot\,,\,\cdot\,]$ & commutator of operators\\
$\to$ & map\\
$\hookrightarrow$ & injective map\\
$\xrightarrow{\,\,\sim\,\,}$ & bijective map\\
$\oplus$ & direct sum\\
$\wedge$ & exterior product\\
$\otimes$ & tensor product\\
$\times$ & Cartesian product\\
$\twist$ & index for twisted objects & \ref{twisted}\\
$\prototimetwist$ & index for objects that are twisted with respect to $\G$ & \ref{gtwisted}, \ref{metricprelim}\\
$\protodoubletwist$ & index for objects that are twisted with respect to $X$ and $\G$ & \ref{gtwisted}\\
$\widetilde{\rule{1ex}{0ex}}$ & indicates Lie-coalgebra-valued objects, like in $\tilde{\df{\kappa}}$ & \ref{psipush}\\
$\sim$ & equivalence relation\\
$\lie{}$ & Lie derivative\\
$\groupderiv$, $\partial_U$ & group derivatives & \ref{groupderiv}, \ref{axialsplit}\\
$\specialwedge$ & notation used for operators like $(\df{\Gamma}\specialwedge\groupderiv)$ & \ref{covderiv}\\
$\partial$ & partial derivative, with respect to a coordinate\\
$\partial_\mu$ & directional derivative along basis vector $\vec{e}_\mu$ & \ref{amoreadapt}\\
$\partial_\tau$ & time derivative, with respect to proper time & \ref{propertime}\\
$\nabla$ & covariant derivative of the Levi-Civita connection on space-time & \ref{kine}\\
$|\,\cdot\,|$ & pointwise norm of (multi-)vector fields and differential forms & \ref{metricprelim}\\
$*_2$ & Hodge operator in two dimensions & \ref{axialsplit}\\
$*_3$ & parametric Hodge operator in three dimensions & \ref{dhodge}\\
$*_4$ & Hodge operator in four dimensions & \ref{dhodge}\\
$\dagger$ & index for reciprocal quantities & \ref{reciproc}\\
$\uparrow$ & indicates positive time orientation, like in $\vec{e}^\uparrow$ & \ref{timeorient}\\
$\mathring{\rule{1ex}{0ex}}$ & indicates objects that are defined with respect to the axial splitting, like in $\mathring{\pi}$ & \ref{axialsplit}\\
$\bar{\rule{1ex}{0ex}}$ & indicates objects that are defined in the meridian plane & \ref{axialsplit}\\
$[\,\,\,]$ & symmetrization of tensor indices & \ref{amoreadapt}, \ref{akine}\\
$\{\,\,\}$ & anti-symmetrization of tensor indices & \ref{amoreadapt}, \ref{akine}\\
$,j$ & partial derivative with respect to tensor index $j$ & \ref{akine}\\
$;j$ & covariant derivative with respect to tensor index $j$ & \ref{akine}\\
\end{longtable}\end{small}\clearpage
%%%%%%%%%%%%%%%%%%%%%%%%%%%%%%%%%%%%%%%%%%%%%%
%\bibliographystyle{plain}
%\bibliography{lit1000,local,papers}
%\end{document}
%
% New references, as of 03.05.2014:
% carlson, schouten, fleischmann, gourgbook, Hehl2005a, iso, Matolcsi1998, sommerfeld, tontibook, wallot
%

\end{document}